\documentclass[aps,pre,preprint,superscriptaddress]{revtex4-1}

\usepackage{subfigure}
\usepackage{amsmath}

\usepackage{tikz}
\usetikzlibrary{positioning,calc}
\usetikzlibrary{arrows,shapes}
\tikzstyle{block} = [rectangle,rounded corners,thin,align=center,fill=green!20,draw=black!20]
\tikzstyle{line} = [-latex]

\begin{document}

\title{An investigation of non-equilibrium heat transport in a gas system under external force field}

\author{Tianbai Xiao}
\email[Email:]{xiaotianbai@pku.edu.cn}
\affiliation{Department of Mechanics and Engineering Science, College of Engineering, Peking University, Beijing 100871, China}

\author{Kun Xu}
\email[Email:]{makxu@ust.hk}
\affiliation{Department of Mathematics, Hong Kong University of Science and Technology, Hong Kong}
\affiliation{Department of Mechanical and Aerospace Engineering, Hong Kong University of Science and Technology, Hong Kong}

\author{Qingdong Cai}
\email[Email:]{caiqd@pku.edu.cn}
\affiliation{Department of Mechanics and Engineering Science, College of Engineering, Peking University, Beijing 100871, China}

\author{Tiezheng Qian}
\email[Email:]{maqian@ust.hk}
\affiliation{Department of Mathematics, Hong Kong University of Science and Technology, Hong Kong}
	
\date{\today}
	
	\begin{abstract}
		The gas dynamics under external force field is essentially associated with multiple scale nature due to the large variations of density and local Knudsen number.
		Single scale fluid dynamic equations, such as the Boltzmann and Navier-Stokes equations, are valid in their respective modeling scales,
		and it is challenging for the modeling and computation of a multiple scale problem across different regimes and capture the corresponding non-equilibrium flow physics.
		Based on the direct modeling of conservation laws in the discretized space, 
		a well-balanced unified gas-kinetic scheme (UGKS) for multiscale flow transport under external force field has been developed and is used in the current study of non-equilibrium gaseous flow under external force field.
		With the variation of modeling scale, i.e., the cell size and time step, the UGKS is able to recover cross-scale flow physics from particle transport to hydrodynamic wave propagation.
		Theoretical analysis based on the kinetic model equation is presented to conceptually illustrate the effects of external force on the non-equilibrium heat transport.
		The heat conduction problem in the near-equilibrium regime due to the external forcing term is quantitatively investigated.
		In the lid-driven cavity flow study, the stratified flow is observed under external force field.
		With the increment of external force,
		the flow topological structure changes dramatically, and the temperature gradient, shearing stress, and external force play
		different roles in the determination of the heat flux in different layers corresponding to different flow regimes.
		As a typical non-Fourier's heat effect in the transition regime, 
		the additional external force enhances the heat flux significantly along the forcing direction, 
		and the relationship $\Delta q\propto \nabla \Phi$, where $q$ is the heat flux and $\Phi$ is the external force potential, is fully confirmed in the flow regimes with non-vanishing effect of particle mean free path.
		Through the numerical experiment, it is clear that the external force plays an important role
		in the dynamic process of non-equilibrium flow transport and heat transfer.
	\end{abstract}
	
	
	\maketitle
	
\section{Introduction}

The gas dynamics under external force field is usually associated with multiple scale nature due to the possible large variation of gas density and local Knudsen number along the direction of force.
If we look into the flow field downwards to the mesoscopic level, the kinetic theory can be employed to illustrate the physical effect of external force.
In the kinetic scale, the Boltzmann equation follows the evolution of velocity distribution function $f$ to describe the particle transport and collision assembly \cite{chapman1970mathematical,vincenti1965introduction}.
With an external acceleration $\phi_i$ acting on the particles, the evolving process of $f$ is modeled in Eq.(\ref{Boltzmann equation}) with separate operators: the free flight of the particles (left hand terms) and their collisions (right hand term), i.e.,
\begin{equation}
\frac{\partial f}{\partial t} + u_i \frac{\partial f}{\partial x_i} + \phi_i \frac{\partial f}{\partial u_i} = Q(f).
\label{Boltzmann equation}
\end{equation}
Here $u_i$ is the particle velocity and $Q(f)$ is the collision term.
For a real gas dynamic system, even with the initial Maxwellian of a  barometric distribution under external force, the free transport of particles between two successive collisions always evolves the system towards non-equilibrium state.
Under external force field, the particle acceleration or deceleration process during this time interval results in a distortion of the distribution function in the velocity space.
The deviation from equilibrium distribution is restricted by the particle collision time $\tau$.
On the other hand, the particle collision takes effect to drive the system back to equilibrium state.
In the continuum limit, the deviation from equilibrium is weak due to intensive intermolecular collisions, and thus the non-equilibrium transport is well described with viscosity and heat conductivity in the constitutive relationship.
However, in rarefied regime, the particle free transport and collision are loosely coupled due to a large particle collision time.
Much complicated nonlinear dynamics due to external force can emerge and present a peculiar non-equilibrium flow behavior.
Usually the strong non-equilibrium effects are expected in the highly dissipative regions,
such as the shock and boundary layers. However, with the existence of external force field
non-equilibrium gas evolution may spread to the whole flow system in a  large scale, such as the gravitational system, where the gravity will result in an observable variance of density, so is the variation of the particle mean free path and the local Knudsen number.

The study of gas dynamics is mostly based on the governing equations constructed on different modeling scales, such as the Boltzmann and Navier-Stokes equations.
The Boltzmann equation is defined on the particle mean free path and collision time, i.e., the kinetic scale.
On such a modeling scale, the particle transport and collision can be separately formulated in Eq.(\ref{Boltzmann equation}).
The application of the Boltzmann equation to other scales is to resolve other scales all the way to the mean free path and particle collision time.
Based on the Fourier's law and Newton's stress and strain relationship, the Navier-Stokes-Fourier (NSF) equations are constructed to describe the fluid motion and heat transfer on macroscopic scale.
The fluid element picture is used in the NS modeling and the intensive collisions prevent particle penetration between adjacent fluid elements. 
Although the modeling of the NS equations is obtained from the first physical principle of conservation laws, 
the scale for the validity of the NS equations, i.e., the quantitative description of the fluid element, is not clearly defined, even though it always refers to the hydrodynamic one. 
The domain for the validity of NS equations is unclear, which raises question about its applicability beyond continuum flow regime.

Physically, the valid application of distinguishable gas dynamic equations, such as the NS and Boltzmann individually,
depends on the clear scale separation.
However, for a system under external force field,
the flow physics may vary continuously from the kinetic Boltzmann modeling in the upper rarefied layer
to the hydrodynamic one in the lower dense region.
With the variation of characteristic scale, there should exist a continuous spectrum of dynamics between these two limits. 
The multiple scale governing equation with flexible degrees of freedom is in need to capture the scale-dependent flow physics from the kinetic to the hydrodynamic ones.
It is well known that with a proper normalization of time and space scales, the NS equations can be derived from the Boltzmann equation through the Chapman-Enskog expansion.
In recognition of this, starting from the Boltzmann equation,
many efforts have been devoted to develop the coarse-graining technique to extend the Boltzmann equation to other scales.
However, the mathematical derivation of equations beyond the NS equations is not very successful without specifying the modeling scale physically.
Which scale should be used in the modeling between the above two limits still remains an open problem.

For conventional research of gas dynamics, the modeling and computation are handled separately.
Once the governing equations are constructed, the CFD method serves to get the numerical solution of differential equations.
The Boltzmann equation can be numerically solved through the Direct Simulation Monte Carlo (DSMC) method \cite{bird1994molecular} or the direct Boltzmann solver \cite{cercignani1988boltzmann, cercignani2000rarefied},
and a series of Riemann solvers are used for macroscopic gas dynamic equations.
Without multiple scale governing equations, it becomes difficult to capture multiple scale flow physics with traditional CFD method.
The direct modeling concept is to merge the construction of the governing equations and the development of numerical algorithm together.
Based on the numerical cell size and time step scales, the corresponding discretized multiscale governing equations can be constructed.
Based on this concept, the unified gas-kinetic scheme (UGKS) is proposed \cite{xubook, xu2010unified},
and the corresponding well-balanced scheme under external force field has been developed with theoretical and numerical validations \cite{Xiao2017well}.
Through a coupled treatment of particle transport, collision, and external forcing effect
in the evaluation of flux transport across a cell interface and inner cell flow evolution, the cross-scale flow physics from kinetic particle transport to hydrodynamic wave propagation can be recovered \cite{liu2016unified}.
In this paper, the well-balanced UGKS is employed to investigate the non-equilibrium gas evolution under external force field.

In this paper, the lid-driven cavity flow is used as a typical example for the study of non-equilibrium gas dynamics under external force field.
Even under such a simple geometry,
the cavity flow displays complex fluid mechanical phenomena with multiple scales, including shearing layers, eddies, secondary flows, heat transfer, hydrodynamic instabilities, and laminar-turbulence transition, etc \cite{shankar2000fluid}.
Great efforts have been devoted to the study of the flow physics in different flow regimes as well.
In the continuum regime, the cavity problem is a typical benchmark case for the validation of numerical algorithms for the NS solutions \cite{shankar2000fluid, ghia1982high, SCHREIBER1983310, vanka1986block, hou1994simulation}.
In rarefied regime, the direct simulation Monte Carlo (DSMC) \cite{bird1994molecular} and kinetic Boltzmann solvers
\cite{cercignani1988boltzmann, cercignani2000rarefied} provide the benchmark solutions.
Naris et al. \cite{naris2005driven} discretized a linearized BGK equation to investigate the rarefaction effect
on the flow pattern and dynamics over the whole range of the Knudsen number.
Mizzi et al. \cite{mizzi2007effects} compared the simulation results from the  Navier-Stokes-Fourier equations (NSF) with slip boundary conditions and the DSMC results in  a lid-driven micro cavity case.
John et al. \cite{john2010investigation} applied the DSMC, discovered counter-gradient heat transport in the transition regime,
and investigated the dynamic effect from the expansion cooling and viscous dissipation on the heat transport mechanism.
In all previous work, there is few study about the  cavity flow under external force field.
Due to the external force effect, the cavity flow becomes even more complicated with its non-equilibrium multiple scale evolution.
A few new phenomena, including the connection between the heat transfer and external force, and stratified flow of different regimes,
have been observed through this study.

This paper is organized as follows.
The basic kinetic theory and the analysis of the influence of external force on the macroscopic flow transport are presented in Section 2.
The unified gas kinetic modeling and computation under external force field is briefly summarized in Section 3.
Section 4 presents the numerical experiments and discussion on the non-equilibrium flow transport and heat transfer across different flow regimes.
The last section is the conclusion.

\section{Analysis on physical effect from external force}
In the continuum regime with vanishing effect of Knudsen number, the macroscopic fluid dynamic equations can be used to describe the gas evolution \cite{sone2012kinetic}, i.e.,
\begin{equation}
\begin{aligned}
& \frac{\partial \rho}{\partial t} + \frac{\partial}{\partial x_i}(\rho U_i)=0, \\
& \frac{\partial }{\partial t} (\rho U_i)+ \frac{\partial}{\partial x_j}(\rho U_iU_j+P_{ij})= -\rho \frac{\partial \Phi}{\partial x_i}, \\
& \frac{\partial }{\partial t} \left[\rho \left(e+\frac{1}{2}U_iU_i\right)\right]+ \frac{\partial }{\partial x_j}\left[\rho U_j\left(e+\frac{1}{2}U_iU_i\right)+U_iP_{ij}+q_j\right]=-\rho U_j \frac{\partial \Phi}{\partial x_j} .
\end{aligned}
\label{fluid dynamic equation}
\end{equation}
Here $e$ is the internal energy and the external force potential $\Phi$ is assumed to be independent of time and molecular velocity.
The stress tensor $P_{ij}$ and heat flux $q_i$ are assumed to be related with inhomogeneous spatial distribution of macroscopic variables to close the system in Eq.(\ref{fluid dynamic equation}), i.e.,
\begin{equation*}
P_{ij}=p\delta_{ij},q_i=0,
\end{equation*}
for the Euler equations, and 
\begin{equation*}
P_{ij}=p\delta_{ij}-\mu \left( \frac{\partial U_i}{\partial x_j}+\frac{\partial U_j}{\partial x_i}-\frac{2}{3}\frac{\partial U_k}{\partial x_k}\delta_{ij} \right)-\mu_B \frac{\partial U_k}{\partial x_k}\delta_{ij},q_i=-\kappa \frac{\partial T}{\partial x_i},
\end{equation*}
for the Navier-Stokes-Fourier equations. 
The notation $\delta_{ij}$ is Kronecker's delta, and $\mu$, $\mu_B$ and $\kappa$ are the viscosity, bulk viscosity, and thermal conductivity coefficients of the gas respectively.

The macroscopic flow transport and heat transfer have a close relationship with the particle motion on the micro scale.
The kinetic theory can be employed to describe a gas dynamic system under external force field as well.
In the steady continuum limit, if the distribution function $f$ is the exact Maxwellian due to intensive particle collisions, i.e. the Euler regime, and since the collision term will not modify the equilibrium state, the Boltzmann equation goes to
\begin{equation}
\vec u \cdot \nabla_{x_i} f - \nabla_{x_i} \Phi \cdot \nabla_{u_i} f = 0.
\label{eqn:steady boltzmann equation}
\end{equation}
The general solution of Eq.(\ref{eqn:steady boltzmann equation}) is
\begin{equation*}
f(\vec x,\vec u) = \mathcal{F} (\Phi+\frac{1}{2}\vec u^2),
\end{equation*}
where $\mathcal{F}$ is an arbitrary function. 
In this solution, the temperature becomes a constant multiplier for the function of  $(\Phi+\frac{1}{2}u^2)$ \cite{chapman1970mathematical,luo2011well}, and it indicates the isothermal hydrostatic equilibrium flow field which is the solution of Eq.(\ref{fluid dynamic equation}),
\begin{equation*}
\rho=\rho(\vec x),U=0,\nabla p=-\rho \nabla \Phi.
\end{equation*}
For a constant gravitational acceleration $\phi_x$ in one dimensional case, the corresponding solution is simple,
\begin{equation}
\rho=\rho_0 \exp\left(\frac{\phi_x x}{RT}\right),u=0,p=p_0\exp \left(\frac{\phi_x x}{RT}\right),
\label{hydrostatic solution 1D}
\end{equation}
where $R$ is the gas constant.
Since there is no macroscopic velocity or its derivatives involved, the Euler and Navier-Stokes equations allow the same solution
in Eq.(\ref{hydrostatic solution 1D}).

Now the question becomes the gas dynamics under external force field for a non-vanishing particle mean free path and collision time.
This problem can be qualitatively illustrated by the kinetic theory.
As shown in Fig. $1$, a column of gas is enclosed between two parallel plates.
The upper plate is kept with the constant temperature $T_u$, and $T_d$ at the lower plate.
The gas is static everywhere with no macroscopic flow.
We assume a virtual interface $I$ perpendicular to the $x$ axis somewhere inside the domain.
Three thought experiments can be carried out here to illustrate the effect of external force on the gas dynamics with limited particle mean free path.
\begin{figure}[htb!]
	\centering
	{
		\begin{tikzpicture}[thick]
		\node[rectangle] (bx) {};
		\node at ($(bx)+(+1.5,+0.5)$) {$T_b$};
		\draw ($(bx)+(-1,+0.5)$) -- ($(bx)+(+1,+0.5)$);
		\draw ($(bx)+(-0.9,+0.3)$) -- ($(bx)+(-0.7,+0.5)$);
		\draw ($(bx)+(-0.6,+0.3)$) -- ($(bx)+(-0.4,+0.5)$);
		\draw ($(bx)+(-0.3,+0.3)$) -- ($(bx)+(-0.1,+0.5)$);
		\draw ($(bx)+(0,+0.3)$) -- ($(bx)+(+0.2,+0.5)$);
		\draw ($(bx)+(+0.3,+0.3)$) -- ($(bx)+(+0.5,+0.5)$);
		\draw ($(bx)+(+0.6,+0.3)$) -- ($(bx)+(+0.8,+0.5)$);
		\node at ($(bx)+(+1.5,+7)$) {$T_u$};
		\draw ($(bx)+(-1,+7)$) -- ($(bx)+(+1,+7)$);
		\draw ($(bx)+(-0.8,+7)$) -- ($(bx)+(-0.6,+7.2)$);
		\draw ($(bx)+(-0.5,+7)$) -- ($(bx)+(-0.3,+7.2)$);
		\draw ($(bx)+(-0.2,+7)$) -- ($(bx)+(0.0,+7.2)$);
		\draw ($(bx)+(0.1,+7)$) -- ($(bx)+(+0.3,+7.2)$);
		\draw ($(bx)+(+0.4,+7)$) -- ($(bx)+(+0.6,+7.2)$);
		\draw ($(bx)+(+0.7,+7)$) -- ($(bx)+(+0.9,+7.2)$);
		\draw[line] ($(bx)+(-2,+0.5)$) -- ($(bx)+(-2,+3)$);
		\node at ($(bx)+(-2.4,+3)$) {$x$};
		\draw[line] ($(bx)+(-3,+3)$) -- ($(bx)+(-3,+0.5)$);
		\node at ($(bx)+(-3.4,+0.5)$) {$g$};
		\node at ($(bx)+(0.5,+5.5)$) {$A$};
		\node at ($(bx)+(0.5,+1.8)$) {$B$};
		\node[shape=circle,fill=gray] (b) at (0,5.5) {};
		\node[shape=circle,fill=gray] (b) at (0,1.8) {};
		\draw[line] ($(bx)+(0,+5.2)$) -- ($(bx)+(0,+3.6)$);
		\draw[line] ($(bx)+(0,+2.1)$) -- ($(bx)+(0,+3.4)$);
		\node at ($(bx)+(-0.5,+4.4)$) {$\ell_A$};
		\node at ($(bx)+(-0.5,+2.7)$) {$\ell_B$};
		\node at ($(bx)+(+1.8,+3.5)$) {interface $I$};
		\draw ($(bx)+(-0.7,+3.5)$)[dashed] -- ($(bx)+(+0.7,+3.5)$);
		\end{tikzpicture}
	}
	\label{particle theory}
	\caption{Schematic of gas enclosed between two plates.}
\end{figure}
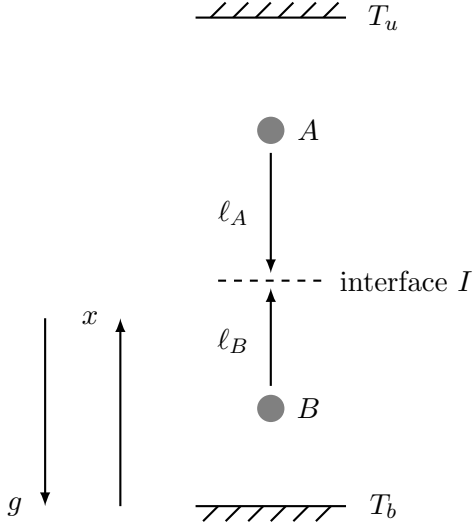

\textbf {\emph{Problem A: homogeneous distribution}}

Assume all the distributions of density, velocity, pressure and temperature are uniform everywhere between the plates.
The temperatures of the two plates and the gas are equal with $T_u=T_d=T$.
Now we consider the molecules traveling across the interface $I$ from the upside and the downside.
A class of molecules transporting downwards are called $A$, which is located a particle mean free path $\ell_A$ away from the interface.
According, the class traveling upwards is named $B$ with $\ell_B$.
Since the flow field is homogeneous, there are same probable amounts of molecules traveling across the interface from $A$ and $B$, with the same mean velocity 
$c \propto \sqrt{RT}$ where $R$ is the gas constant and $T$ is the temperature.
Therefore, across the interface $I$, there is no net particle flux, and all the hydrodynamic fluxes are zero.

\textbf {\emph{Problem B: hydrostatic distribution under external force field}}

In this problem, we include the constant external force $\Phi_x=-g$ along the negative $x$-direction into the above gas dynamic system.
The initial distribution is assumed to be the exact Maxwellian with respect to the barometric solution in Eq.(\ref{hydrostatic solution 1D}),
in which the pressure gradient is balanced by the external force.
Due to the density stratification, now the particle density at $B$ is larger than $A$.
According to the hydrostatic equilibrium solution, the temperature is uniform with $T_A=T_B$, and thus the two classes of molecules share the same mean velocity $c_A=c_B \propto \sqrt{RT}$ initially.
However, from the point $A/B$ to the interface $I$, the particles from class $A$ ($B$) get accelerated (decelerated) due to external force $g$ within a collision time.
Therefore, at the interface there is less molecules $A$ transporting with high speed from the upside, while more molecules $B$ with low speed come from the downside.
In this way, the net particle flux across the interface can be ensured to be zero, and there is no macroscopic flow velocity.
However, if we check the energy flux carried along with the molecules $A$ and $B$, we can find that the formal one is larger since the energy flux is related with higher-order contributions of particle velocity.
Since there is no macroscopic transport phenomenon in this case, the energy transfer rate is equal to the heat flux $q\propto c^3$, 
and the temperature downwards becomes higher than the one upwards.
The existence of external force field breaks the symmetry of thermal motion of molecules, resulting in the inhomogeneous distribution of temperature.
During the particle's traveling, there is a dynamic energy transformation between the potential, kinetic and thermal energy.

\textbf {\emph{Problem C: pure heat conduction under external force field}}

In this problem, we consider the static heat conduction where the boundary temperature is set as $T_b>T_u$.
The initial distribution is set as the Fourier flow solution, in which the macroscopic flow is absent and a steady heat flux is along the negative direction of thermal gradient.
In the absence of external force, the pressure is uniform between the two plate.
The temperature decreases along the positive $x$ direction, and the density increases correspondingly.
The particle density satisfies $\rho_A>\rho_B$,
and the mean velocity $c_A<c_B$.
Thus, the requirement of zero net particle flux across the interface is satisfied.
Now we add the external force field $\Phi_x=-g$ suddenly into the heat conduction system. 
Due to the same acceleration and deceleration mechanism described in Problem B, at the interface the mean particle speed from $A$ increases with $c'_A>c_A$, and the one from $B$ decreases correspondingly with $c'_B<c_B$.
Therefore, the original Fourier's heat flux is suppressed.
When the magnitude of external force is small, the effect of force field performs a small modification of the original heat flux.
However, if the force is extremely large, it may even play the dominant role over thermal gradient in the determination of heat flux.
Much complicated non-equilibrium dynamics would happen in such a strong external field.
Analogously, if we change the direction of force $\phi_x$ into positive $x$-direction, aligned with the original heat flux, then it results in the particle velocity $c'_A<c_A$ and $c'_B>c_B$, and thus there would be an increment of heat flux correspondingly.
So far, we have qualitatively illustrated the mechanism of particle transport and high-order energy transfer under external force field.
It is noted that the above analysis is based on the particle free transport mechanism within a particle mean free path, and this non-equilibrium effect appears in all flow regimes with non-vanishing particle mean free path.
In the NS regime, the viscosity and heat conductivity coefficient correspond to non-vanishing particle mean free path in the relationship $\mu,\kappa \propto \rho c \ell$, where $\rho$ is the density, $c$ is the mean particle velocity and $\ell$ is the particle mean free path \cite{chapman1970mathematical},
and the effect of external force on the heat transport will appear definitely.
With vanishing effect of particle mean free path and collision time, the hydrostatic solution given in Eq.(\ref{hydrostatic solution 1D}) can be applied.
But, this ideal Euler system does not exist in reality.

The above particle-based analysis can be demonstrated as well using partial differential equations. 
Due to the complexity of the nonlinear integro-differential Boltzmann equation,
let us go to the BGK kinetic equation first to discuss the influence of external force on a near-equilibrium gas dynamic system.
For brevity, the one dimensional case is considered, i.e.,
\begin{equation*}
\frac{\partial f}{\partial t} + u \frac{\partial f}{\partial x} + \phi_x \frac{\partial f}{\partial u} = \frac{f^+-f}{\tau},
\end{equation*}
where $\tau=1/\nu$ is the collision time.
The Maxwellian distribution $f^+$ is
\begin{equation}
f^+=\rho\left(\frac{\lambda}{\pi}\right)^\frac{K+1}{2} e^{-\lambda \left[(u-U)^2+\xi^2\right]} ,
\label{Maxwellian}
\end{equation}
where $\lambda=m/2k_BT$ with $m$ the particle mass, $k_B$ the Boltzmann constant,
and $K$ is the internal degree of freedom.
For a monatomic gas in one-dimensional flow, it sets $K=2$ to account the random motion in $y$ and $z$ directions.
Following the strategy of Chapman-Enskog expansion \cite{chapman1970mathematical}, to solve the BGK equation, the distribution function can be expanded into series with respect to a tiny factor,
\begin{equation*}
f=f^{(0)}+f^{(1)}\epsilon+f^{(2)}\epsilon^2+\cdots.
\end{equation*}

In the near-equilibrium region, the expansion is carried out based on the collision time as well as the corresponding Knudsen number conventionally,
which can be written as the following successive BGK equation \cite{ohwada2004kinetic},
\begin{equation}
f=f^+-\tau \frac{D}{Dt} f^++\tau^2 \frac{D^2}{Dt^2}f^++\cdots,
\label{BGK successive}
\end{equation}
where $D/Dt$ is the total derivative of both physical and velocity space.
If we consider the first order approximation of Eq. (\ref{BGK successive}) with respect to the collision time $\tau$, the distribution function $f$ has the corresponding expansion,
\begin{equation}
f=f^+-\tau(f^+_t+uf^+_x)-\tau \phi_x f^+_u.
\label{f expansion}
\end{equation}
The first two terms in Eq. (\ref{f expansion}) describe the free transport of particles during the traveling time between two successive collisions.
This expression is consistent with the Chapman-Enskog expansion for the Navier-Stokes solutions,
and a dynamic viscous coefficient $\mu=\tau p$ can be obtained \cite{vincenti1965introduction, xu2001gas}.
At the same time, the force acceleration distorts the distribution function in the velocity space with the following contributions
to the macroscopic flow variables.
As macroscopic variables are related with particle distribution function through velocity moments,
\begin{equation*}
\mathbf{W}=\left(
\begin{matrix}
\rho \\
\rho U \\
\rho E
\end{matrix}
\right)=\int \psi f d\Xi,
\end{equation*}

\begin{equation*}
p=\frac{1}{3}\int \left((u-U)^2+\xi^2\right)fd\Xi,
\end{equation*}

\begin{equation*}
q=\frac{1}{2}\int (u-U)\left((u-U)^2+\xi^2\right)fd\Xi,
\end{equation*}
where $d\Xi=dud\xi$, $p$ is pressure, $q$ is heat flux and $\psi=\left(1,u,\frac{1}{2}(u^2+\xi^2) \right)^T$ is the vector of moments for collision invariants,
if a sudden external force field is added to this gas dynamic system, then the net contribution of macroscopic variables from the external forcing term can be evaluated as
\begin{equation*}
\Delta \mathbf{W}=\left(
\begin{matrix}
\Delta \rho \\
\Delta \rho U \\
\Delta \rho E
\end{matrix}
\right)=\int \psi (-\tau \phi_x f^+_u) d\Xi,
\end{equation*}

\begin{equation*}
\Delta p=\frac{1}{3}\int \left((u-U)^2+\xi^2\right) (-\tau \phi_x f^+_u)d\Xi,
\end{equation*}

\begin{equation*}
\Delta q=\frac{1}{2}\int (u-U) \left((u-U)^2+\xi^2\right) (-\tau \phi_x f^+_u)d\Xi.
\end{equation*}

For the moments of Maxwellian distribution $\int u^\alpha \xi^\beta f^+ d\Xi = \rho <u^\alpha \xi^\beta>$, it has the property that
\begin{equation*}
<u^\alpha \xi^\beta>=<u^\alpha> <\xi^\beta>,
\end{equation*}
and the moments of Maxwellian distribution function are
\begin{equation*}
\begin{aligned}
& <u^0>=1 , \\
& <u^1>=U , \\
& <u^2>=U^2+\frac{1}{2\lambda} , \\
& <u^3>=U^3+\frac{3U}{2\lambda}  , \\
& <u^4>=U^4+\frac{3U^2}{\lambda}+\frac{3}{4\lambda^2} , \\
& <\xi^2>=\frac{K}{2\lambda}.
\end{aligned}
\label{Maxwellian moments}
\end{equation*}
Thus if the forcing term $\phi_x$ and the collision time $\tau$ are viewed as local constants, after integration by parts, we have the following relations,
\begin{equation}
\Delta \mathbf{W}=\left(
\begin{matrix}
\Delta \rho \\
\Delta \rho U \\
\Delta \rho E
\end{matrix}
\right) = \tau \phi_x \rho \left(
\begin{matrix}
0 \\
1 \\
U
\end{matrix}
\right),
\label{analysis w increment}
\end{equation}

\begin{equation}
\Delta p = \frac{1}{3} \tau \phi_x \rho \left( 2<u^1> - 2U<u^0>\right)=0,
\label{analysis p increment}
\end{equation}

\begin{equation}
\Delta q = \frac{1}{2} \tau \phi_x \rho \left(3<u^2> - 6U<u> + 3U^2<u^0> + <\xi^2> \right) = \frac{K+3}{4} \frac{\tau \phi_x \rho}{\lambda}.
\label{analysis q increment}
\end{equation}
The Eq. (\ref{analysis w increment}), (\ref{analysis p increment}) and (\ref{analysis q increment}) present a qualitative contribution of external force field on the macroscopic flow variables.
The contribution of external forcing term to conservative variables is exactly the source term in the conservation law during the mean traveling time between two successive particle collisions.
The isotropic pressure is not affected by the external field in the current order of expansion.
However, it is clear that there exists contribution to the heat flux from the external forcing term under current order of expansion.
The modification of heat flux is along the positive direction of external force acceleration $\phi_x$.
In other words, the external force will drive the heat flux in its direction.

The analysis above supplements the particle-based thought experiments.
The result is consistent with the one given in \cite{tij2000influence}, where a first-order modification on the heat flux from gravity is illustrated using asymptotic perturbation method.
At a limited particle collision time $\tau$ corresponding to non-vanishing viscosity and heat conduction coefficients, the external forcing term does affect
the transport, especially in the transition and free molecular regimes.
It is noted that the above PDE-based analysis is for the case of near-equilibrium and small external force.
With the increment of external force and degree of rarefaction, even strong non-equilibrium effect is expected to appear, and the cross-scale algorithm is needed to investigate the non-equilibrium gas dynamics under external force field.

\section{Gas kinetic modeling}

Based on the direct modeling on the cell size and time step, 
the unified scheme is a combination of gas kinetic modeling and computation, where the governing equations are constructed in the discretized space and then solved in the numerical algorithm.
Here we give a brief introduction of the principle of the UGKS.
With the notation of cell averaged distribution function in the control volume,
\begin{equation*}
f_{x_i,y_j,t^n,u_k,v_l}=f_{i,j,k,l}^n=\frac{1}{\Omega_{i,j}(\vec x)\Omega_{k,l}(\vec u)} \int_{\Omega_{i,j}} \int_{\Omega_{k,l}}f(x,y,t^n,u,v)d\vec x d\vec u,
\end{equation*}
the update of macroscopic conservative variables and the particle distribution function are coupled in the following way,
\begin{equation}
\textbf{W}_{i,j}^{n+1}=\textbf{W}_{i,j}^n+\frac{1}{\Omega_{i,j}}\int_{t^n}^{t^{n+1}}\sum_{r}\Delta \textbf{L}_r\cdot {\textbf{F}}_rdt+\frac{1}{\Omega_{i,j}}\int_{t^n}^{t^{n+1}}\textbf{G}_{i,j}dt,
\label{eqn:macro update}
\end{equation}
\begin{equation}
\begin{aligned}
f_{i,j,k,l}^{n+1}=&f_{i,j,k,l}^n+\frac{1}{\Omega_{i,j}}\int_{t^n}^{t^{n+1}} \sum_{r} u_r\hat f_r(t)\Delta L_r dt\\
&+\frac{1}{\Omega_{i,j}}\int_{t^n}^{t^{n+1}}\int_{\Omega_{i,j}}Q(f)d\vec x dt+\frac{1}{\Omega_{i,j}}\int_{t^n}^{t^{n+1}}\int_{\Omega_{i,j}}G(f)d\vec x dt,
\end{aligned}
\label{eqn:distribution update}
\end{equation}
where  $\textbf{F}_r$ is the flux of conservative variables, $f_r$ is the time-dependent gas distribution function at cell interface and $\Delta L_r$ is the cell interface length.
The $\textbf{G}_{i,j}$ and $G(f)$ are the external forcing sources of macroscopic conservative variables and particle distribution function, and $Q(f)$ is the collision term respectively,
\begin{equation}
\textbf{G}_{i,j}=\int_{\Omega_{k,l}}\left(-\phi_x \Delta t \frac{\partial}{\partial u} f_{i,j,k,l}^{n+1/2}-\phi_y \Delta t\frac{\partial}{\partial v} f_{i,j,k,l}^{n+1/2}\right)\psi dudvd\xi,
\end{equation}
\begin{equation}
\begin{aligned}
&Q(f)=\frac{f_{i,j,k,l}^{+}-f_{i,j,k,l}^{n+1/2}}{\tau}, \\
&G(f)=-\phi_x \frac{\partial}{\partial u} f_{i,j,k,l}^{n+1/2}-\phi_y \frac{\partial}{\partial v} f_{i,j,k,l}^{n+1/2}.
\end{aligned}
\label{distribution function source}
\end{equation}
Here $\vec \phi=\phi_x \vec i+\phi_y \vec j$ is the external force acceleration, and $f^+$ is the equilibrium state. 
The implementation of the full Boltzmann collision term $Q(f)$ can be done as well when the time step
is on the order of particle collision time \cite{liu2016unified}. However, if the time step is a few times of the local particle collision time,
the use of kinetic relaxation model is accurate enough because the accumulating physical effect in a multiple particle collision time scale is not sensitive to the individual particle collision anymore.

In the numerical algorithm, the conservative variables are updated first in Eq. (\ref{eqn:macro update}),
and the updated macroscopic variables can be used for the construction of the equilibrium state in $Q(f)$ at $t^{n+1}$ time step for an implicit treatment.
The derivatives of particle velocity in $G(f)$ are evaluated via upwind finite difference method in the discretized velocity space.

For the gas kinetic modeling in a control volume framework, the key point is to construct a multiscale evolving flux function.
In the UGKS, the flux function is derived through interface distribution function $f_r$, which can be evaluated through the evolving solution of kinetic model equation.
The model equation with external force term in the two-dimensional Cartesian coordinate system is
\begin{equation}
f_t+uf_x+vf_y+\phi_xf_u+\phi_yf_v=\frac{f^+-f}{\tau},
\label{BGK 2D}
\end{equation}
where $\tau=\mu/p$ is the particle collision time.
For the BGK equation, $f^+$ is exactly the Maxwellian distribution
\begin{equation*}
f^+=\rho\left(\frac{\lambda}{\pi}\right)^\frac{K+2}{2}e^{-\lambda[(u-U)^2+(v-V)^2+\xi^2]},
\end{equation*}
where $\lambda={\rho}/{(2p)}$ and $K$ is the dimension of internal degree of freedom $\xi$.
For the Shakhov model equation, $f^+$ takes the following form to provide correct Prandtl number,
\begin{equation*}
f^+=g\left[1+(1-\mathrm{Pr} )(\vec c \cdot \vec q )\left(\frac{c^2}{RT}-5\right)/(5pRT)\right],
\end{equation*}
where $\vec c=(u-U)\vec i+(v-V)\vec j$ is the peculiar velocity, $\vec q$ is heat flux, and Pr is Prandtl number.

In the unified scheme, at the center of a cell interface $(x_{i+1/2},y_j)$ the solution $f_{i+1/2,j,k,l}$ is constructed from the integral solution of Eq. (\ref{BGK 2D}).
With the notations $x_{i+1/2}=0,y_j=0$ at $t^n=0$, the time-dependent interface distribution function writes
\begin{equation}
\begin{aligned}
f(0,0,t,u_k,v_l,\xi)=&\frac{1}{\tau}\int_{0}^t f^+(x',y',t',u_k',v_l',\xi)e^{-(t-t')/\tau}dt' \\
&+e^{-t/\tau}f_0(x^0,y^0,0,u_k^0,v_l^0,\xi),
\end{aligned}
\label{eqn:bgk characteristic}
\end{equation}
where $x'=-u_k'(t-t')-\frac{1}{2}\phi_x(t-t')^2,y'=-v_l'(t-t')-\frac{1}{2}\phi_y(t-t')^2,u_k'=u_k-\phi_x (t-t')$, and $v_l'=v_l-\phi_y (t-t')$ are the particle trajectories in physical and velocity space, and $(x^0,y^0,u_k^0,v_l^0)=(-(u_k-\phi_xt)t-\frac{1}{2}\phi_x t^2,-(v_l-\phi_y t)t-\frac{1}{2}\phi_y t^2,u_k-\phi_x t,v_l-\phi_y t)$ is the initial location in physical and velocity space for the particle which passes through the cell interface at time $t$.
The time accumulating effect from the external forcing term on the time evolution of the particle distribution function is explicitly taken into consideration.
The above scale-dependent integral solution plays the most important role for the construction of the well-balanced UGKS, where the contributions from both equilibrium hydrodynamic and non-equilibrium kinetic flow physics are considered.

In the detailed numerical scheme, to the second order accuracy, the initial gas distribution function $f_0$ is reconstructed as
\begin{equation*}
f_0(x,y,0,u_k,v_l,\xi)=\left\{
\begin{aligned}
&f_{i+1/2,j,k,l}^L+\sigma_{i,j,k,l}x+\theta_{i,j,k,l}y, \quad x\le 0, \\
&f_{i+1/2,j,k,l}^R+\sigma_{i+1,j,k,l}x+\theta_{i+1,j,k,l}y, \quad x> 0,
\end{aligned}
\right.
\end{equation*}
where $f_{i+1/2,j,k,l}^L$ and $f_{i+1/2,j,k,l}^R$ are the reconstructed initial distribution functions at the left and right hand sides of a cell interface,
and $\sigma$ and $\theta$ are the slopes of distribution function along $x$ and $y$ directions.

The equilibrium distribution function around a cell interface is constructed as
\begin{equation*}
f^+=f^+_0\left[1+(1-H[x]){a}^Lx+H[x]{a}^Rx+by+{A}t\right],
\end{equation*}
where $f^+_0$ is the equilibrium distribution at $(x=0,t=0)$. 
The coefficients above can be evaluated from the spatial distribution of conservative variables on both sides of the cell interface and the compatibility condition.
After all the coefficients are determined, the time dependent interface distribution function becomes
\begin{equation}
\begin{aligned}
f(0,0,t,u_k,v_l,\xi)=&\left(1-e^{-t/\tau}\right) f^+_0\\
&+\left(\tau(-1+e^{-t/\tau})+te^{-t/\tau}\right)a^{L,R}u_kf^+_0 \\
&-\left[\tau\left(\tau(-1+e^{-t/\tau})+te^{-t/\tau}\right)+\frac{1}{2}t^2e^{-t/\tau}\right] a^{L,R}\phi_x f^+_0 \\
&+\left(\tau(-1+e^{-t/\tau})+te^{-t/\tau}\right)bv_lf^+_0 - \left[\tau\left(\tau(-1+e^{-t/\tau})+te^{-t/\tau}\right)+\frac{1}{2}t^2e^{-t/\tau}\right] b\phi_y f^+_0 \\
&+\tau \left(t/\tau-1+e^{-t/\tau}\right){A}f^+_0 \\
&+e^{-t/\tau}\left[\left(f_{i+1/2,k^0,l^0}^L+\left(-(u_k-\phi_xt)t-\frac{1}{2}\phi_xt^2\right)\sigma_{i,k^0,l^0}\right.\right.\\
&\left.\left. +\left(-(v_l-\phi_yt)t-\frac{1}{2}\phi_yt^2\right)\theta_{i,k^0,l^0}\right)H\left[u_k-\frac{1}{2}\phi_x t\right] \right.\\
&\left.+\left(f_{i+1/2,k^0,l^0}^R+\left(-(u_k-\phi_xt)t-\frac{1}{2}\phi_xt^2\right)\sigma_{i+1,k^0,l^0}\right. \right. \\
&\left. \left. +\left(-(v_l-\phi_yt)t-\frac{1}{2}\phi_yt^2\right)\theta_{i+1,k^0,l^0}\right)(1-H\left[u_k-\frac{1}{2}\phi_x t\right])\right] \\
=&\widetilde f^+_{i+1/2,j,k,l}+\widetilde f_{i+1/2,j,k,l},
\end{aligned}
\end{equation}
where $\widetilde f^+_{i+1/2,j,k,l}$ is related to equilibrium state integration and $\widetilde f_{i+1/2,j,k,l}$ is related to the initial non-equilibrium distribution.
With the variation of the ratio between evolving time $t$ (i.e., the time step in the computation) and collision time $\tau$, the interface distribution function above provides self-conditioned multiple scale flow physics across different flow regimes.
The corresponding flux of conservative variables can be constructed as
\begin{equation*}
{\textbf{F}}_{i+1/2,j}=\int_{\Omega_{k,l}} u_k f(0,0,t,u_k,v_l,\xi) \psi d\Xi.
\end{equation*}

\section{Non-equilibrium flow studies}
In this section, we are going to present and discuss several numerical experiments to
illustrate the non-equilibrium flow dynamics under external force field.
The well-balanced UGKS is employed in all test cases.

\subsection{Poiseuille-type flow}
In the first numerical experiment, we investigate the steady flow of dilute gas between two infinite parallel plates driven by a unidirectional external force \cite{esposito1994hydrodynamic,tij1994perturbation,mansour1997validity,aoki2002poiseuille,xu2003super,uribe1999burnett,hess1999temperature}.
This case serves as a supplementary validation of the current numerical algorithm besides the cases presented in Ref. \cite{Xiao2017well}.
The two plates at rest are located at $y=\pm L/2$ and kept at temperature $T_0$. 
The gas initially at rest with uniform density $\rho_0$ and temperature $T_0$ is subject to a uniform external force in the positive $x$ direction, i.e., in the direction parallel to the plates. 
The initial particle distribution function is set as the Maxwellian everywhere in the flow domain.
There is no pressure gradient in the $x$ direction.
If we consider this problem in the framework of the Navier-Stokes equations, then it is a simple one-dimensional example.
In the framework of kinetic theory,
the steady BGK model equation for Maxwell molecules under external force field is used to describe the gas evolution in this system,
\begin{equation}
v\frac{\partial f}{\partial y} + \phi_x\frac{\partial f}{\partial u} = A_c \rho (f^+-f),
\label{eqn:bgk aoki}
\end{equation}
where $A_c$ is a constant and the collision frequency is $1/\tau=A_c \rho$.
The Maxwellian diffuse reflection boundary is assumed in the simulation.

With the dimensionless variables defined as
\begin{equation*}
\begin{aligned}
& \hat{x}=\frac{x}{L_0}, \hat{y}=\frac{y}{L_0}, \hat{\rho}=\frac{\rho}{\rho_0}, \hat{T}=\frac{T}{T_0}, \\
& \hat{u}_i=\frac{u_i}{(2RT_0)^{1/2}}, \hat{U}_i=\frac{U_i}{(2RT_0)^{1/2}}, \hat{f}=\frac{f}{\rho_0 (2RT_0)^{3/2}}, \\
& \hat{P}_{ij}=\frac{P_{ij}}{\rho_0 (2RT_0)}, \hat{q}_i=\frac{q_i}{\rho_0 (2RT_0)^{3/2}}, \hat{\phi}_i=\frac{\phi_i}{2RT_0/L_0},
\end{aligned}
\end{equation*}
where $u_i$ is the particle velocity, $U_i$ is the macroscopic flow velocity, $P_{ij}$ is the stress tensor, $q_i$ is the heat flux and $\phi_i$ is the external force acceleration, the dimensionless BGK equation writes
\begin{equation*}
\hat{v} \frac{\partial \hat{f}}{\partial \hat{y}}+\hat{\phi}_x \frac{\partial \hat{f}}{\partial \hat{u}} = \frac{2}{\sqrt{\pi}} \frac{1}{Kn} \hat{\rho} (\hat{f^+}-\hat{f}),
\end{equation*}
where $Kn$ is the Knudsen number in the reference state.
The collision constant is absorbed with unit value $A_c=1$.
For simplicity, we will drop the hat notation henceforth to denote dimensionless variables.

To describe this basic system, 
Aoki and his co-workers used the asymptotic analysis \cite{aoki2002poiseuille, sone2012kinetic} for small Knudsen numbers and derived a system of fluid-dynamic-type equations and their boundary conditions up to the second order. 
The Hilbert expansion and its Knudsen layer correction are carried out with respect to $\epsilon=(\sqrt{2}/\pi)  Kn$, and the external force acceleration is set as $\phi_x = \alpha  Kn$.
When the Knudsen number is relatively large where the asymptotic theory fails, they also performed a numerical analysis by means of a finite difference method to solve the BGK equation Eq. (\ref{eqn:bgk aoki}) with respect to different $ Kn$ and $\alpha$.
Using the well-balanced UGKS with 100 physical cells and 41 velocity points, we simulate the case with $ Kn=0.02, 0.05, 0.1$ and $\alpha=1, 2, 3$ and compare the results with the asymptotic solutions and the finite difference ones \cite{aoki2002poiseuille}.

\begin{figure}
	\centering
	\subfigure[Density]{
		\includegraphics[width=5cm]{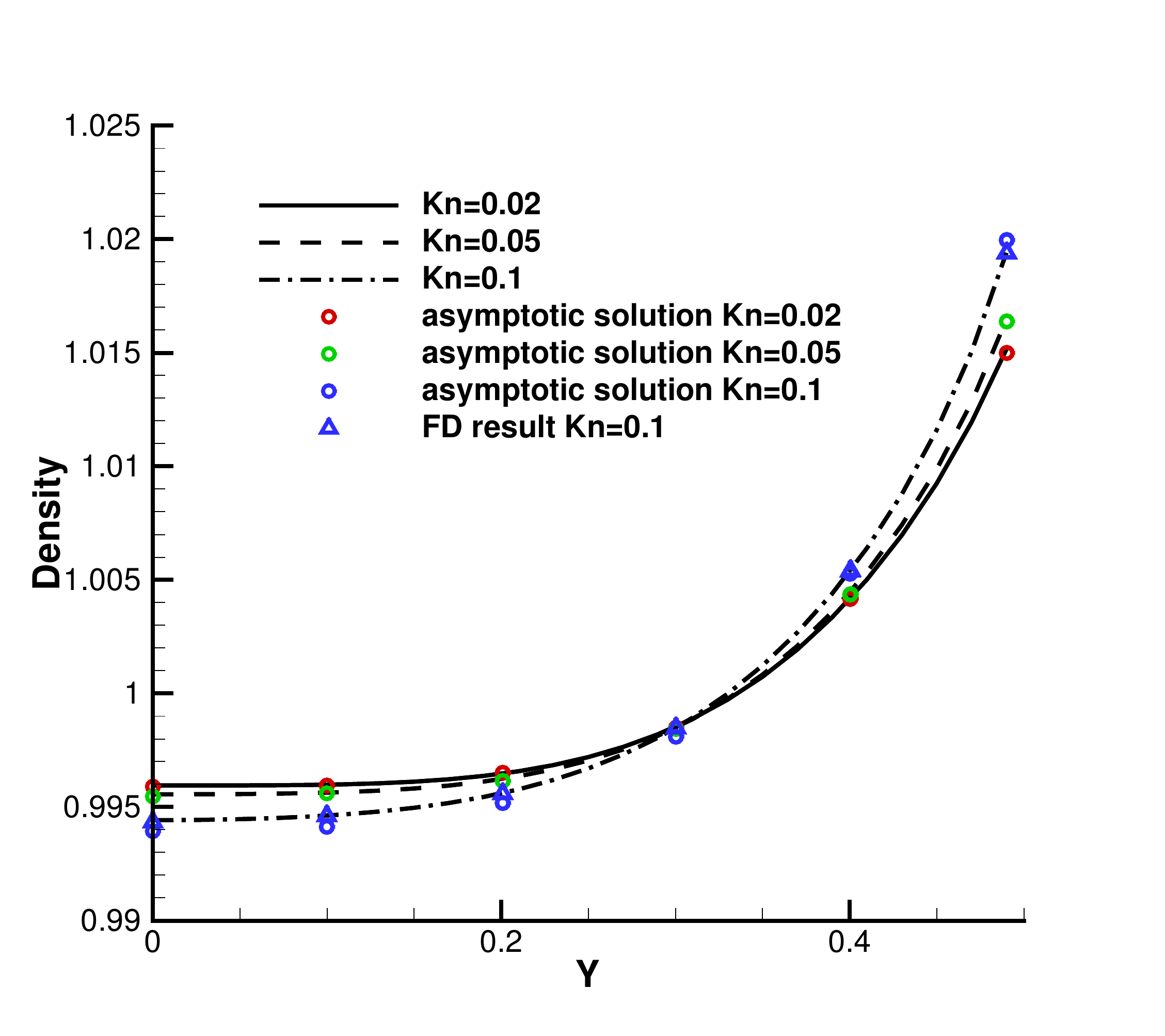}
	}
	\subfigure[U-velocity]{
		\includegraphics[width=5cm]{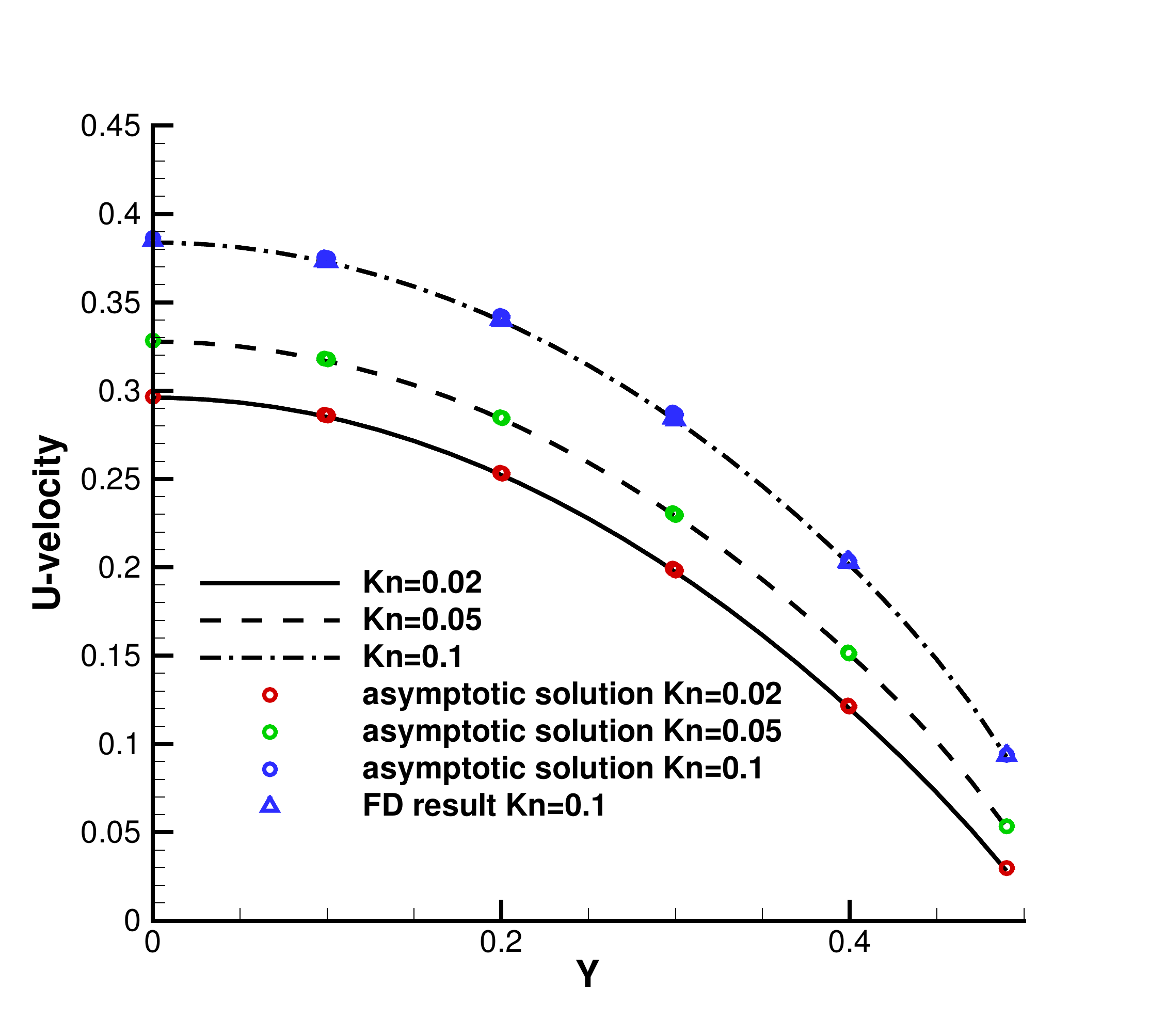}
	}
	\subfigure[Temperature]{
		\includegraphics[width=5cm]{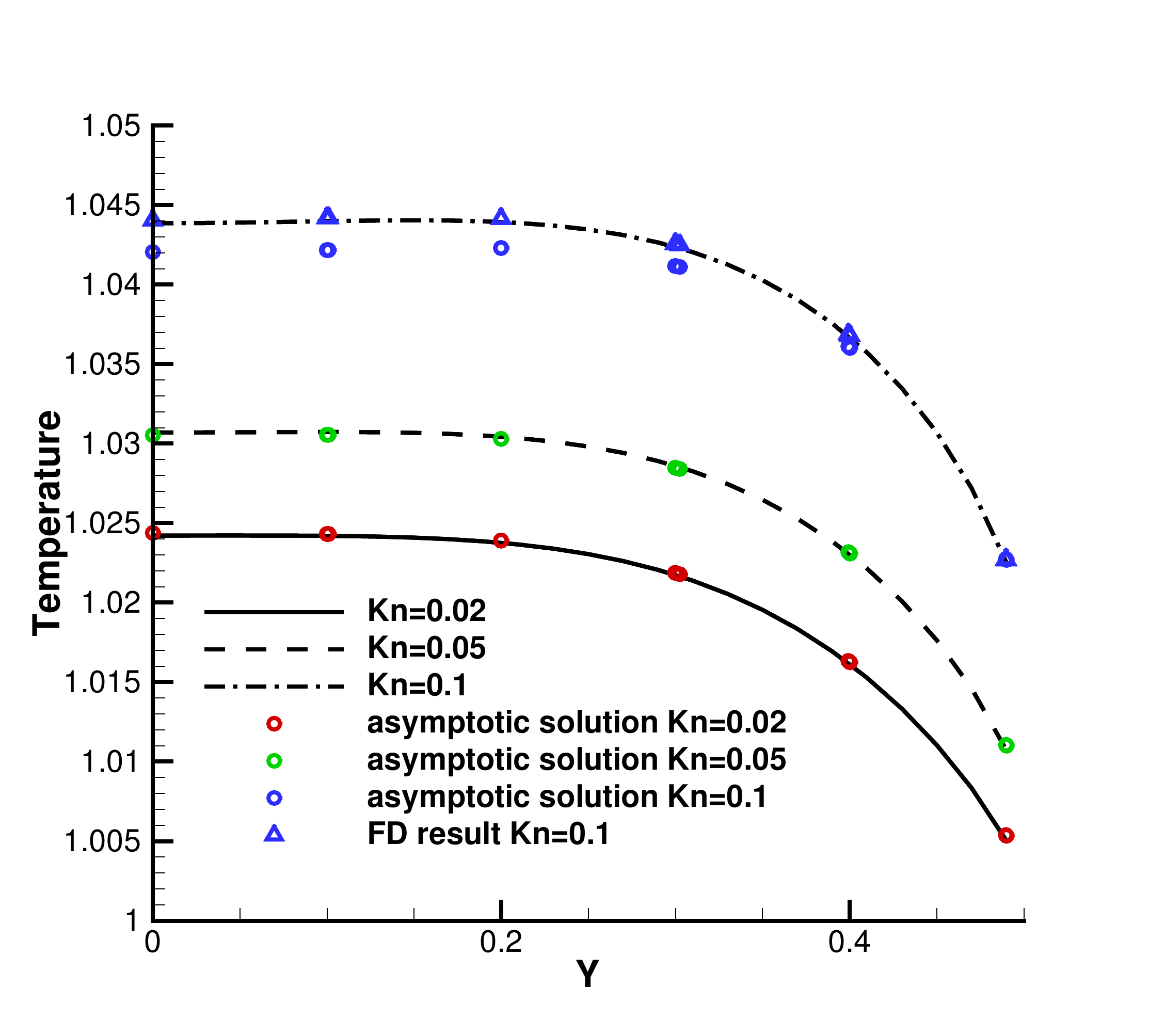}
	}
	\caption{The profiles of density, U-velocity and temperature with $\alpha=1$.}
	\label{pic:ap poiseuille a=1}
\end{figure}

\begin{figure}
	\centering
	\subfigure[Density]{
		\includegraphics[width=5cm]{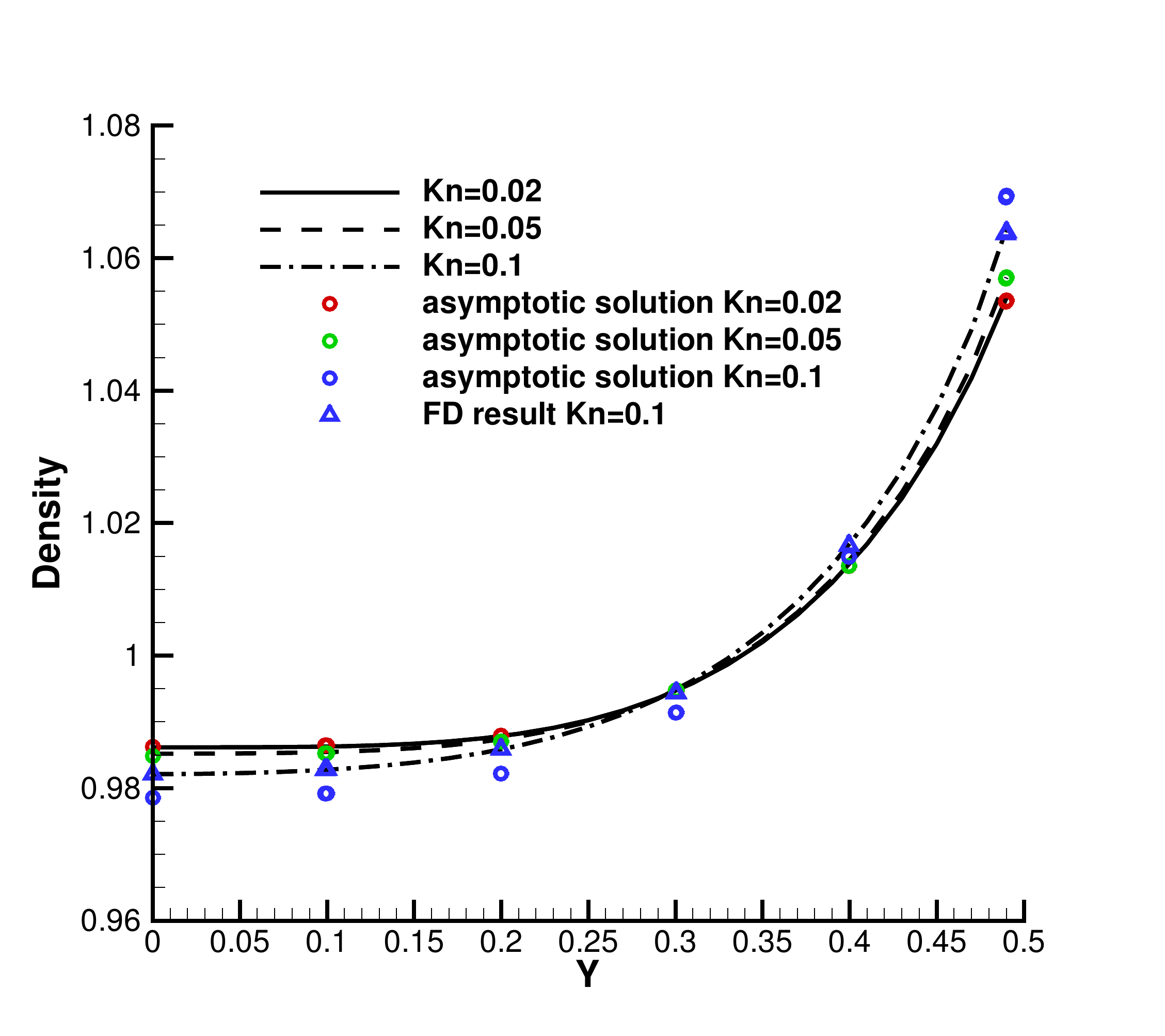}
	}
	\subfigure[U-velocity]{
		\includegraphics[width=5cm]{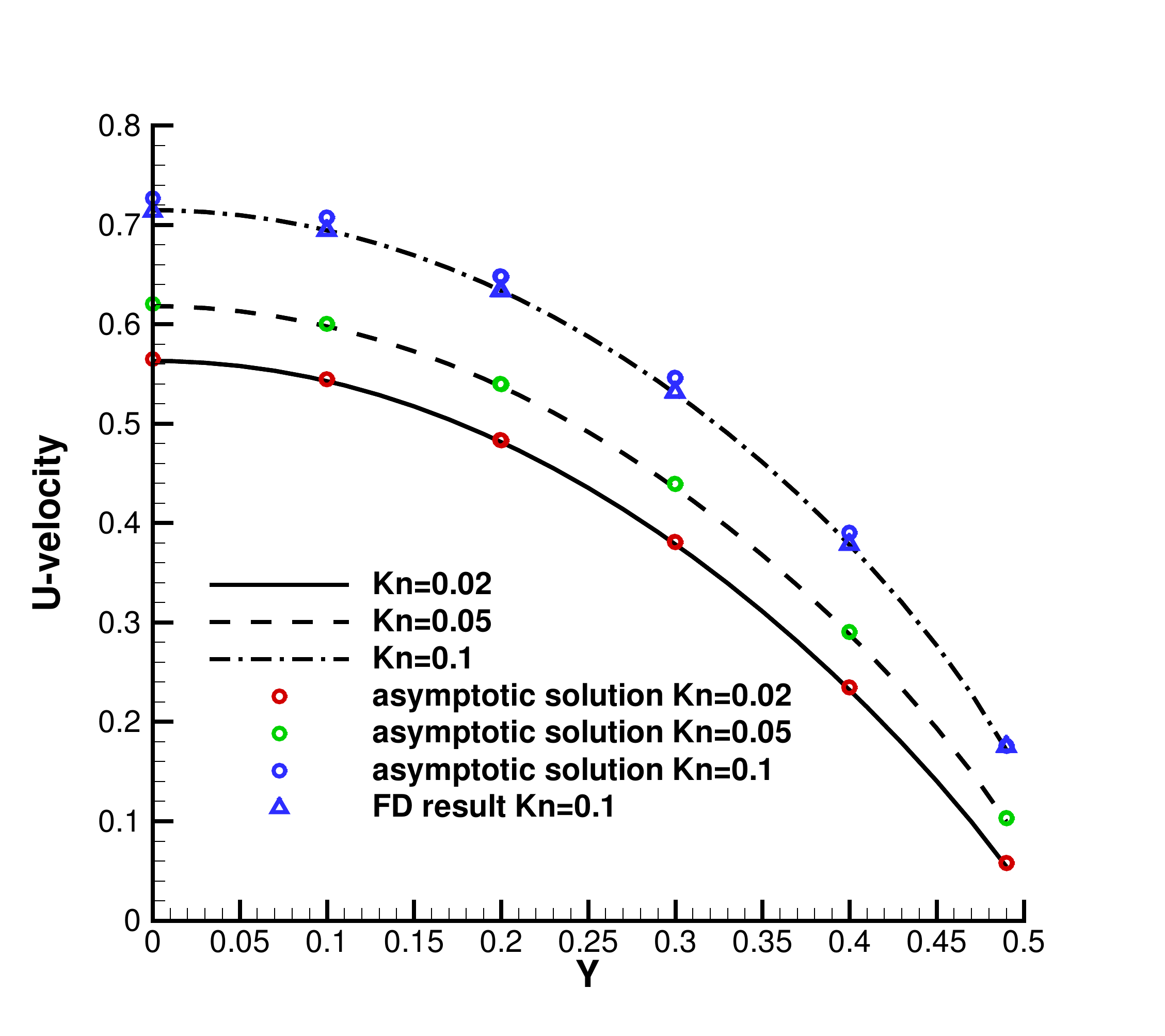}
	}
	\subfigure[Temperature]{
		\includegraphics[width=5cm]{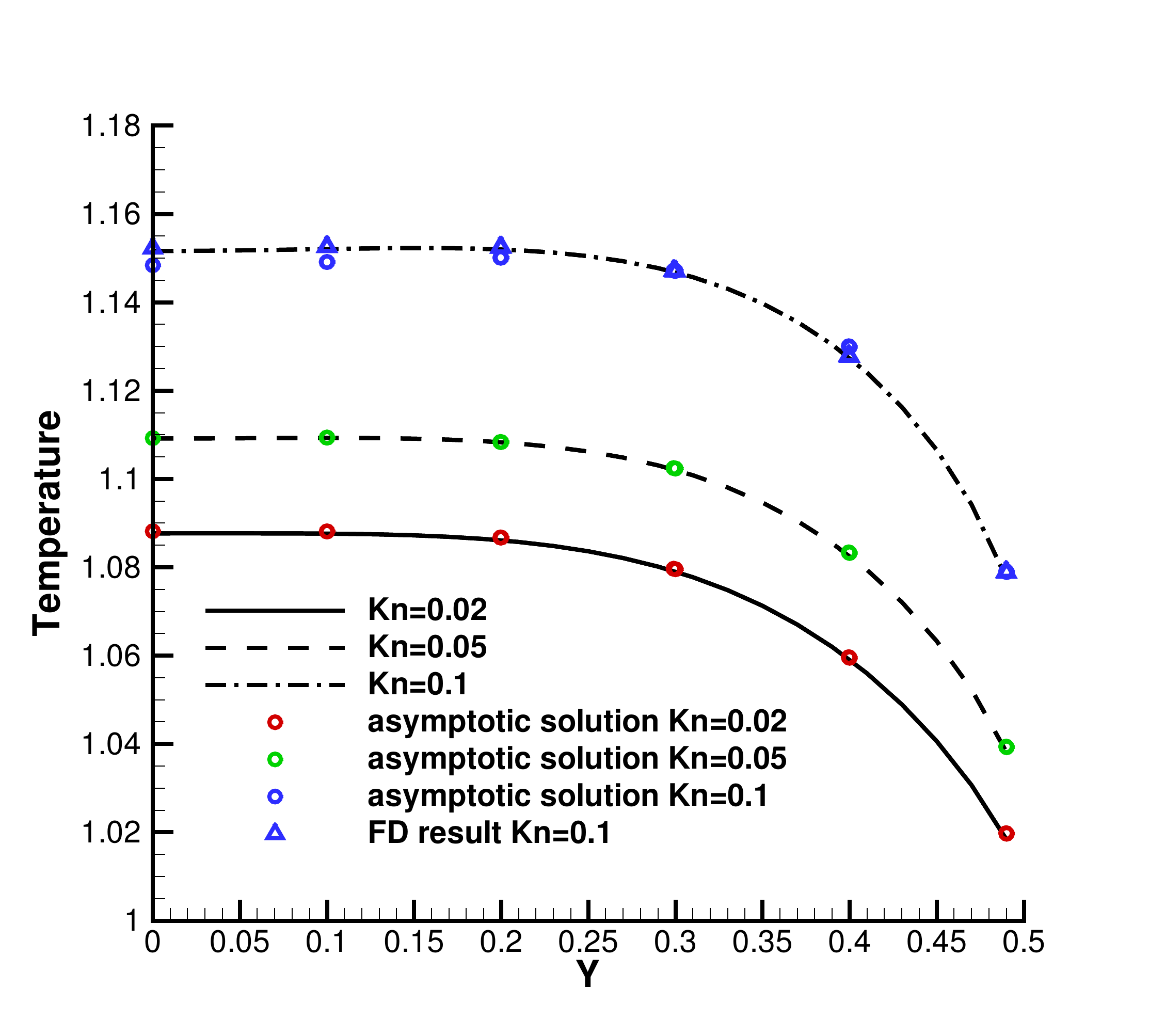}
	}
	\caption{The profiles of density, U-velocity and temperature with $\alpha=2$.}
	\label{pic:ap poiseuille a=2}
\end{figure}

\begin{figure}
	\centering
	\subfigure[Density]{
		\includegraphics[width=5cm]{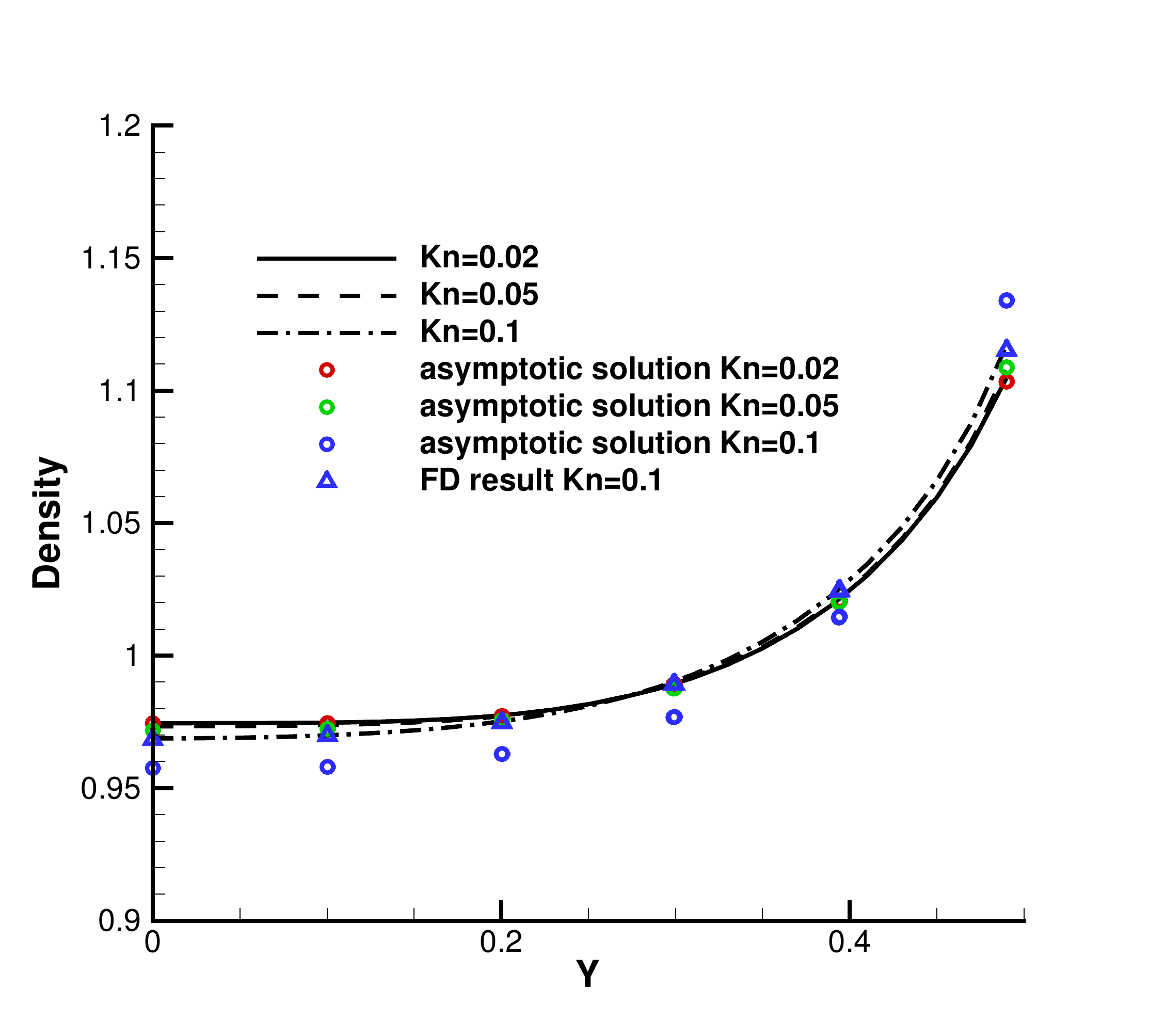}
	}
	\subfigure[U-velocity]{
		\includegraphics[width=5cm]{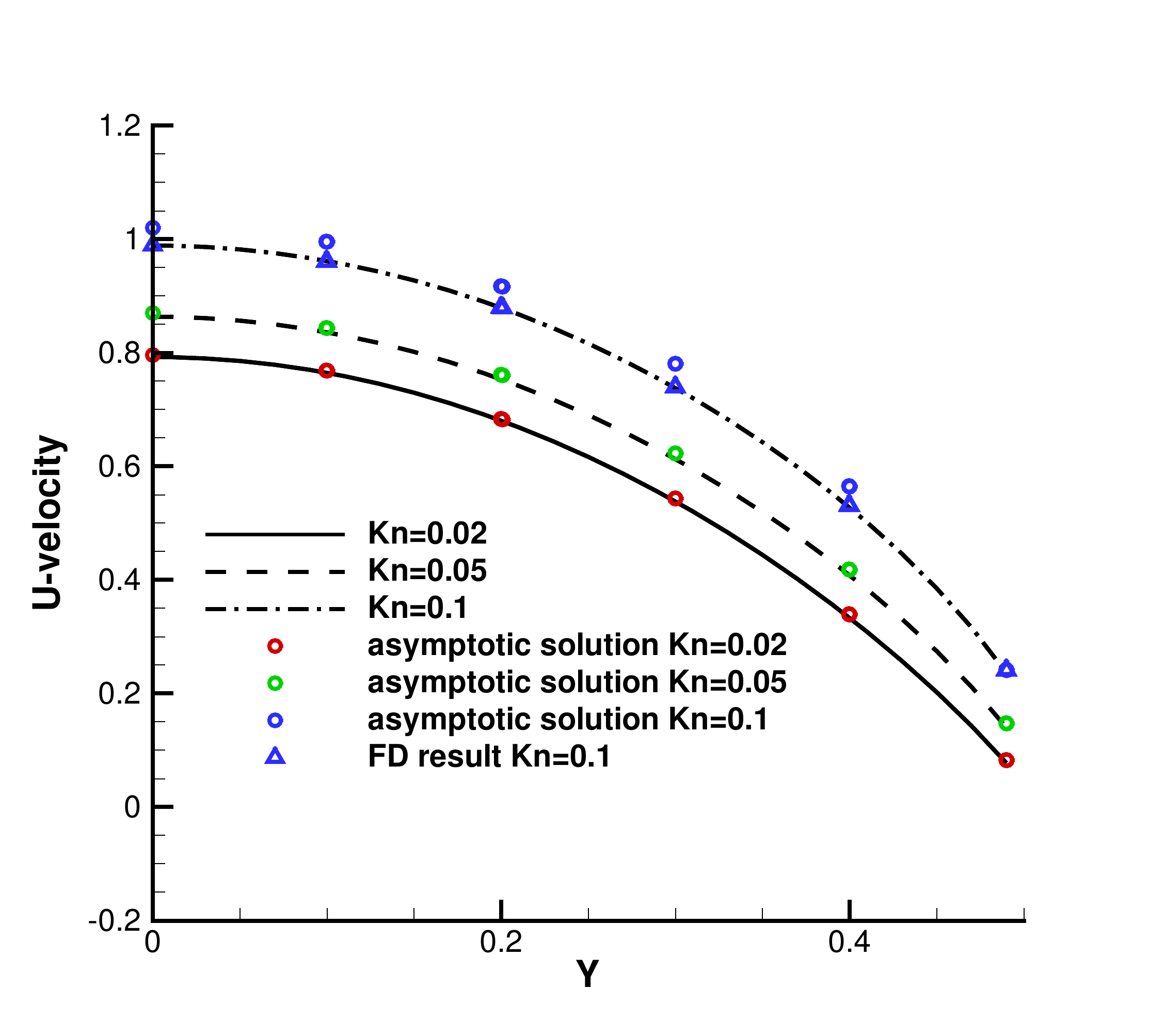}
	}
	\subfigure[Temperature]{
		\includegraphics[width=5cm]{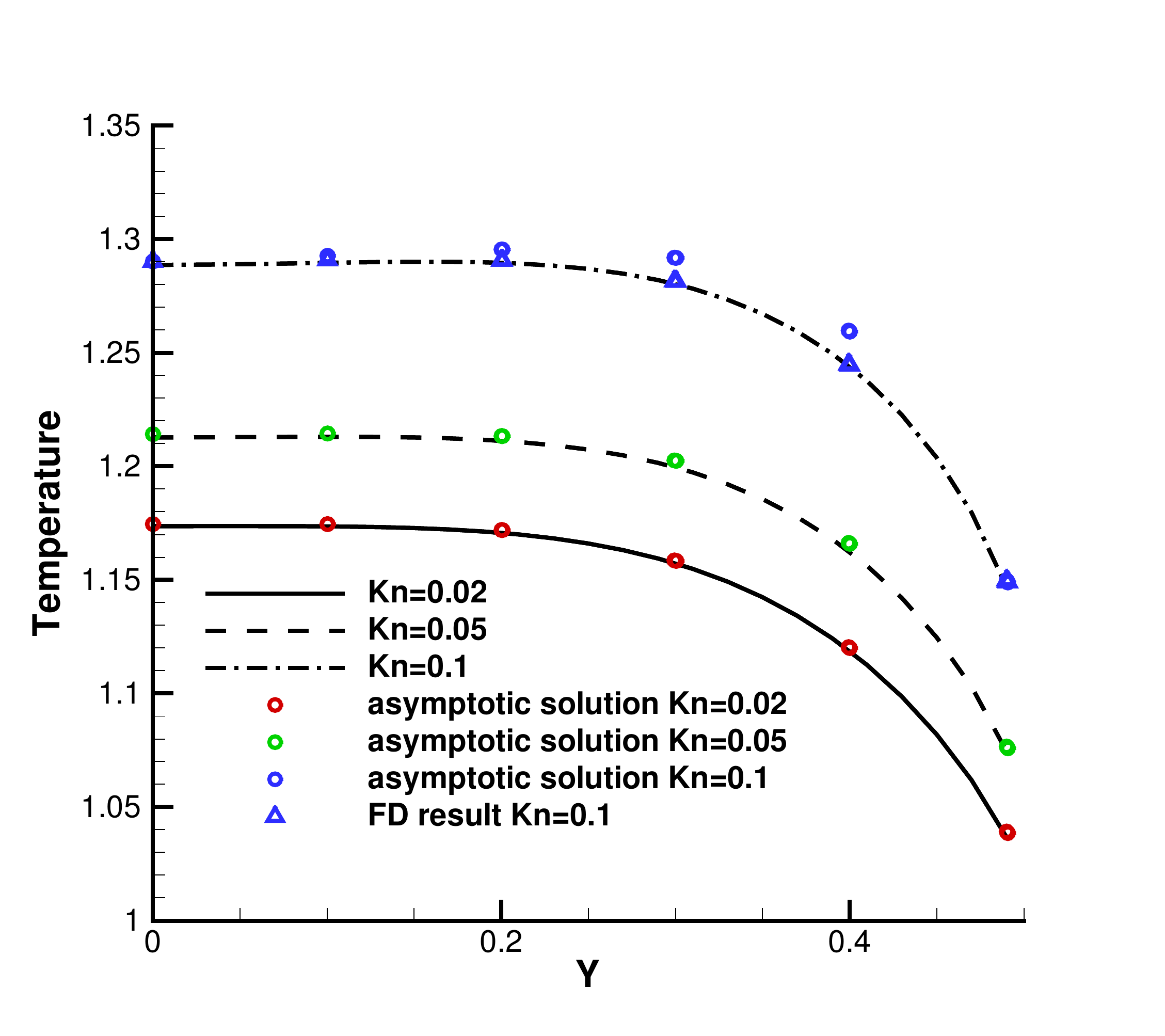}
	}
	\caption{The profiles of density, U-velocity and temperature with $\alpha=3$.}
	\label{pic:ap poiseuille a=3}
\end{figure}

\begin{figure}
	\centering
	\subfigure[$P_{xx}$]{
		\includegraphics[width=5cm]{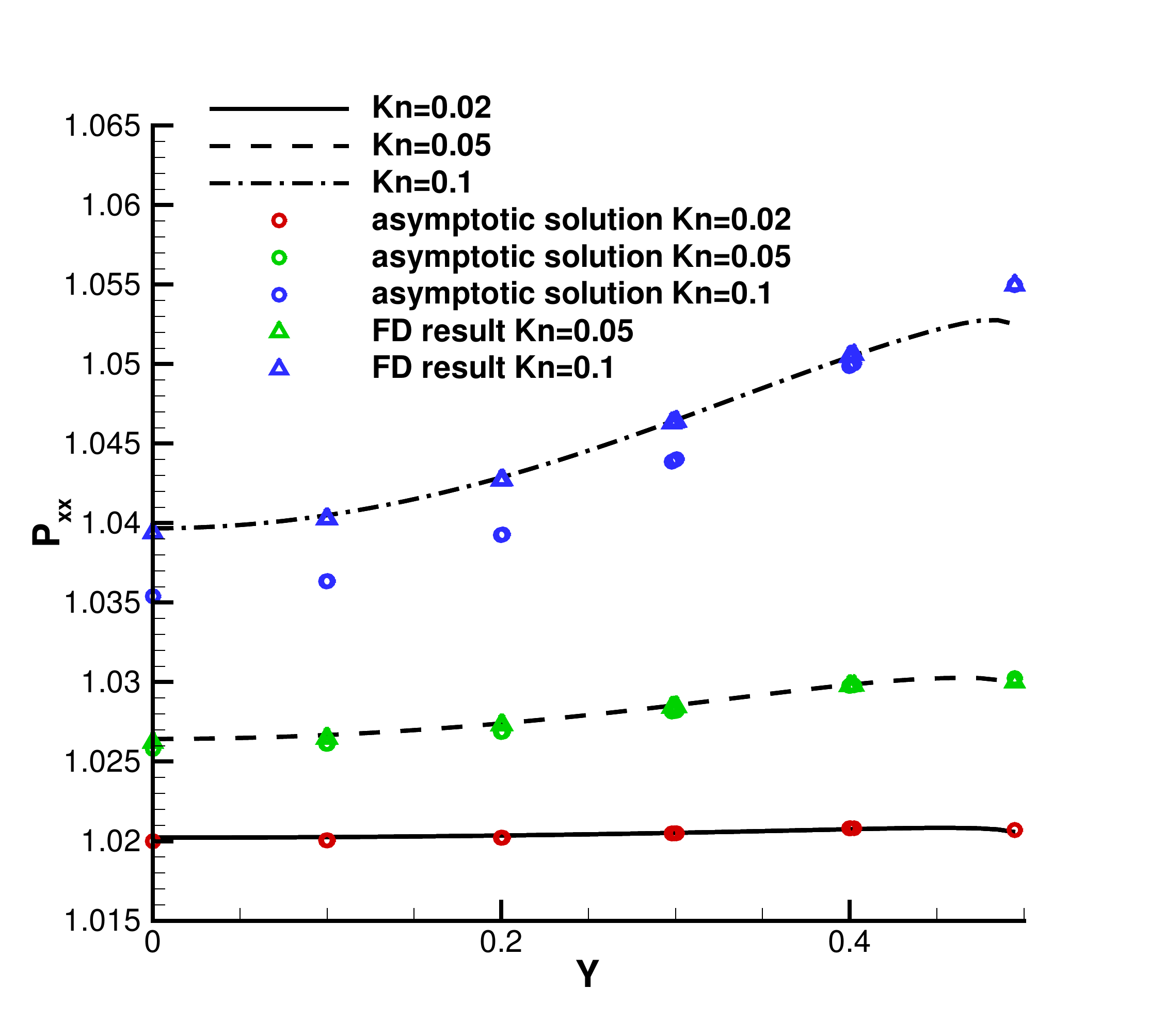}
	}
	\subfigure[$P_{xy}$]{
		\includegraphics[width=5cm]{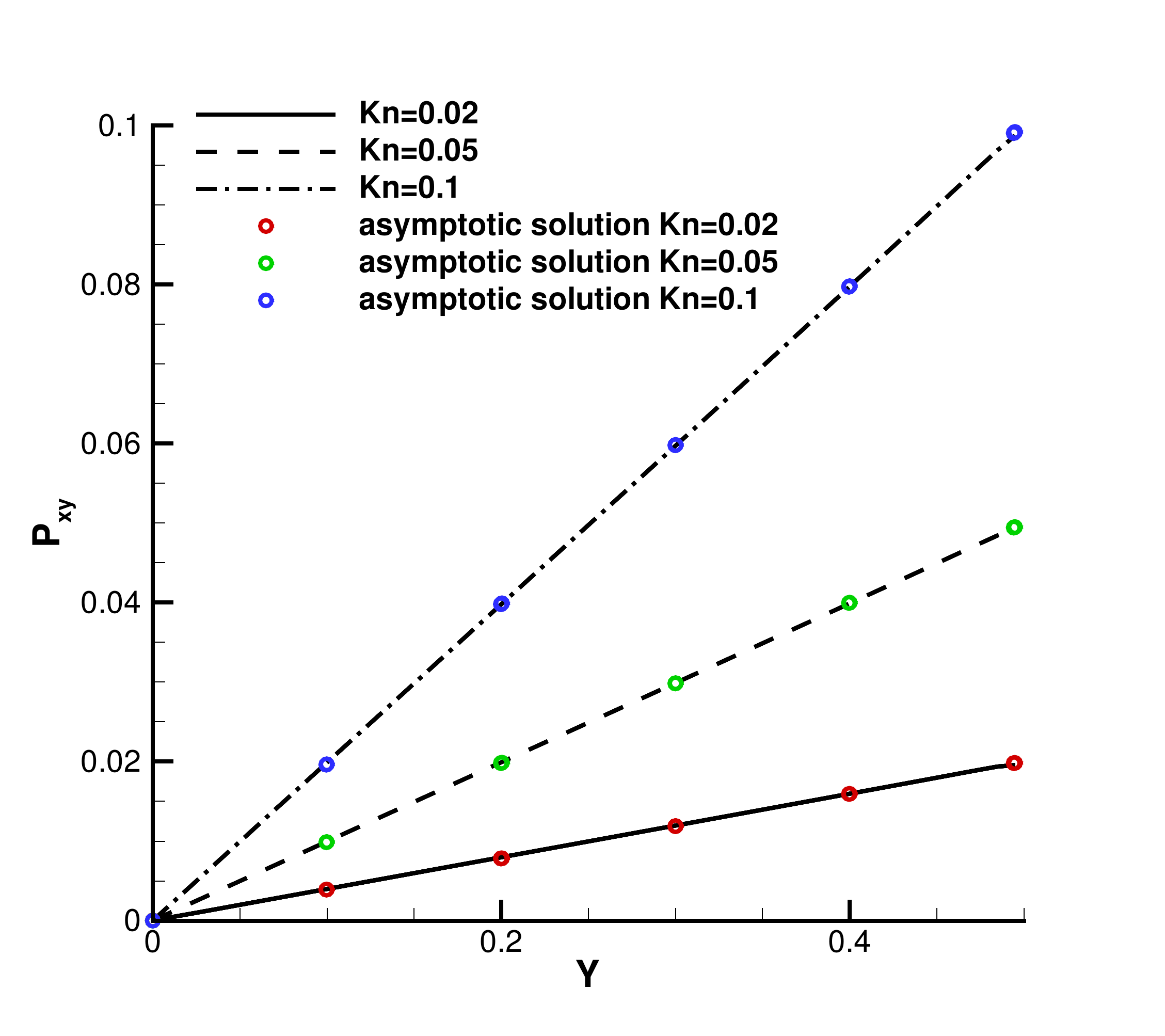}
	}
	\subfigure[$P_{yy}$]{
		\includegraphics[width=5cm]{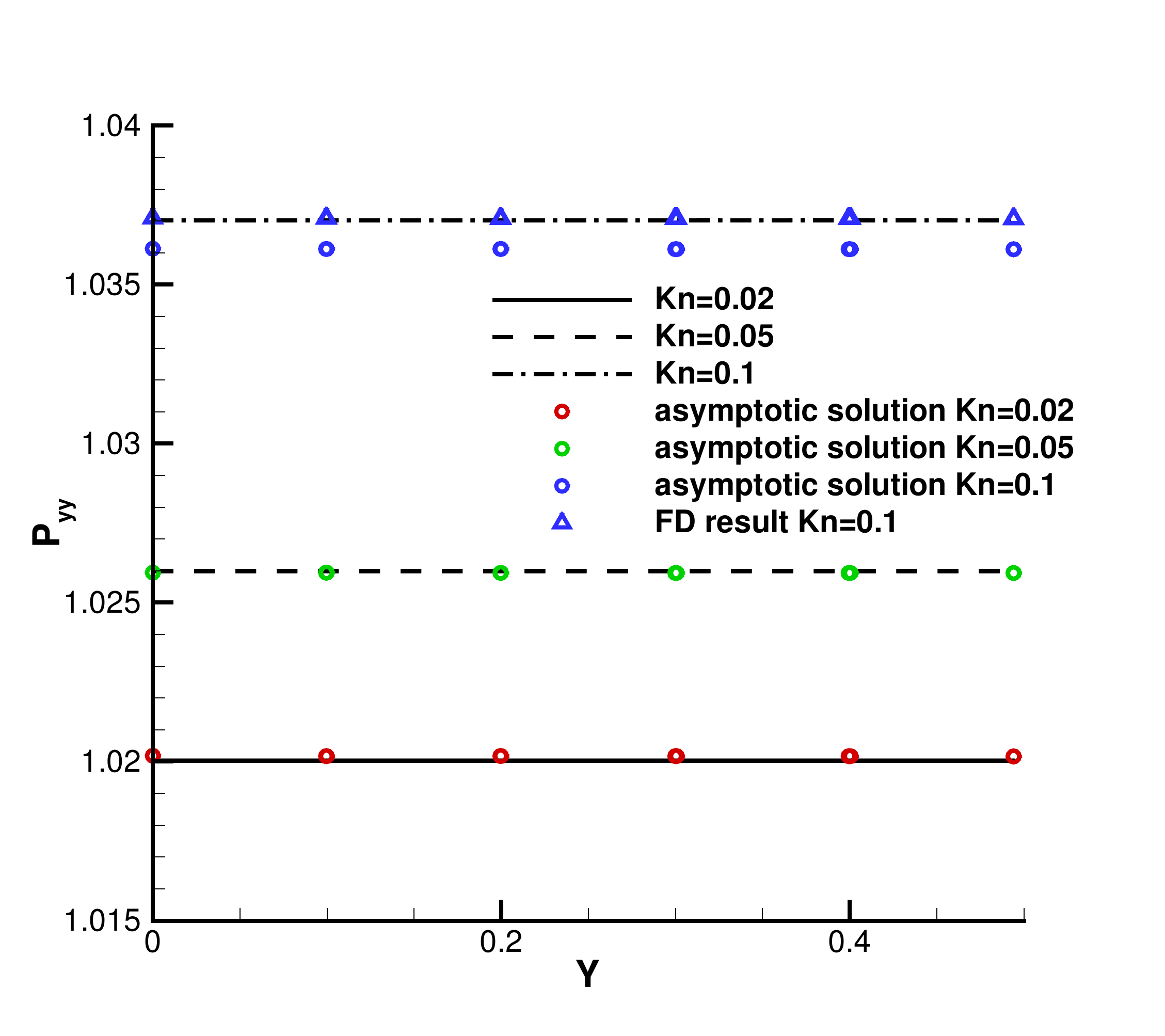}
	}
	\subfigure[$q_{x}$]{
		\includegraphics[width=5cm]{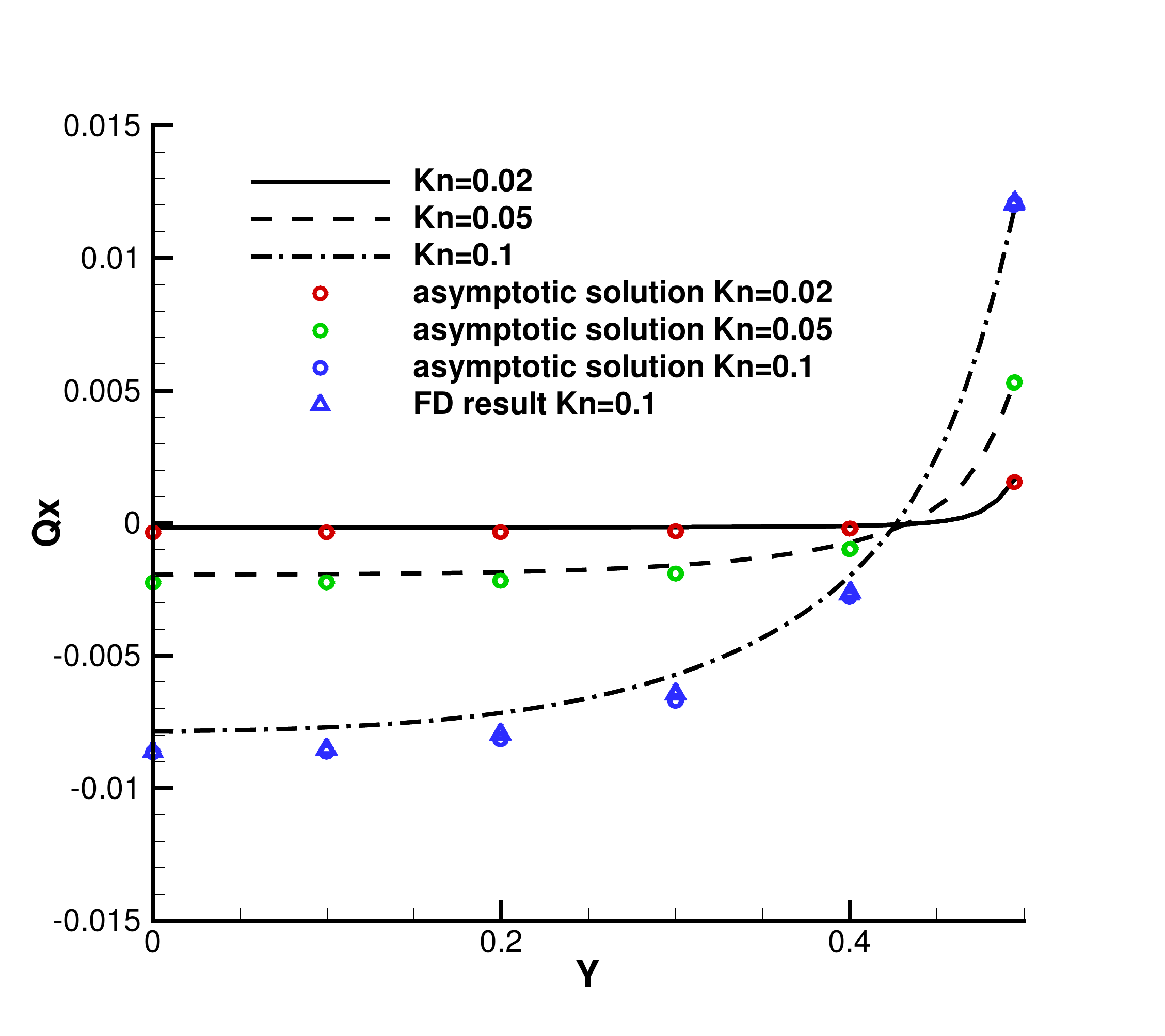}
	}
	\subfigure[$q_y$]{
		\includegraphics[width=5cm]{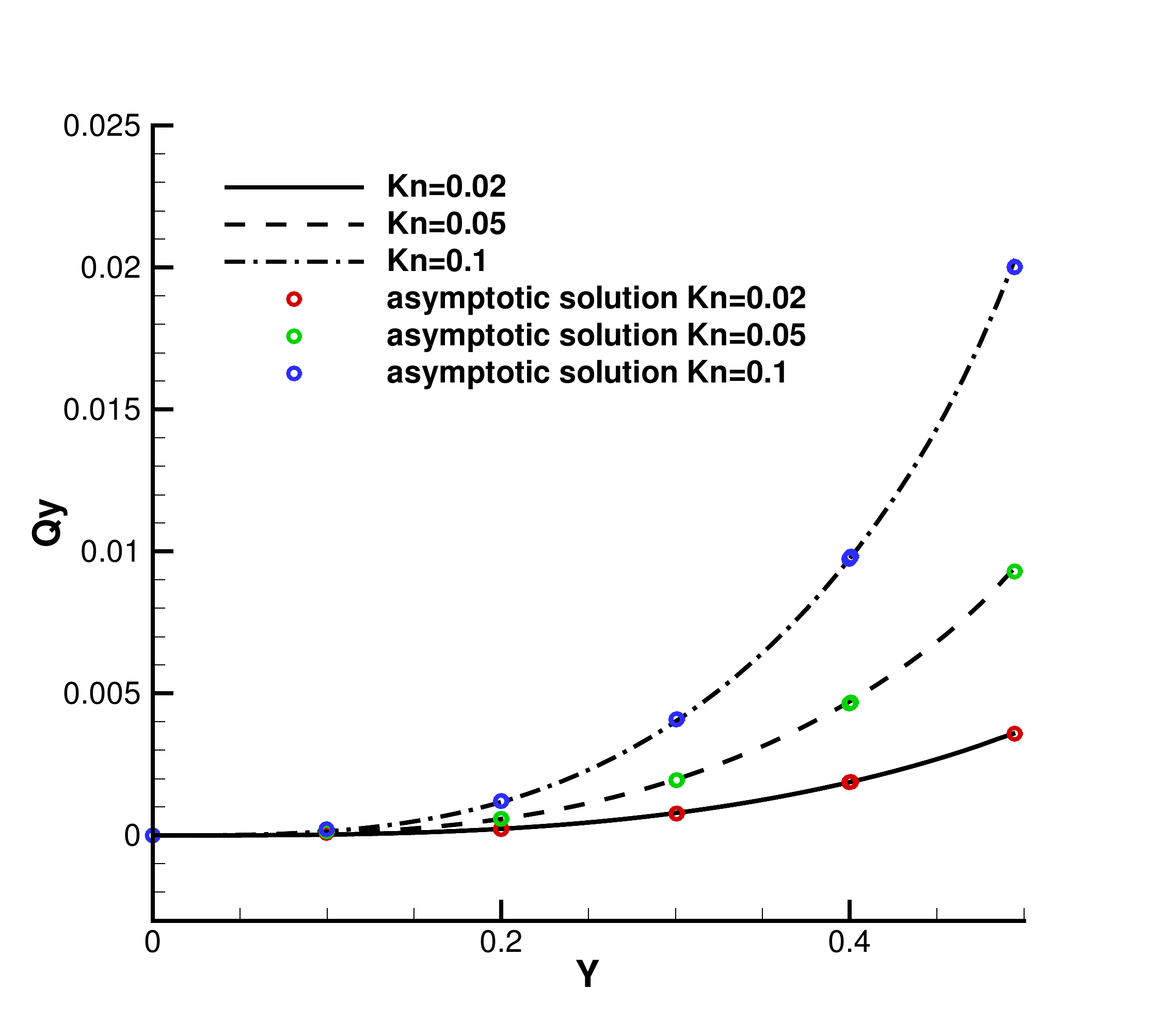}
	}
	\caption{The profiles of stress and heat flux with $\alpha=1$.}
	\label{pic:ap poiseuille second a=1}
\end{figure}

\begin{figure}
	\centering
	\subfigure[$P_{xx}$]{
		\includegraphics[width=5cm]{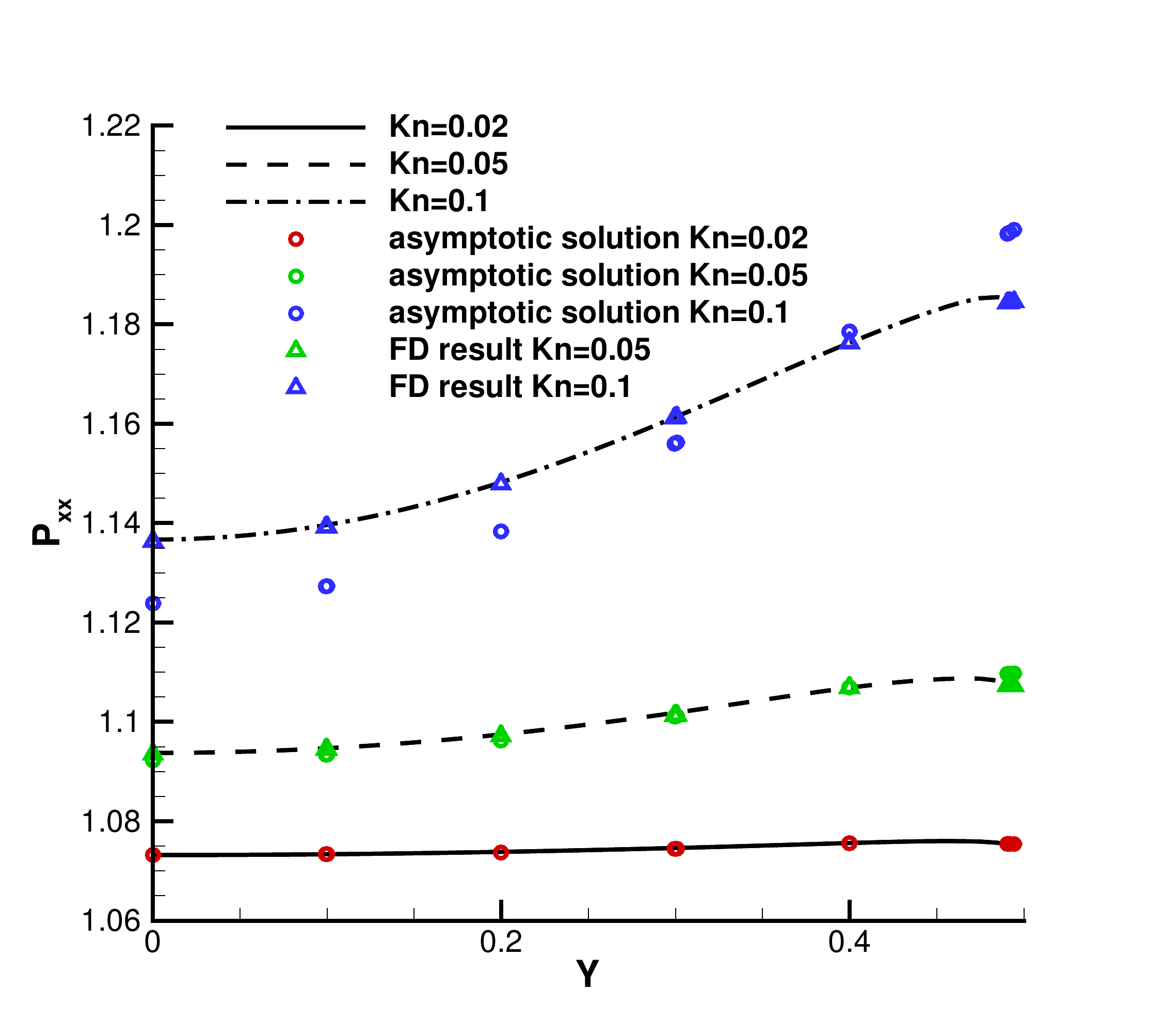}
	}
	\subfigure[$P_{xy}$]{
		\includegraphics[width=5cm]{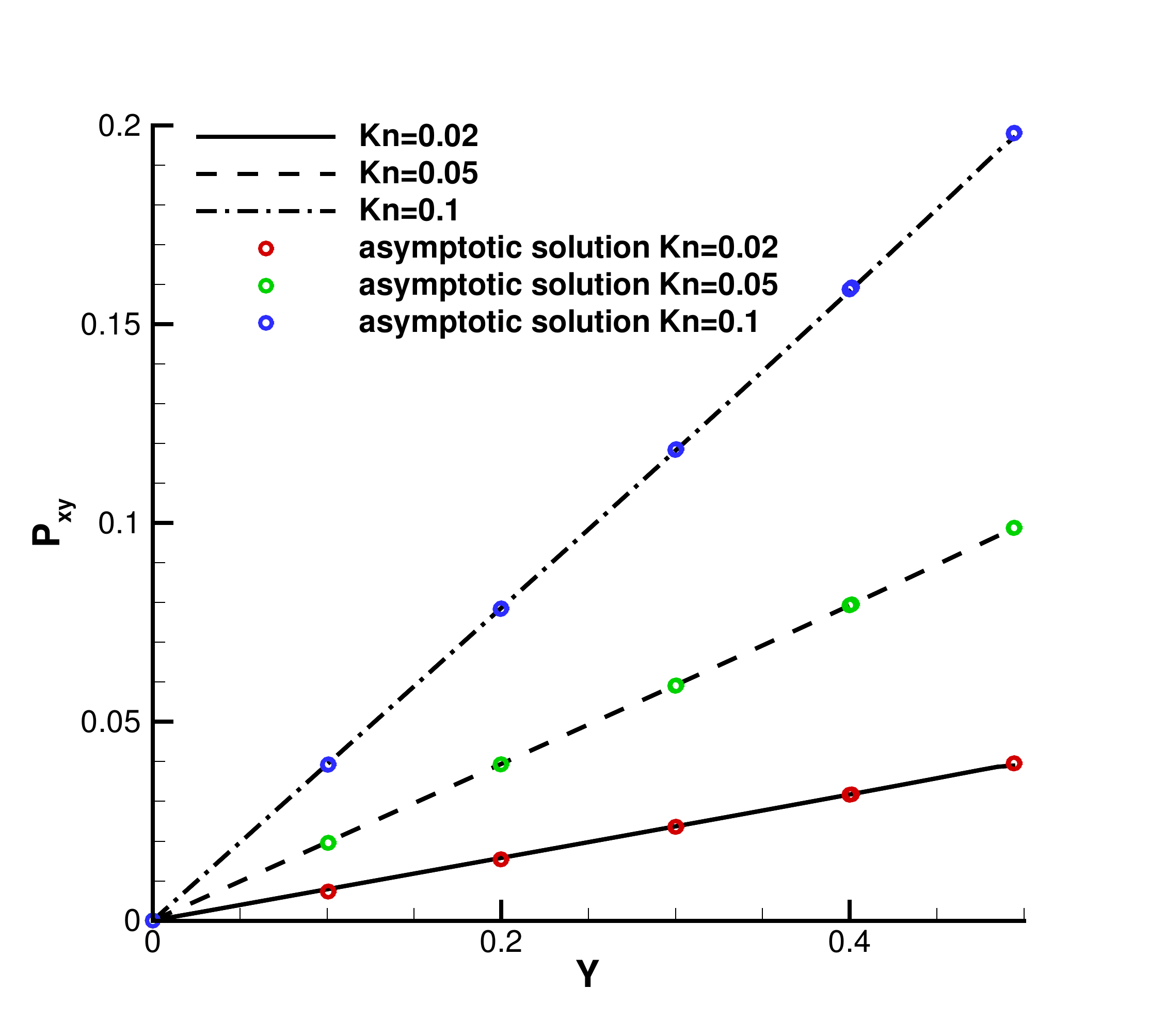}
	}
	\subfigure[$P_{yy}$]{
		\includegraphics[width=5cm]{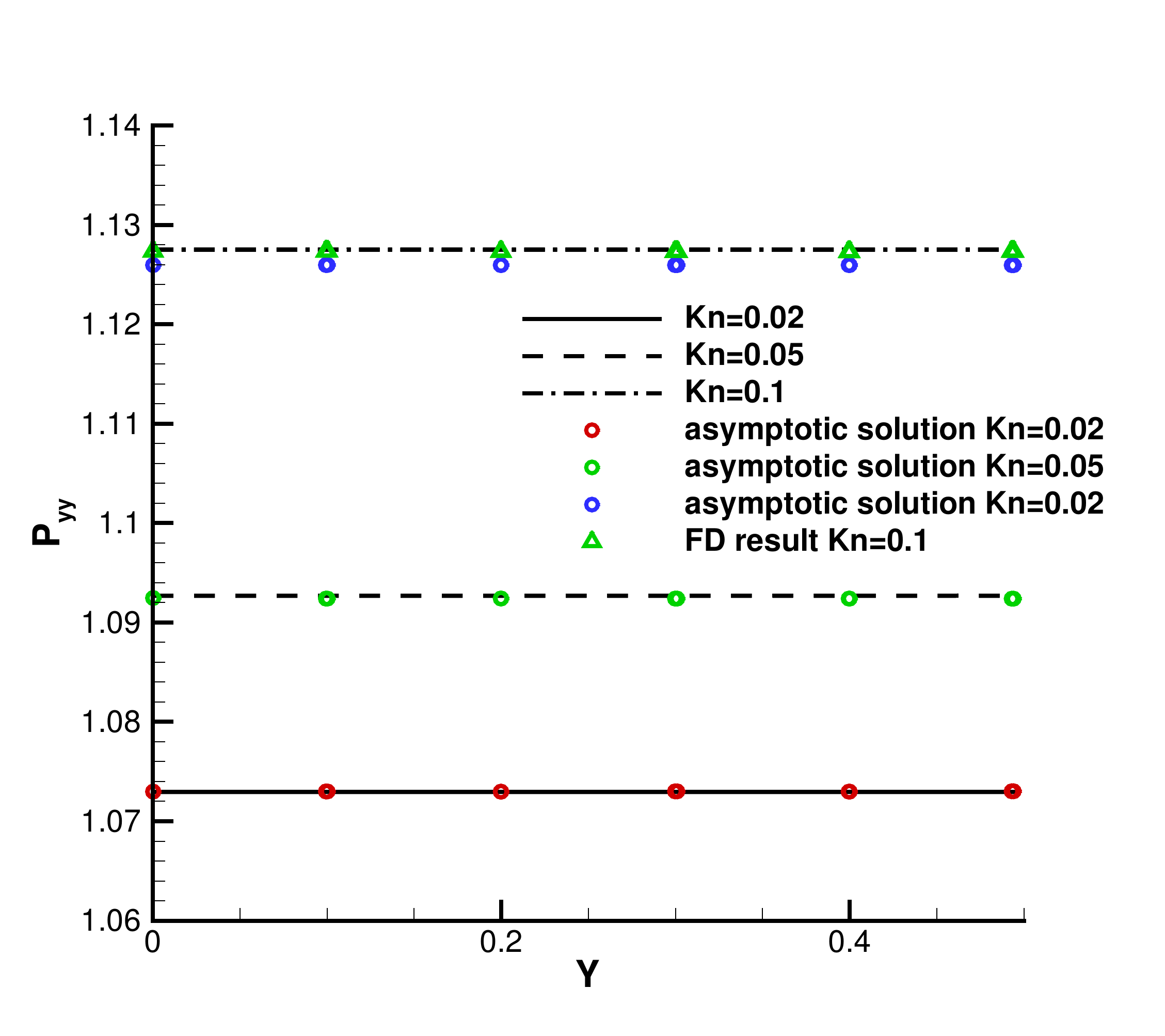}
	}
	\subfigure[$q_{x}$]{
		\includegraphics[width=5cm]{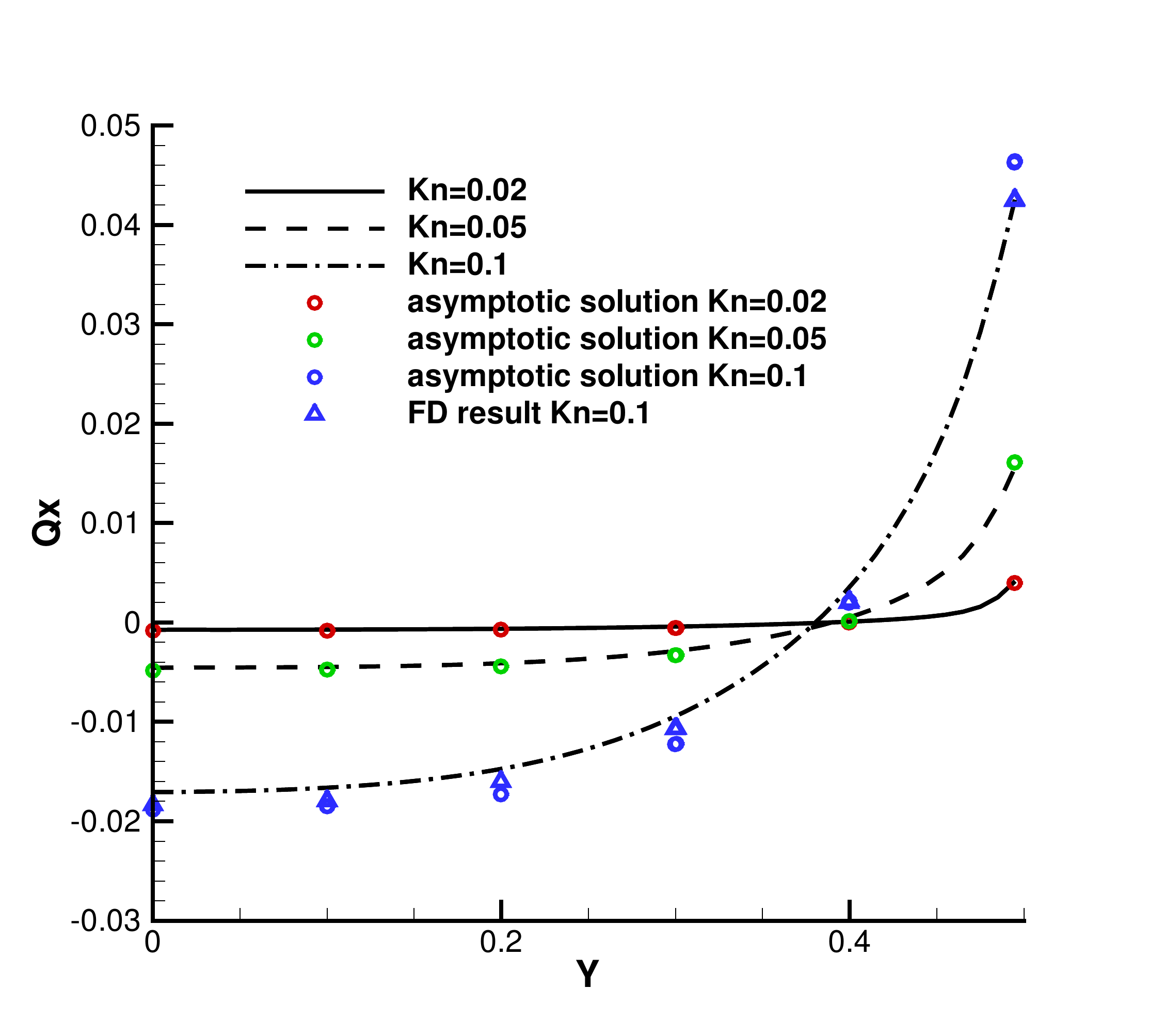}
	}
	\subfigure[$q_y$]{
		\includegraphics[width=5cm]{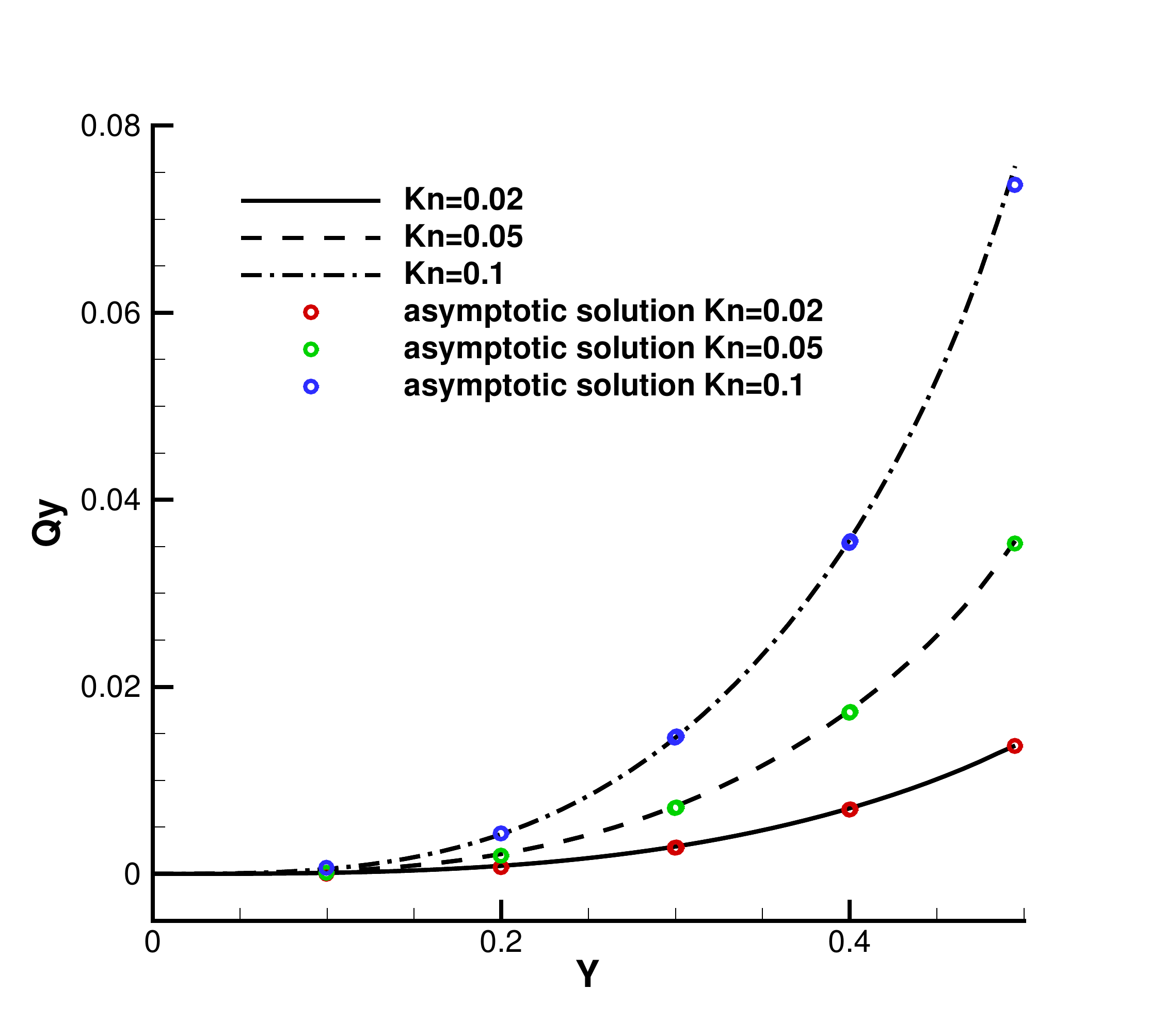}
	}
	\caption{The profiles of stress and heat flux with $\alpha=2$.}
	\label{pic:ap poiseuille second a=2}
\end{figure}

\begin{figure}
	\centering
	\subfigure[$P_{xx}$]{
		\includegraphics[width=5cm]{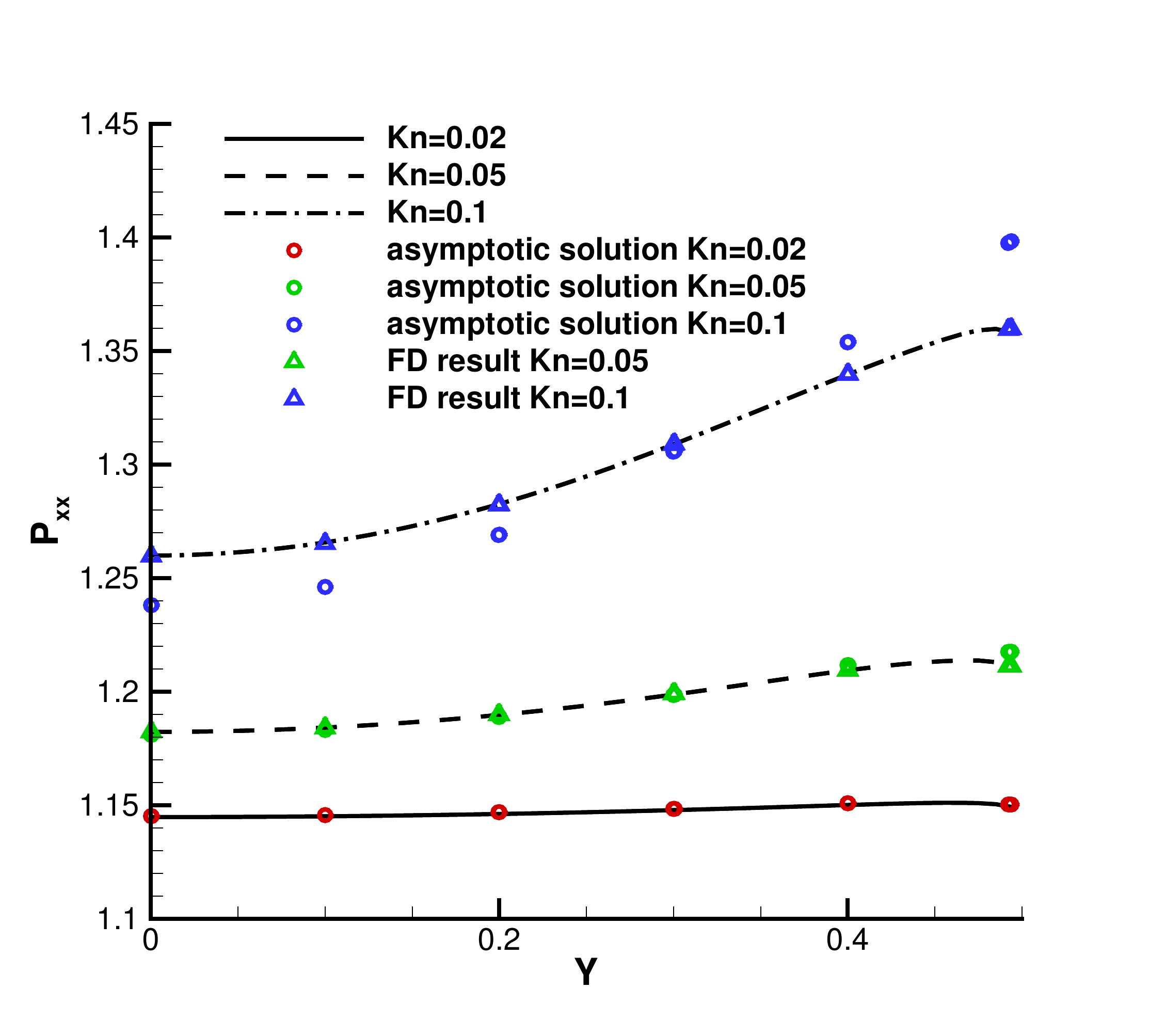}
	}
	\subfigure[$P_{xy}$]{
		\includegraphics[width=5cm]{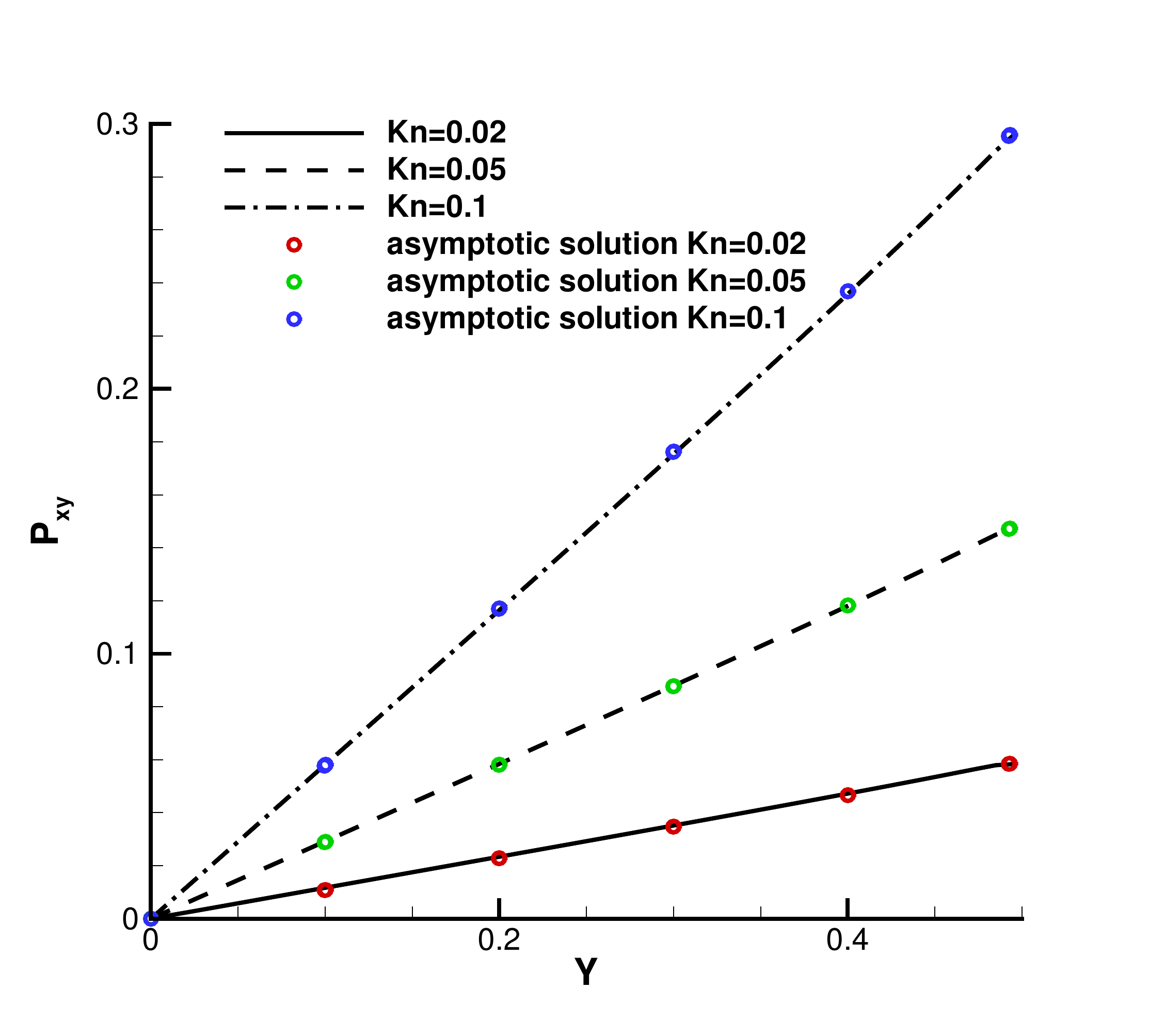}
	}
	\subfigure[$P_{yy}$]{
		\includegraphics[width=5cm]{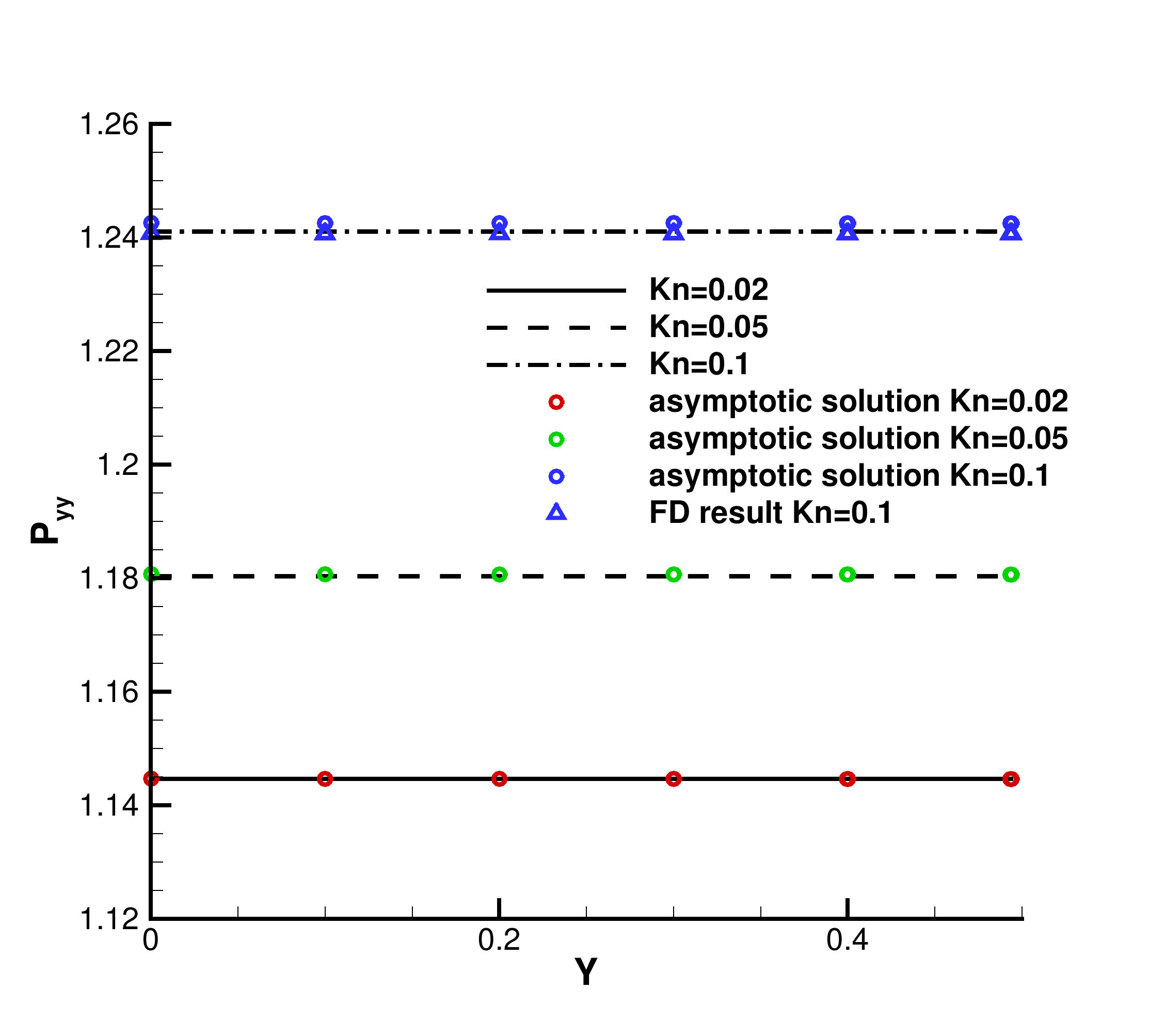}
	}
	\subfigure[$q_{x}$]{
		\includegraphics[width=5cm]{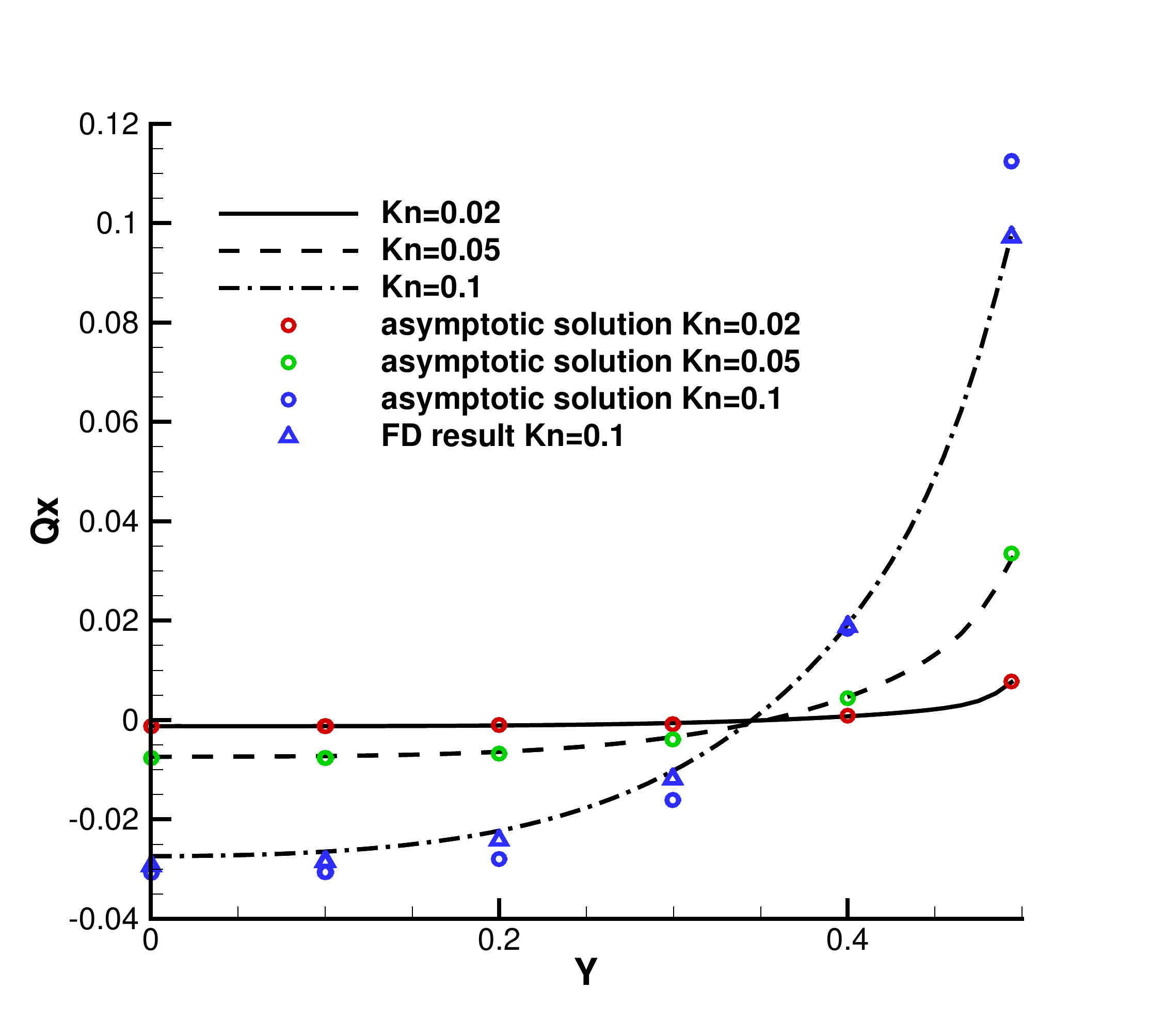}
	}
	\subfigure[$q_y$]{
		\includegraphics[width=5cm]{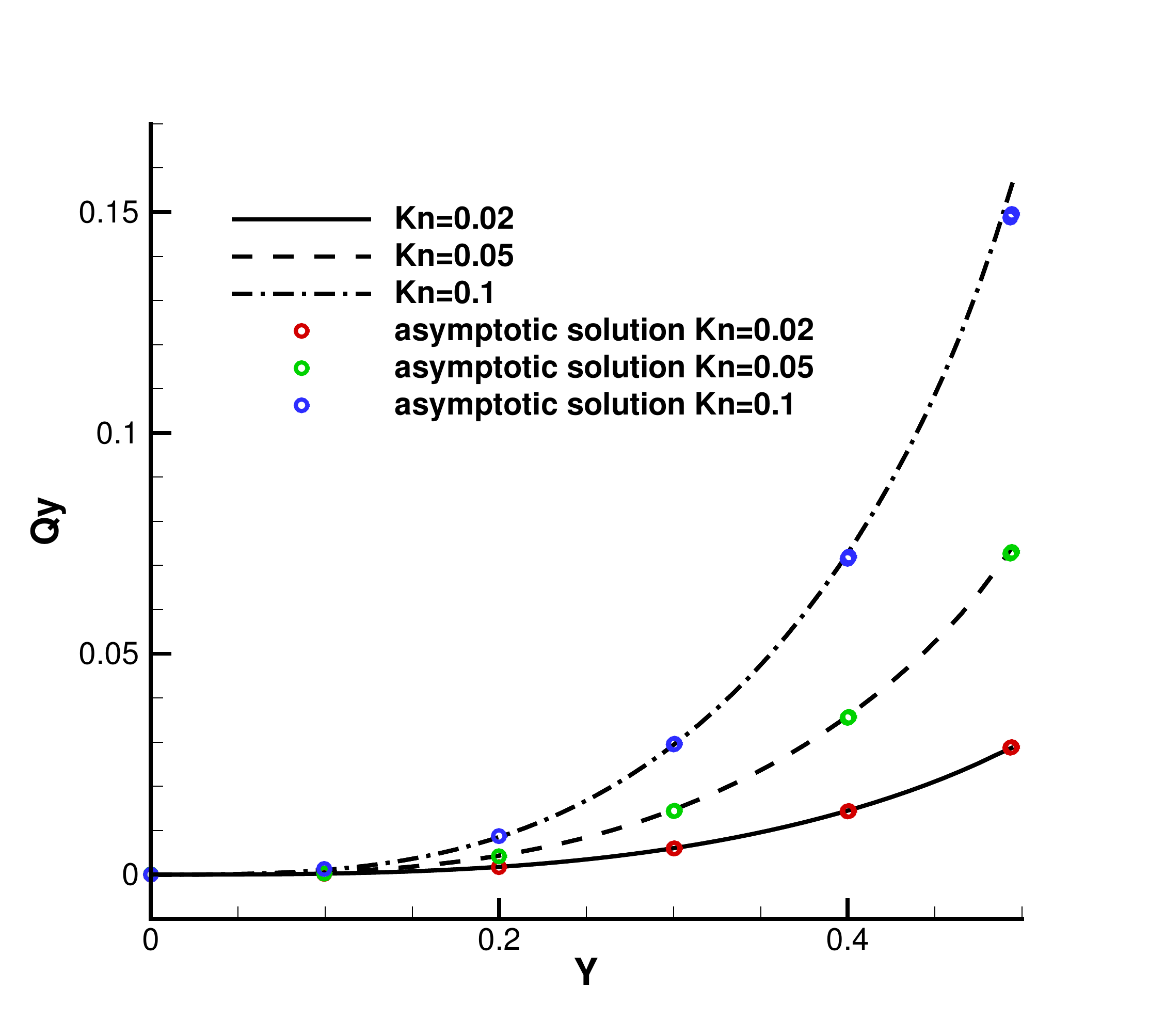}
	}
	\caption{The profiles of stress and heat flux with $\alpha=3$.}
	\label{pic:ap poiseuille second a=3}
\end{figure}

Fig. \ref{pic:ap poiseuille a=1},  \ref{pic:ap poiseuille a=2}, and \ref{pic:ap poiseuille a=3} show the profiles of the density, $U$-velocity and temperature for $\alpha=1$, 2, and 3 in the upper half ($0\leq Y \leq0.5$) of the flow domain.
Fig. \ref{pic:ap poiseuille second a=1},  \ref{pic:ap poiseuille second a=2}, and \ref{pic:ap poiseuille second a=3} are the results of the stress tensor components $P_{xx}$, $P_{xy}$ and $P_{yy}$ of the stress tensor and the heat flux $q_x$ and $q_y$.
The lines (solid, dashed and dash dot) are the results calculated by the well-balanced UGKS, and the circle indicates asymptotic solutions, whereas the delta denotes the reference results given by Aoki's finite difference method.
It can be seen that in the cases with low Knudsen number, the UGKS solutions correspond well with the asymptotic results.
For $\alpha=2$ and 3, there is a small discrepancy between numerical solutions and asymptotic ones.
The reason for such a deviation is that the asymptotic analysis is confined to the second order in the Knudsen number.
When the Knudsen number and external force are relative large, the truncated high-order Hilbert solutions will become significant in comparison with the leading and first-order ones, and the validity of corresponding Knudsen layer correction needs to be reconsidered as well.
Therefore, in the cases with $Kn=0.1$, the finite difference solutions (denoted by "FD result" in the figure) are supplemented as the benchmark results, which are consistent with the well-balanced UGKS solutions.

\begin{figure}
	\centering
	\subfigure[Kn=0.02]{
		\includegraphics[width=5cm]{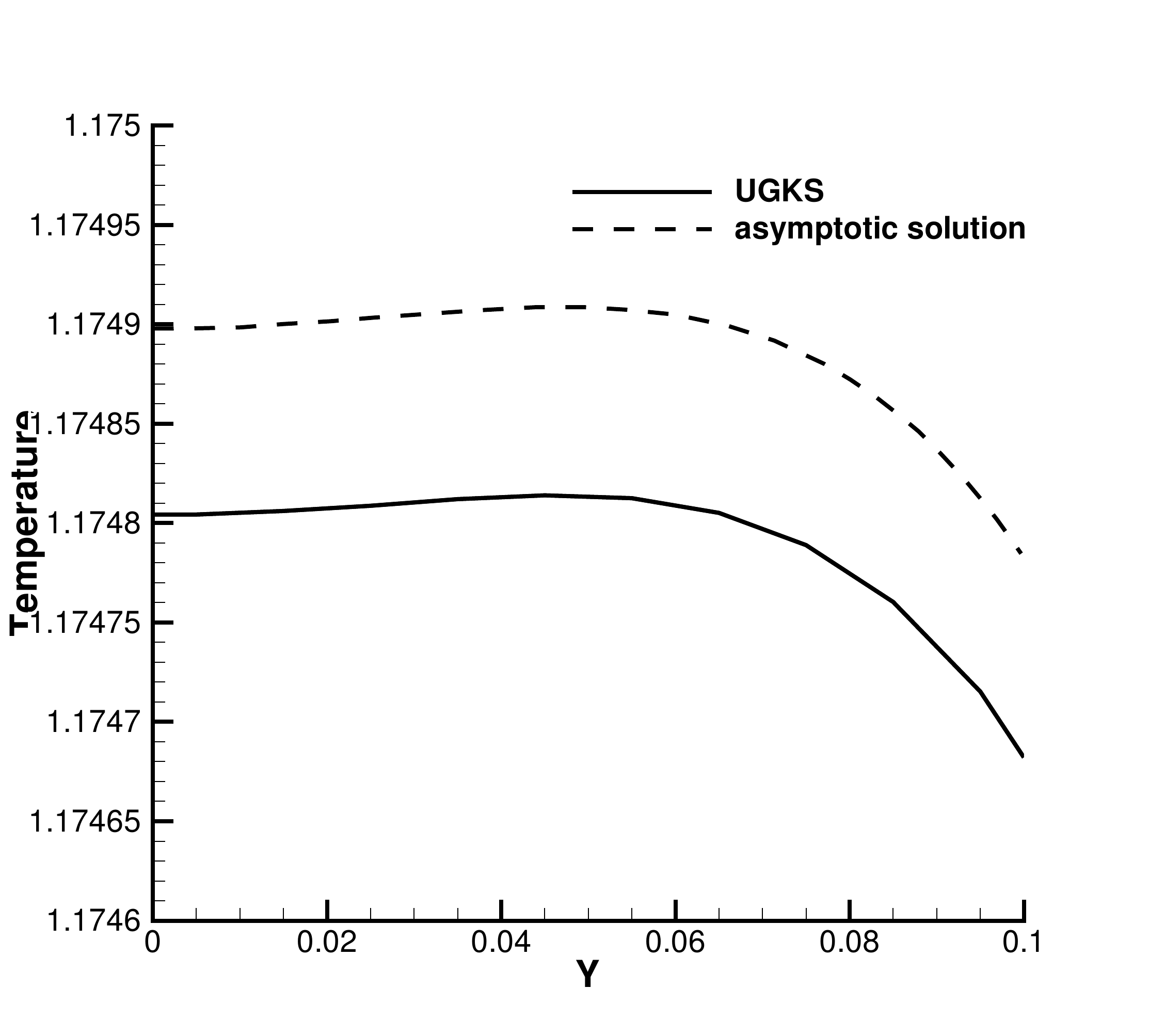}
	}
	\subfigure[Kn=0.05]{
		\includegraphics[width=5cm]{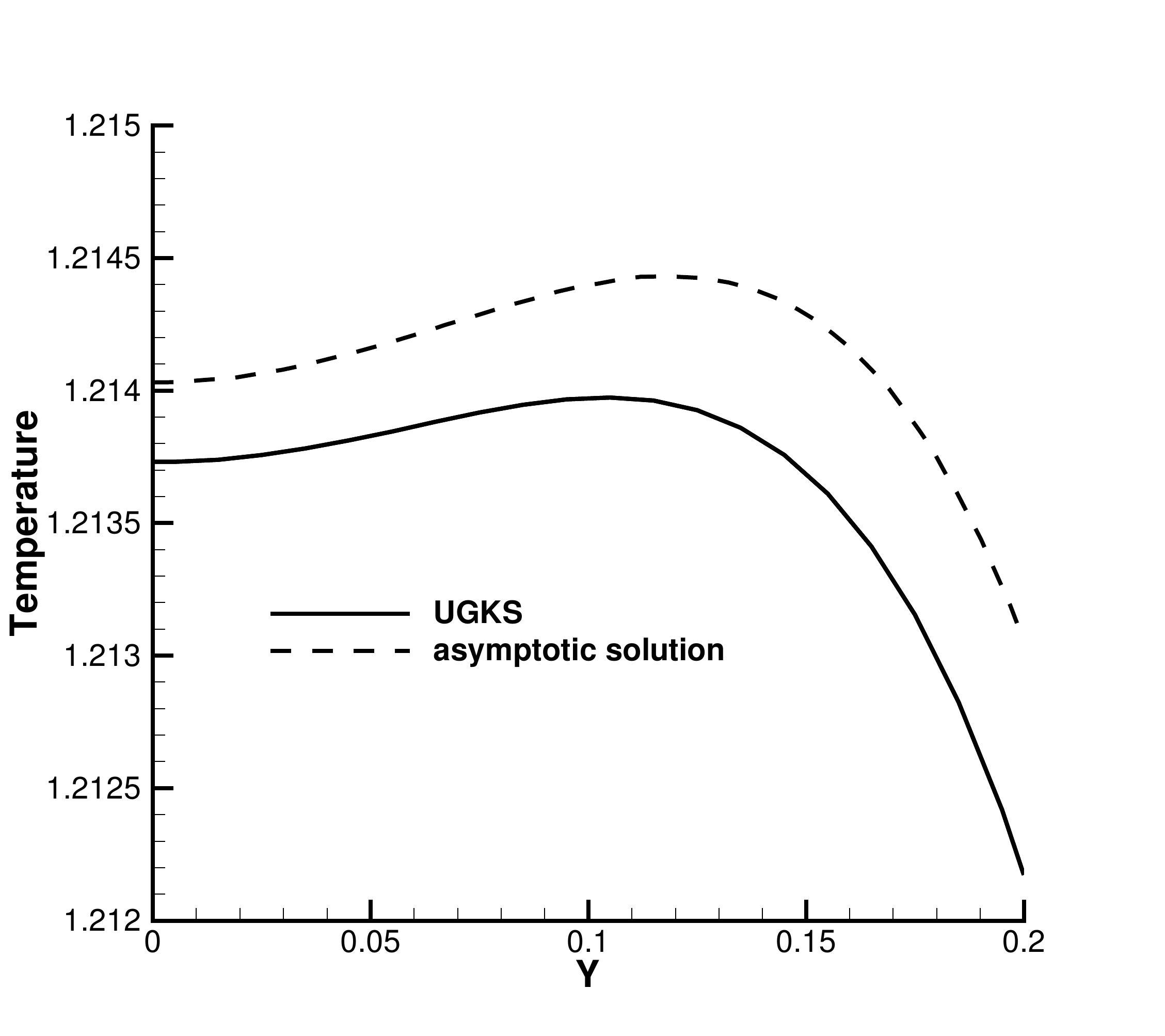}
	}
	\subfigure[Kn=0.1]{
		\includegraphics[width=5cm]{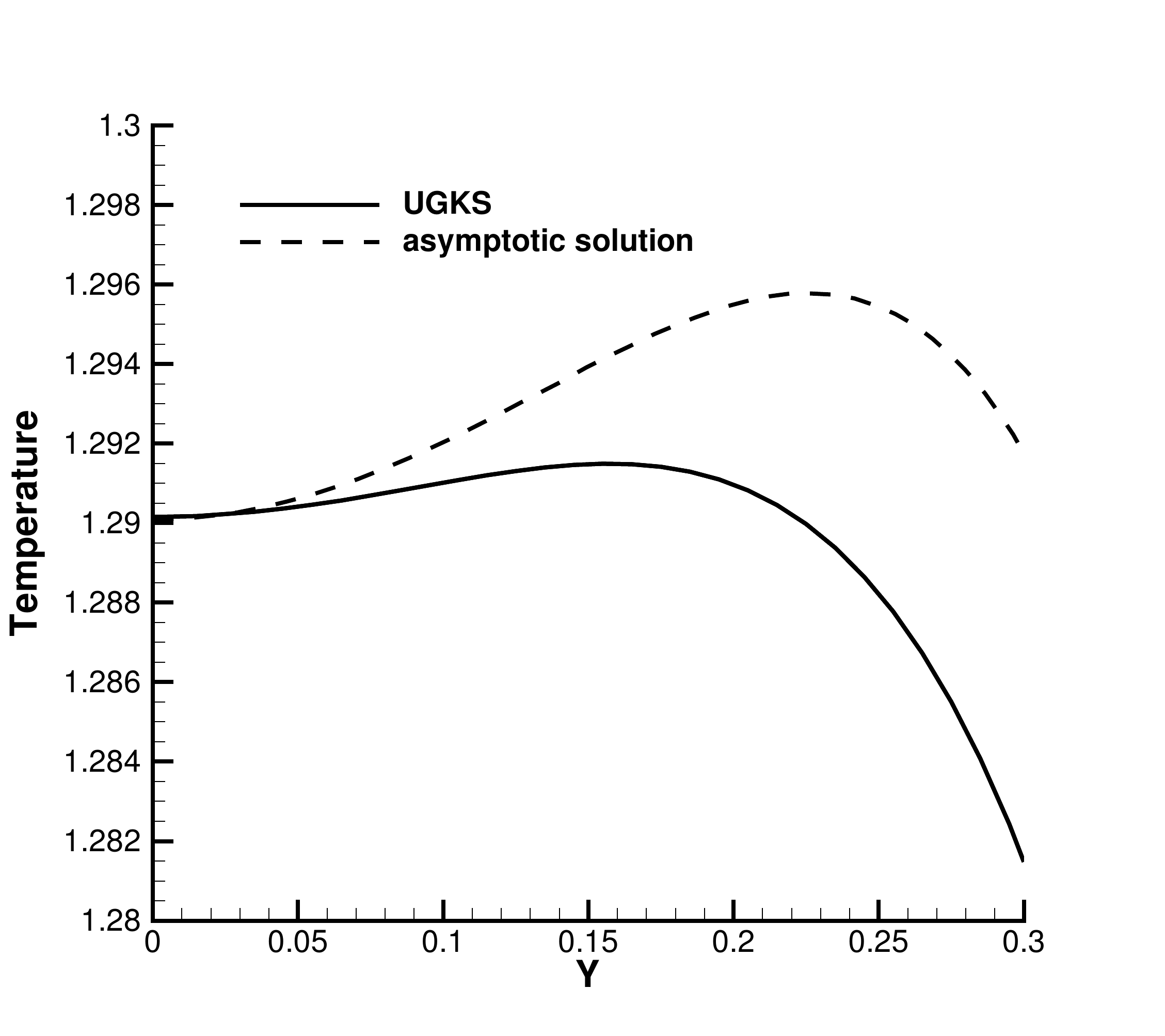}
	}
	\caption{The temperature profile in the central part with $\alpha=3$.}
	\label{pic:ap poiseuille center}
\end{figure}

As reported in the previous research, an interesting non-equilibrium phenomenon in this force driven system is the bimodal temperature profile, with a hollow near the center between two plates.
This effect was first pointed by Malek et al. \cite{mansour1997validity} using the DSMC method, and reproduced by the kinetic simulation and theory \cite{aoki2002poiseuille,xu2003super,uribe1999burnett,hess1999temperature}.
The localized temperature profiles calculated by the UGKS and asymptotic solutions are presented in Fig. \ref{pic:ap poiseuille center}.
In general,  the temperature minimum in this case can be attributed to the contribution of higher order terms, and can be resolved with higher-order macroscopic equations, such as the super-Burnett one \cite{xu2003super}.
Although in Fig. \ref{pic:ap poiseuille center} the second-order asymptotic analysis overestimates the strength of the temperature hollow, it is obvious that even for this simple case with rectangular geometry and uniform, weak external force, the Navier-Stokes equations fail to describe the accurate gas evolution, especially for the non-equilibrium effects, and kinetic modeling and computation may become necessary.
This case validates the capacity of the well-balanced UGKS to simulate non-equilibrium gas dynamics under external force field.

\subsection{Static heat conduction}
Consider a column of gas enclosed between two infinite parallel plates at $x=0$ and $x=1$, both maintained with different temperature under a constant external force field perpendicular to the plates.
Instead of studying the Rayleigh-B\'enard convection \cite{benard1901tourbillons, rayleigh1916lix, sone1997benard, stefanov2002rayleigh}, we confine us to a static heat conduction
problem to evaluate the correlation between heat flux and external force field.
This is possible if the governing dynamic parameters don't approach the critical values, for example, the Rayleigh number satisfies $Ra<Ra_c\simeq 1700$ in the continuum, incompressible limit.

Two ways can be used to describe the flow dynamics in this example.
If we use the Navier-Stokes-Fourier equations to study this static system,  with the external force along $x$-direction, the simple one-dimensional example indicates the following relations,
\begin{equation}
\frac{\partial}{\partial y} p=\rho \phi_x,
\end{equation}
\begin{equation}
\frac{\partial}{\partial x} \left(\kappa(y) \frac{\partial}{\partial x} T\right)=0,
\label{NS heat flux}
\end{equation}
where $\kappa$ is the heat conductivity coefficient.
At the same time, it is also feasible to turn to steady one-dimensional steady BGK equation, i.e.,
\begin{equation}
u\frac{\partial f}{\partial x} + \phi_x\frac{\partial f}{\partial u} = \frac{2}{\sqrt{\pi}} \frac{1}{ Kn} {\rho} ({f^+}-{f}),
\label{eqn:bgk heat}
\end{equation}
where the collision frequency is $1/\tau = 2\rho/\sqrt{\pi}{Kn}$. 
Here all the flow variables are dimensionless unless special statements.
As is analyzed in Section 2, the external forcing term will influence the heat evolution process,
resulting in a deviation of the profile away from the above theoretical solution given in Eq. (\ref{NS heat flux}).

We use the well-balanced UGKS with 100 physical cells and 101 velocity points to simulate this case.
The temperature ratio of the cold wall to the hot one is set up with $r=T_c/T_h=0.9$.
The initial uniform gas is at rest, with the temperature same as the hot wall.
An external acceleration $\phi_x$ is imposed along the direction of temperature gradient to the system with a series of different values $\phi_x=\pm 0.001, \pm 0.002, \pm 0.003, \pm 0.005, \pm 0.01, \pm 0.02$.
Here the positive direction of external force is aligned with the heat flux, i.e., if the external force points from hot region towards cold one then the force is positive and vice versa.
The temperature ratio and external force are both set up with relative small values to allow the existence of static heat conduction.

It is pointed out in \cite{sone1997benard} that under external force field the characteristic line of BGK-type kinetic equation in Eq. (\ref{eqn:bgk characteristic}) are distorted to be convex to the opposite direction of force, and a discontinuity of the distribution function will be developed near the solid boundary.
Such a discontinuity cannot be described accurately with a Maxwell-type diffusive boundary condition.
However, as illustrated in \cite{sone2012kinetic, Sone1971Flow}, the direct influence of gas-surface interaction on local flow field becomes negligible as far as 10-15 particle mean free paths away from the boundary owing to intermolecular collisions, and an inaccurate treatment of the discontinuity does not seem to disturb the coarse-grained gas behavior in the bulk region.
As we are concerned about the net contribution of external force on heat evolution, thus we confine us to the middle point $x=0.5$ of the flow domain to minimize the effect of boundary discontinuity and study the steady gas behavior in the absence of macroscopic flow.
The reference Knudsen number is set up with $Kn=0.001$ and $0.01$ to reduce the influence of the boundary with relatively small particle mean free path in the near equilibrium regime.

\begin{figure}[htb!]
	\centering
	\subfigure[$Kn_{ref}=0.001$]{
		\includegraphics[width=7cm]{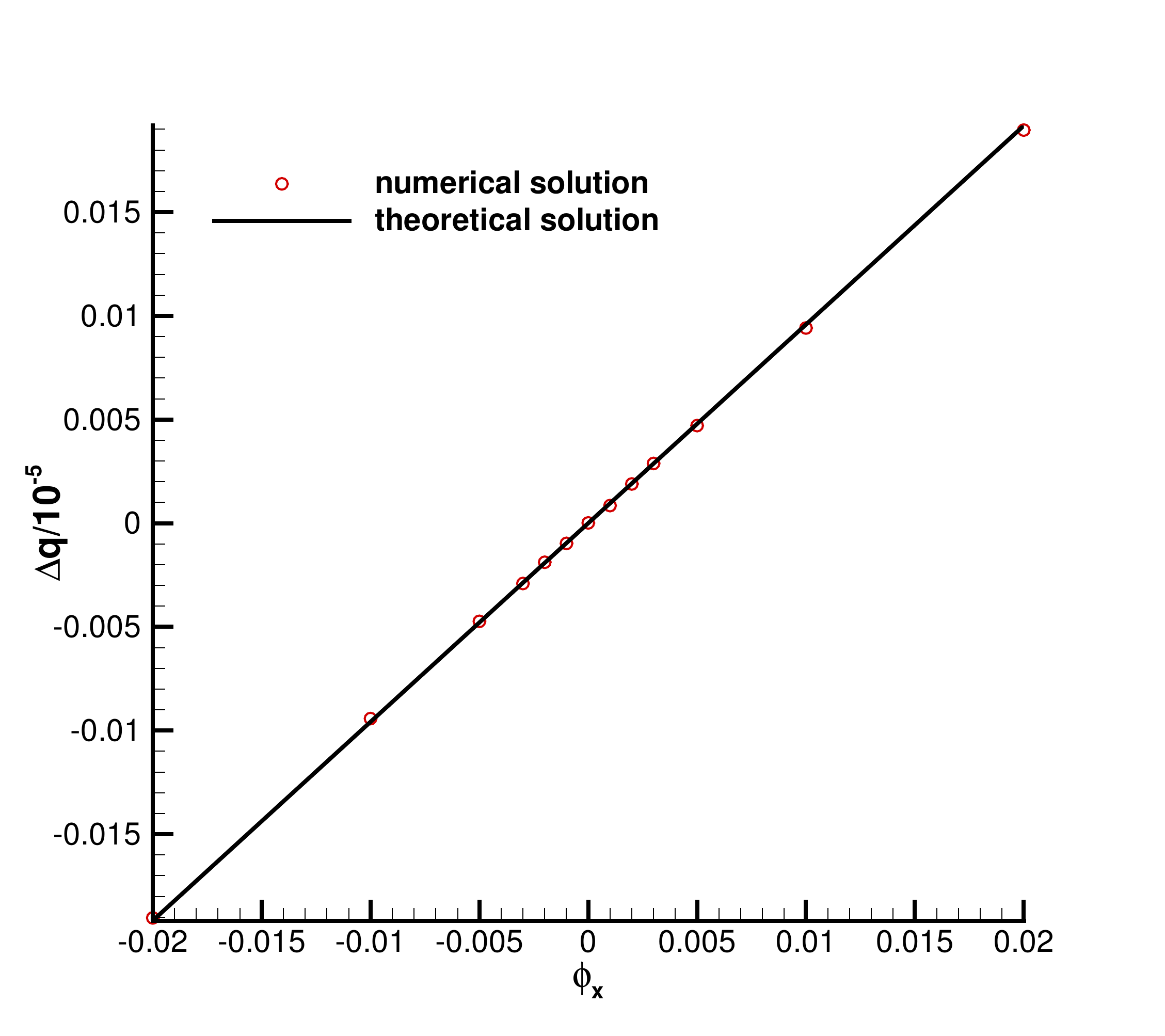}
	}
	\subfigure[$Kn_{ref}=0.01$]{
		\includegraphics[width=7cm]{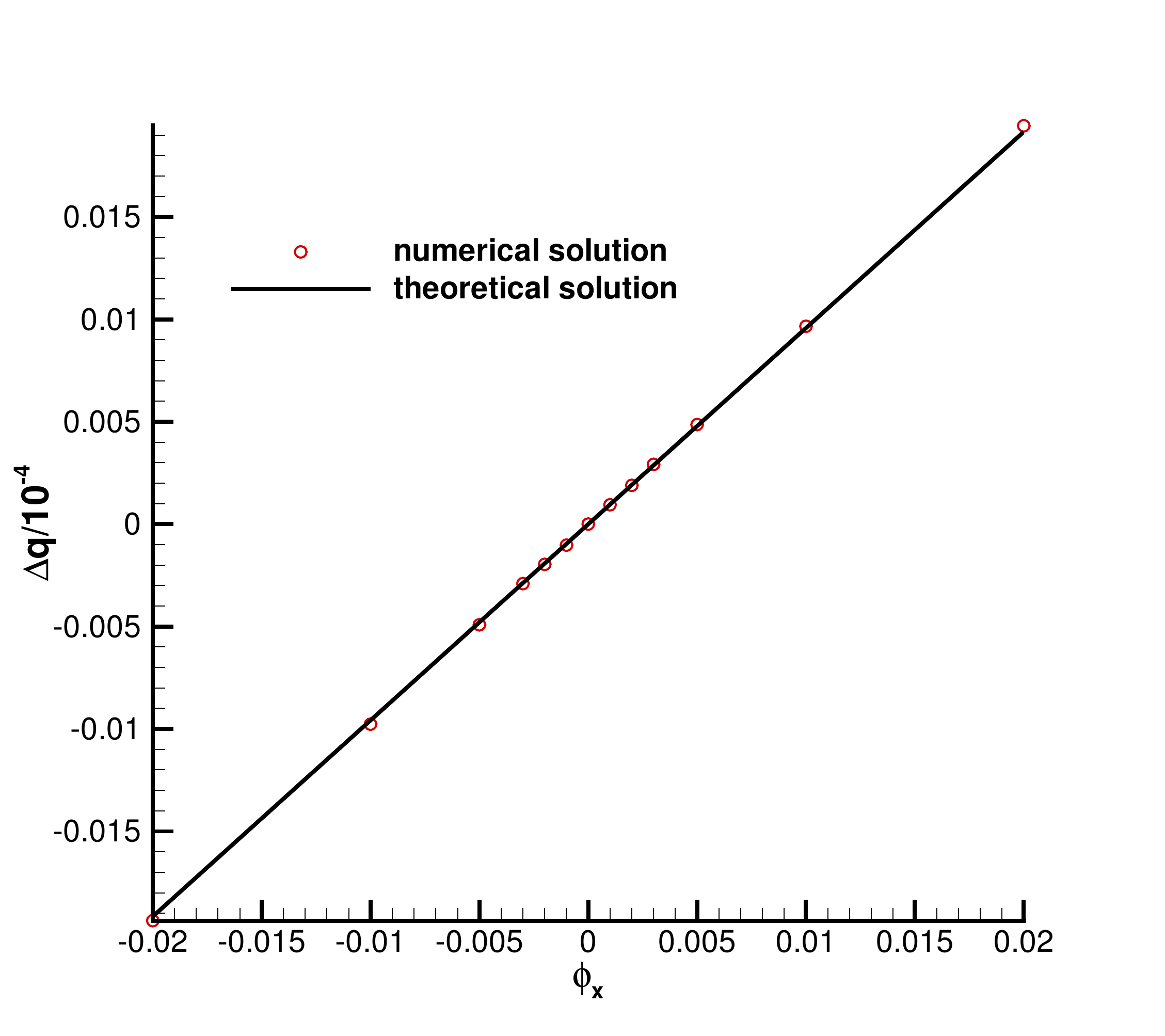}
	}
	\caption{The variation of $\Delta q = q_{UGKS}-q_{Fourier}$ versus external force $\phi_x$. The red circle is computational results, and the black line is the analytical solution.}
	\label{pic:static heat}
\end{figure}

The computational results of convergent state at $x=0.5$ is presented in Table \ref{table:equilibrium} and \ref{table:transition}, with respect to $Kn=0.001$ and $0.01$.
With the unit Prandtl number indicated in the BGK equation, the coefficient of heat conductivity can be calculated via $\kappa=\mu c_p=\tau pc_p$ where $c_p$ is the specific heat, and the heat flux can be obtained through
\begin{equation*}
q_{UGKS}=\frac{1}{2}\int (u-U)\left((u-U)^2+\xi^2\right)fd\Xi, \ q_{Fourier}=-\kappa \nabla T,
\end{equation*}
and the deviation is denoted by $\Delta q=q_{UGKS}-q_{Fourier}$.
It is obvious that with the increase (decrease) of the external force, the heat flux is enhanced (inhabited) along the direction of force.
The deviation between $q_{UGKS}$ and $q_{Fourier}$ versus external force is presented in Fig. \ref{pic:static heat}, denoted with red circles.
At near-equilibrium region with relatively small Knudsen number and weak external force,
the heat flux modification seems to be proportional to the magnitude of the external force in both cases with $Kn_{ref}=0.001$ and $0.01$.
The magnitude of heat flux at $Kn_{ref}=0.01$ is one-order larger than the result at $Kn_{ref}=0.001$,
compatible with the increment of collision time $\tau \propto {Kn}$ and the corresponding non-equilibrium effects.
This simulation result is consistent with the conclusion in Section 2, and here we propose a force-induced heat flux,
\begin{equation*}
q_{force}=C_q \tau \rho T \nabla \Phi,
\end{equation*}
where $C_q$ is a physical parameter of the specific gas, and $\Phi$ is the force potential.
In the current case, it takes the value $C_q=0.0113$.
The theoretical solution based on the above equation is presented in Fig. \ref{pic:static heat}, denoted by the solid line.
It can be seen that the numerical and theoretical solutions agree with each other in the current case with relatively small external force.
Therefore, in the absence of shear stress in one-dimensional case, the net heat flux should be determined via
\begin{equation*}
q=q_{Fourier}+q_{force}+O(\tau^2),
\end{equation*}
where the heat flux comes from the overall contribution of temperature gradient and external force.
With the increment of reference Knudsen number in the transition and free molecular regimes and the magnitude of the external force,
the contribution of heat flux from the forcing term will be much more significant.
Due to the multiple scale effect with large variation of the local Knudsen number and non-equilibrium process,
the additional thermal contribution may be no longer simply proportional to external force, which will be illustrated in the next part.

\begin{table}  
	\caption{Static heat conduction at $x=0.5$ with $Kn_{ref}=0.001$.}  
	\label{table:equilibrium}
	\begin{tabular*}{13cm}{llllll}  
		\hline  
		$\phi_x$ & $q_x/10^{-5}$(UGKS)  & $q_x/10^{-5}$(Fourier) & $\Delta q/10^{-5}$ & $\kappa/10^{-4}$ & $\nabla T/10^{-2}$ \\  
		\hline  
		-0.02  & 5.230964 & 5.250001 & -0.019037 & 5.271195 & 9.959797 \\  
		-0.01  & 5.241005 & 5.250429 & -0.009424 & 5.270205 & 9.962479 \\  
		-0.005  & 5.246134 & 5.250869 & -0.004735 & 5.269701 & 9.964267 \\  
		-0.003  & 5.248152 & 5.251060 & -0.002908 & 5.269498 & 9.965012 \\  
		-0.002  & 5.249229 & 5.251114 & -0.001885 & 5.269395 & 9.965310 \\  
		-0.001  & 5.250274 & 5.251248 & -0.000974 & 5.269292 & 9.965757 \\  
		0  & 5.251317 & 5.251304 & 0.000013 & 5.269192 & 9.966055 \\  
		0.001  & 5.252173 & 5.251324 & 0.000849 & 5.269291 & 9.965906\\  
		0.002  & 5.253001& 5.251110 & 0.001891 & 5.269391 & 9.965310 \\  
		0.003  & 5.253776 & 5.250895 & 0.002881 & 5.269490 & 9.964714 \\  
		0.005  & 5.255487 & 5.250781 & 0.004706 & 5.269691 & 9.964118 \\  
		0.01  & 5.259658 & 5.250244 & 0.009414 & 5.270177 & 9.962181 \\  
		0.02  & 5.267780 & 5.248820 & 0.018960 & 5.271113 & 9.957711 \\  
		\hline  
	\end{tabular*}  
\end{table}  

\begin{table}  
	\caption{Static heat conduction at $x=0.5$ with $Kn_{ref}=0.01$.}  
	\label{table:transition}
	\begin{tabular*}{12cm}{llllll}  
		\hline  
		$\phi_x$ & $q_x/10^{-4}$(UGKS)  & $q_x/10^{-4}$(Fourier) & $\Delta q/10^{-4}$ & $\kappa/10^{-3}$ & $\nabla T/10^{-2}$ \\
		\hline  
		-0.02  & 5.129215 & 5.148576 & -0.019361 & 5.269082	& 9.771297 \\  
		-0.01  & 5.138495 & 5.148263 & -0.009768 & 5.268923	& 9.771000\\  
		-0.005  & 5.143113 & 5.148028 & -0.004915 & 5.268844 & 9.770701\\  
		-0.003  & 5.144938 & 5.147840 & -0.002902 & 5.268812 & 9.770403\\  
		-0.002  & 5.145862 & 5.147825  & -0.001963 & 5.268797 & 9.770403 \\  
		-0.001  & 5.146787 & 5.147810  & -0.001023 & 5.268781 & 9.770403 \\  
		0  & 5.147720 & 5.147716 & 0.000004 & 5.268766 & 9.770254 \\  
		0.001  & 5.148651 & 5.147704  & 0.000947 & 5.268753 & 9.770254 \\  
		0.002  & 5.149587& 5.147692 & 0.001895 & 5.268740 & 9.770254 \\  
		0.003  & 5.150518 & 5.147599 & 0.002919 & 5.268727 & 9.770105 \\  
		0.005  & 5.152359 & 5.147494 & 0.004865 & 5.268699 & 9.769956 \\  
		0.01  & 5.156859 & 5.147195 & 0.009664 & 5.268633 & 9.769509 \\  
		0.02  & 5.166055 & 5.146598 & 0.019457 & 5.268504 & 9.768615 \\  
		\hline  
	\end{tabular*}  
\end{table}  

\subsection{Lid-driven cavity flow under external forcing field}
Different from the purely longitudinal interaction between the temperature gradient and external force in Part. B,
the horizontal moving upper surface of the cavity will induce dissipative shear structure into the dynamic system under
vertical external force field.
In this case, the square cavity has four walls with length  $L=1$.
The upper wall moves in tangential direction (positive $x-$direction) with a velocity $U_w=0.15$.
A series of external forcing acceleration $\phi_y$ is set up in the negative $y$-direction.
The magnitude of force $\phi_y$ is denoted by $g$.
Non-dimensional Froude number can be defined in this system to quantify the relative
importance of the upper wall's driving velocity and the effect of external force,
\begin{equation}
Fr=\frac{U_w}{\sqrt{gL}}.
\end{equation}
The initial density and pressure are defined with barometric balance state,
\begin{equation*}
\rho(x,y,t=0)=\exp(2\phi_y y),p(x,y,t=0)=\exp(\phi_y y),U(x,y,t=0)=0.
\end{equation*}
The initial particle distribution function is set as Maxwellian everywhere with respect to stratified density in the cavity.
The wall temperature is kept with $T_w=1$, and the Maxwell full accommodation boundary is used in the simulation.
The Prandtl number of the gas in this case is $\rm Pr=0.67$ with Shakhov model.
The reference Knudsen number is selected as $Kn_{ref}=0.001, 0.075$, which is defined by reference state at bottom of the cavity $\rho_{\rm ref}=1.0$ and $p_{\rm ref}=1.0$.
The dynamic viscosity in the reference state is calculated via variable hard sphere (VHS) model,
\begin{equation}
\mu_{ref}=\frac{5(\alpha+1)(\alpha+2)\sqrt{\pi}}{4\alpha(5-2\omega)(7-2\omega)}Kn_{ref}.
\label{eqn: vss viscosity}
\end{equation}
Here we choose $\alpha=1.0$ and $\omega=0.5$ to recover a hard sphere monatomic gas, 
and its viscosity model is
\begin{equation}
\mu=\mu_{ref}\left(\frac{T}{T_{ref}}\right)^\theta,
\label{eqn: hs model}
\end{equation}
where $T_{ref}$ is the reference temperature and $\theta$ is the index related to HS model.
In this case we adopt the value $\theta=0.72$.
The local collision time is evaluated with the relation $\tau=\mu/p$.
The computational domain is divided into $45\times45$ uniform cells, and $28\times28$ Gaussian points in velocity space.

\subsubsection{Near equilibrium regime}
\begin{figure}[htb!]
	\centering
	\subfigure[U-velocity along the vertical center line]{
		\includegraphics[width=7cm]{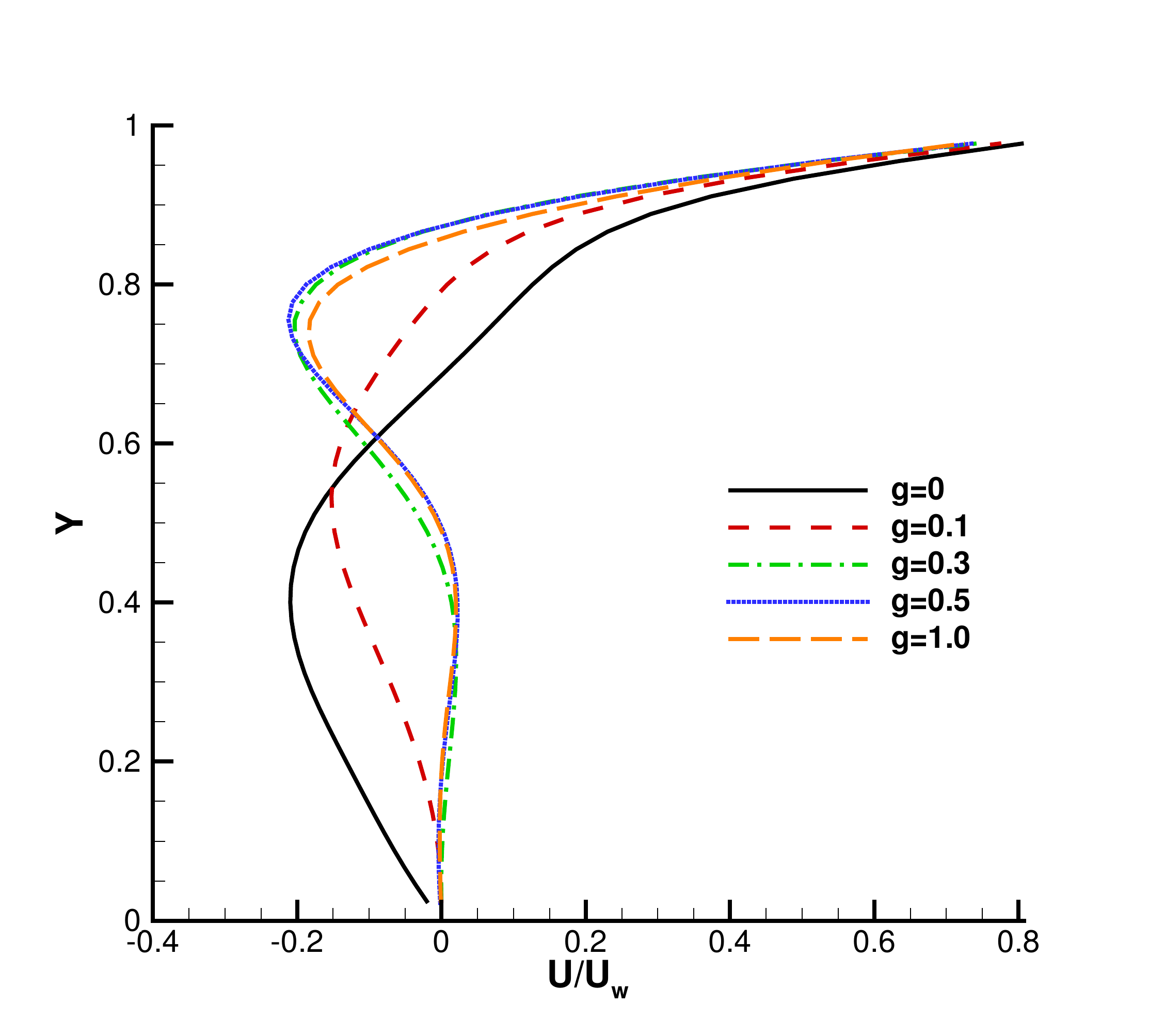}
	}
	\subfigure[V-velocity along the horizontal center line]{
		\includegraphics[width=7cm]{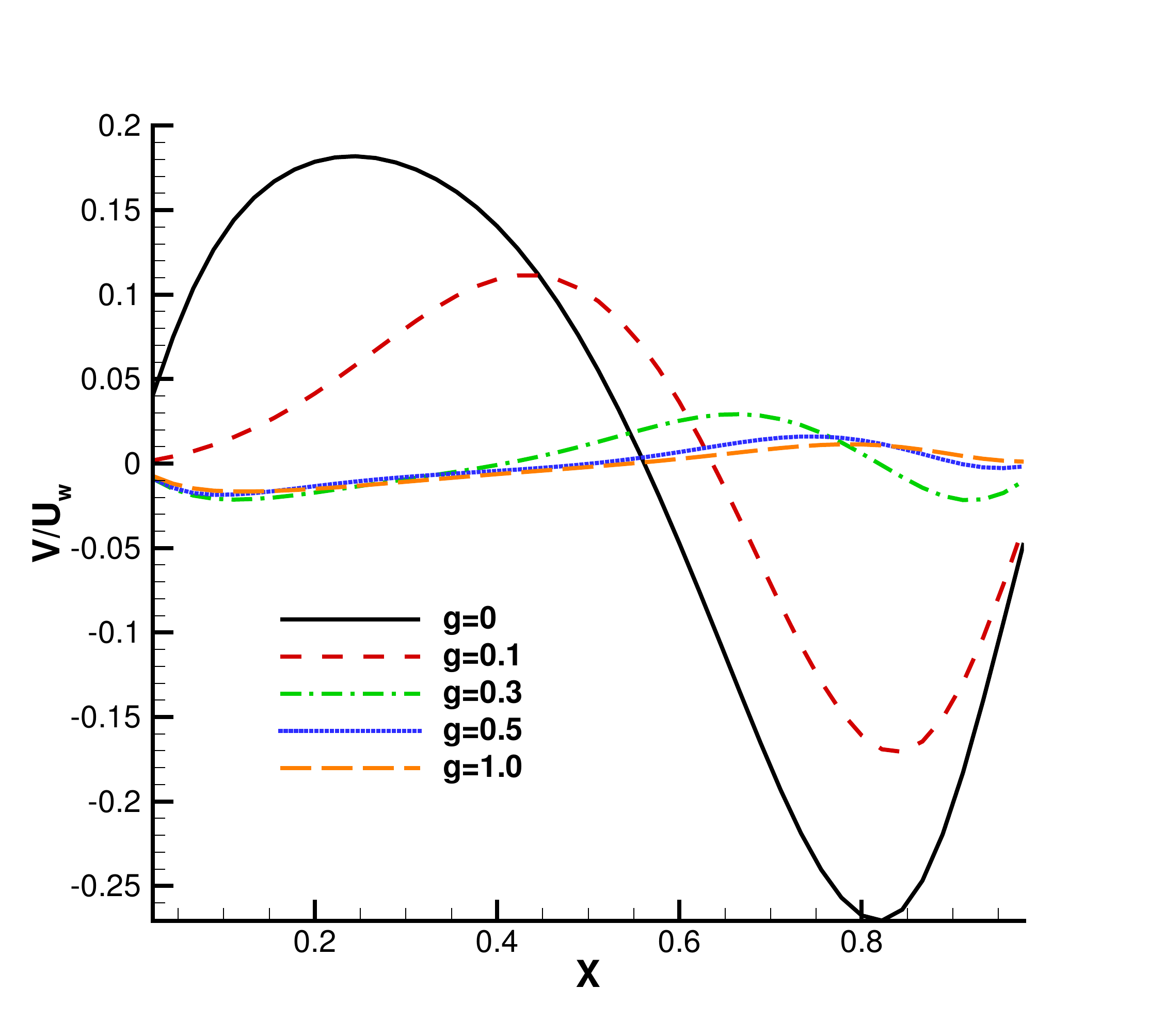}
	}
	\caption{Velocity distribution along the center line with $Kn_{ref}=0.001$.}
	\label{cavity vline continuum}
\end{figure}

\begin{figure}[htb!]
	\centering
	\subfigure[U-velocity and streamline at $\phi_y=0$]{
		\includegraphics[width=5cm]{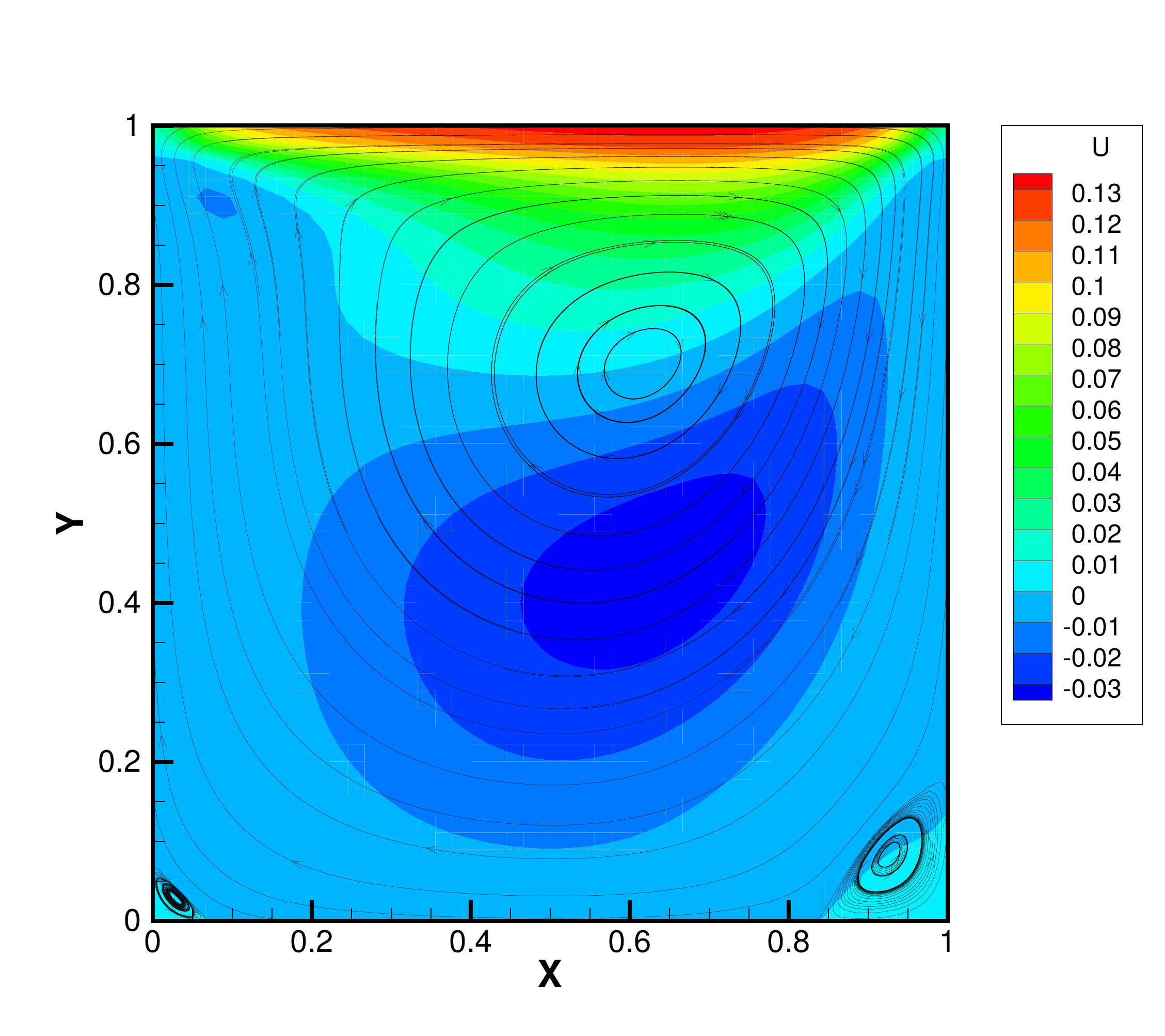}
	}
	\subfigure[U-velocity and streamline at $\phi_y=-0.1$]{
		\includegraphics[width=5cm]{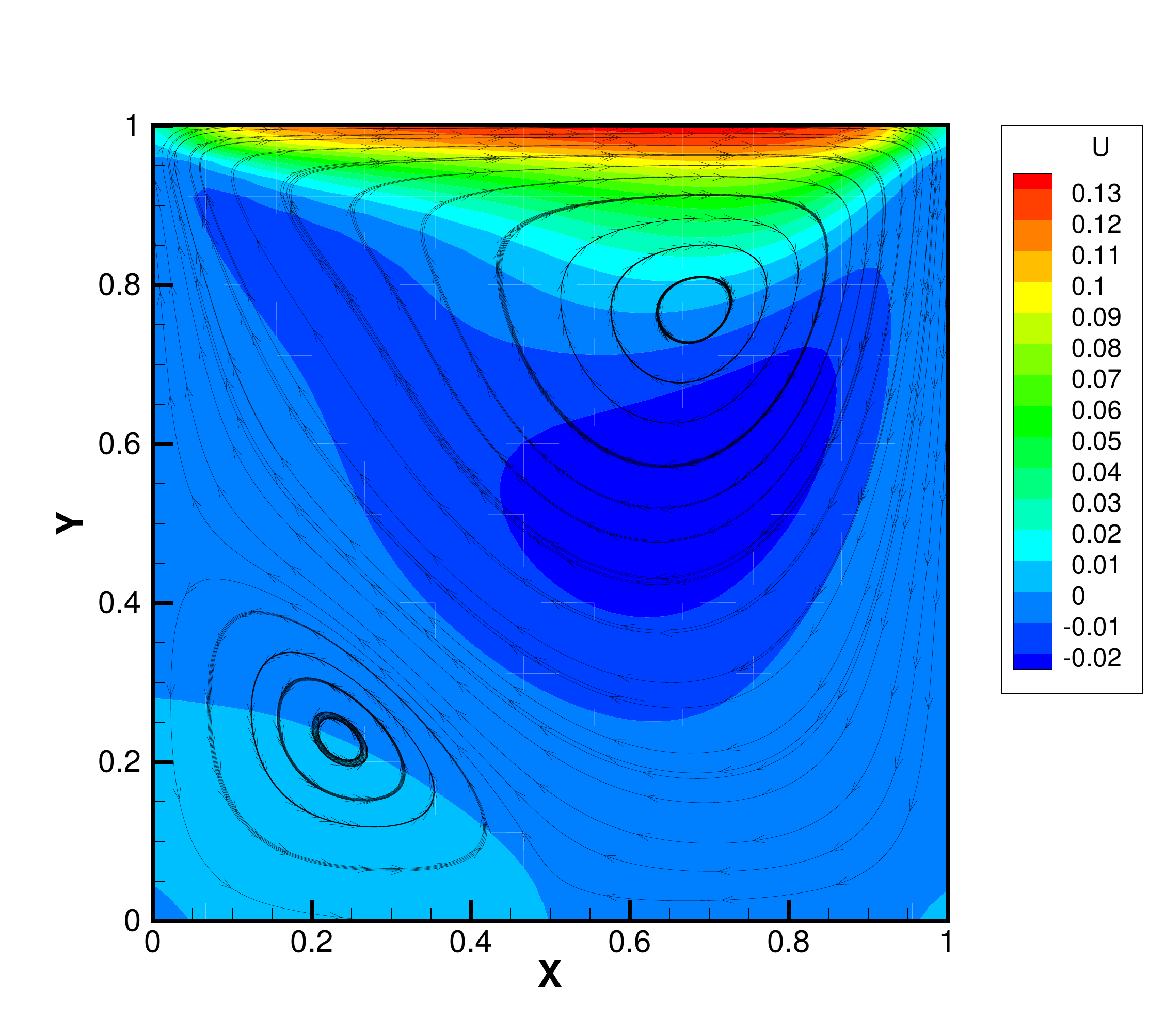}
	}
	\subfigure[U-velocity and streamline at $\phi_y=-0.3$]{
		\includegraphics[width=5cm]{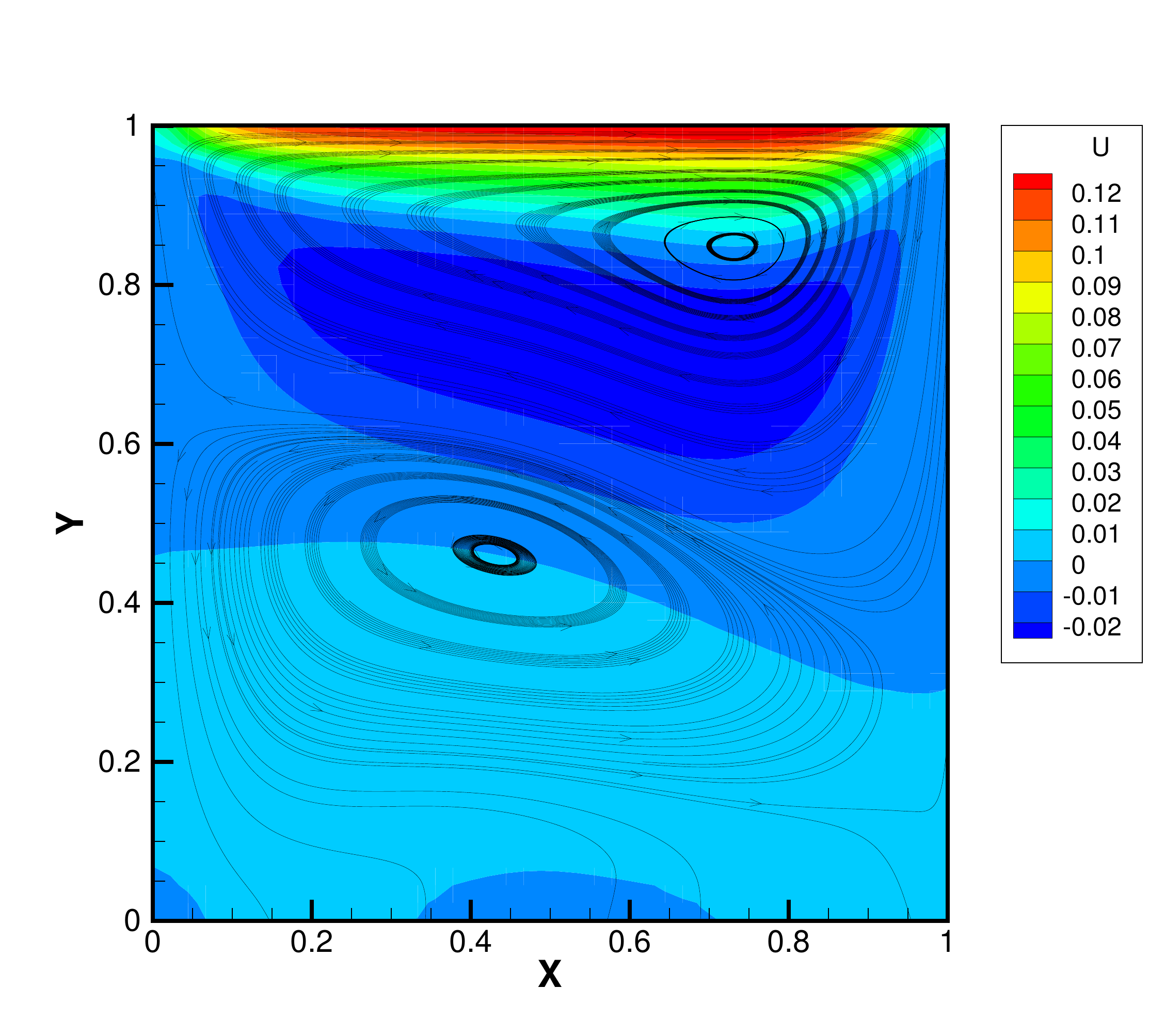}
	}
	\subfigure[V-velocity and velocity vector at $\phi_y=0$]{
		\includegraphics[width=5cm]{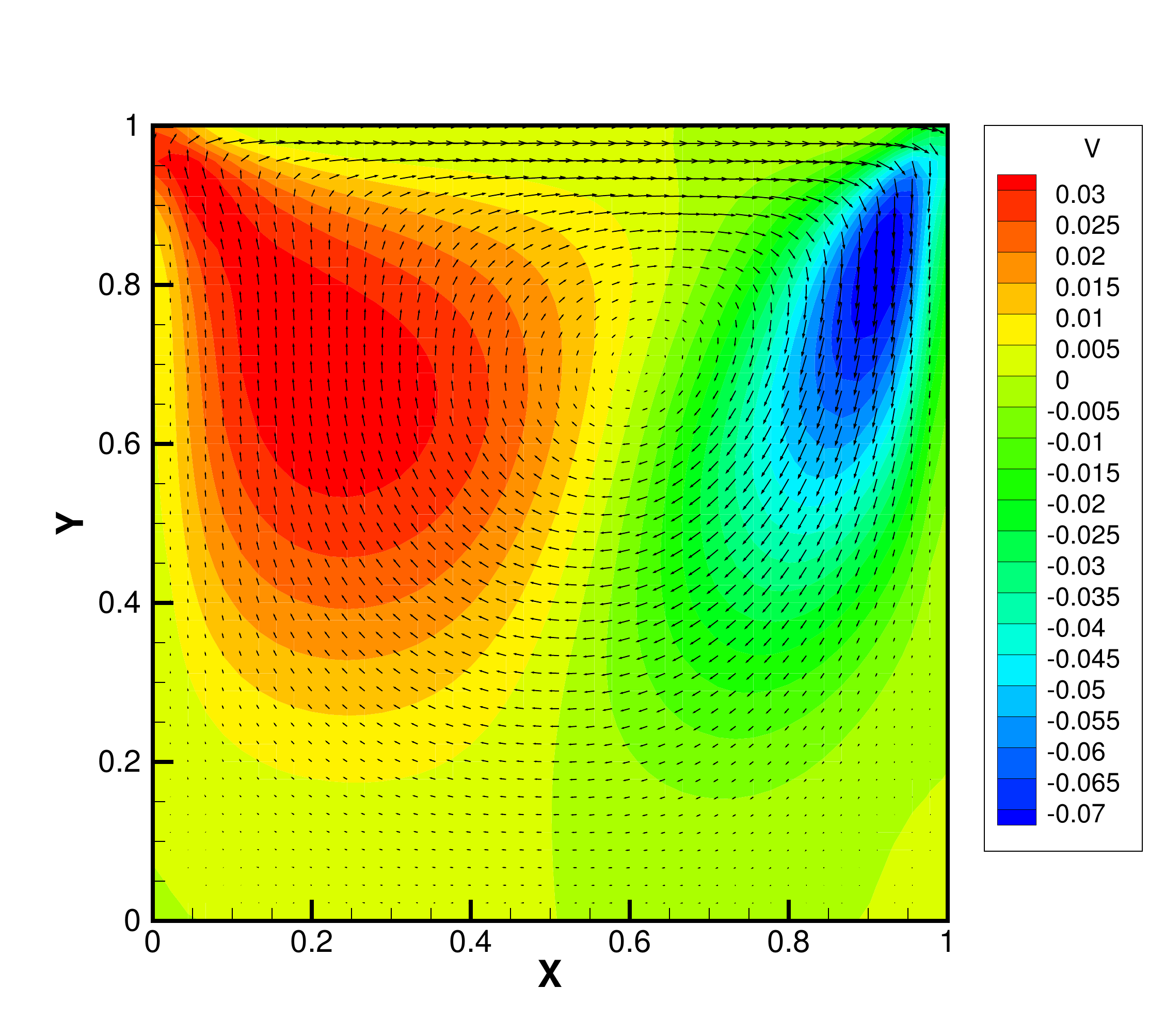}
	}
	\subfigure[V-velocity and velocity vector at $\phi_y=-0.1$]{
		\includegraphics[width=5cm]{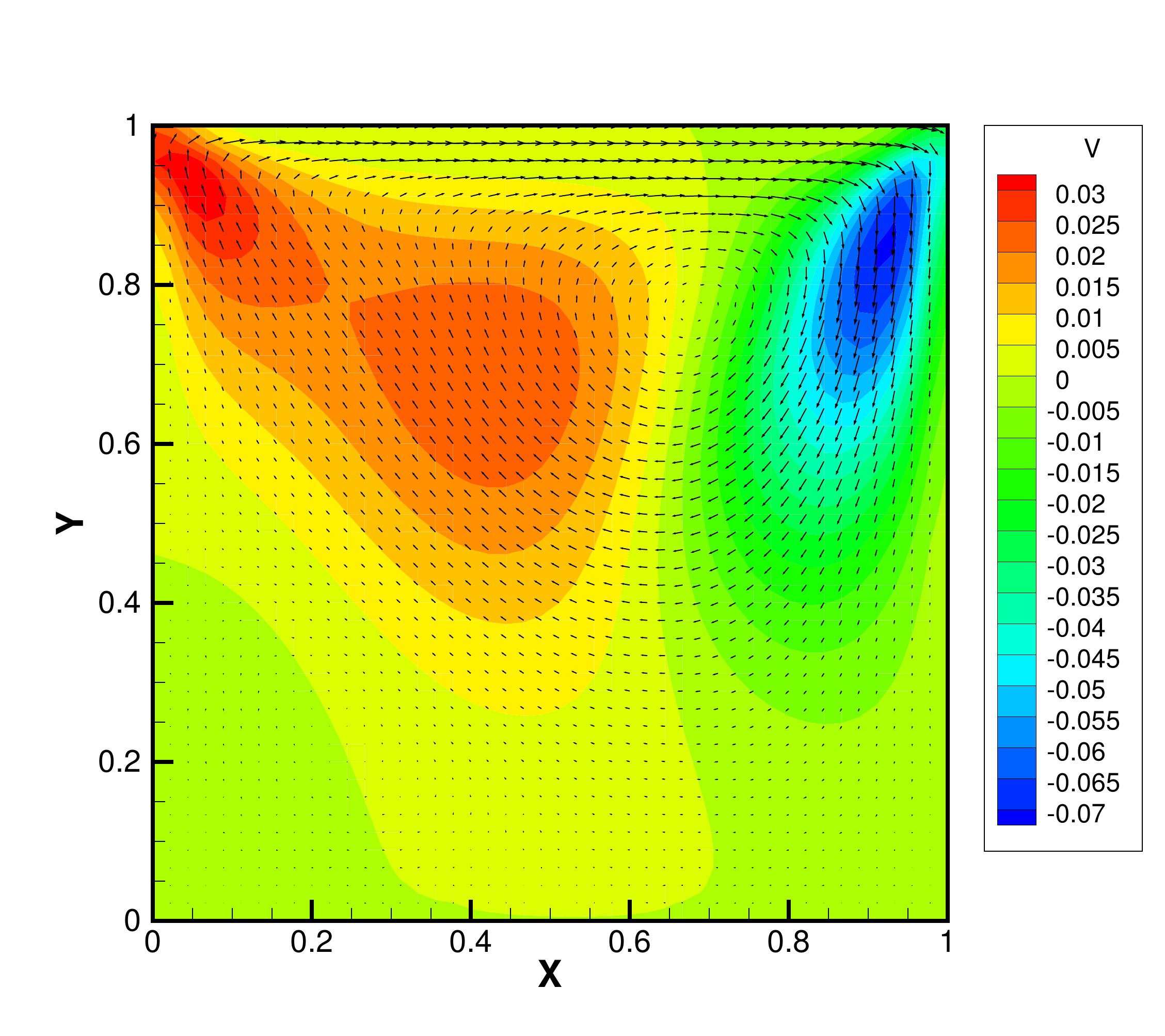}
	}
	\subfigure[V-velocity and velocity vector at $\phi_y=-0.3$]{
		\includegraphics[width=5cm]{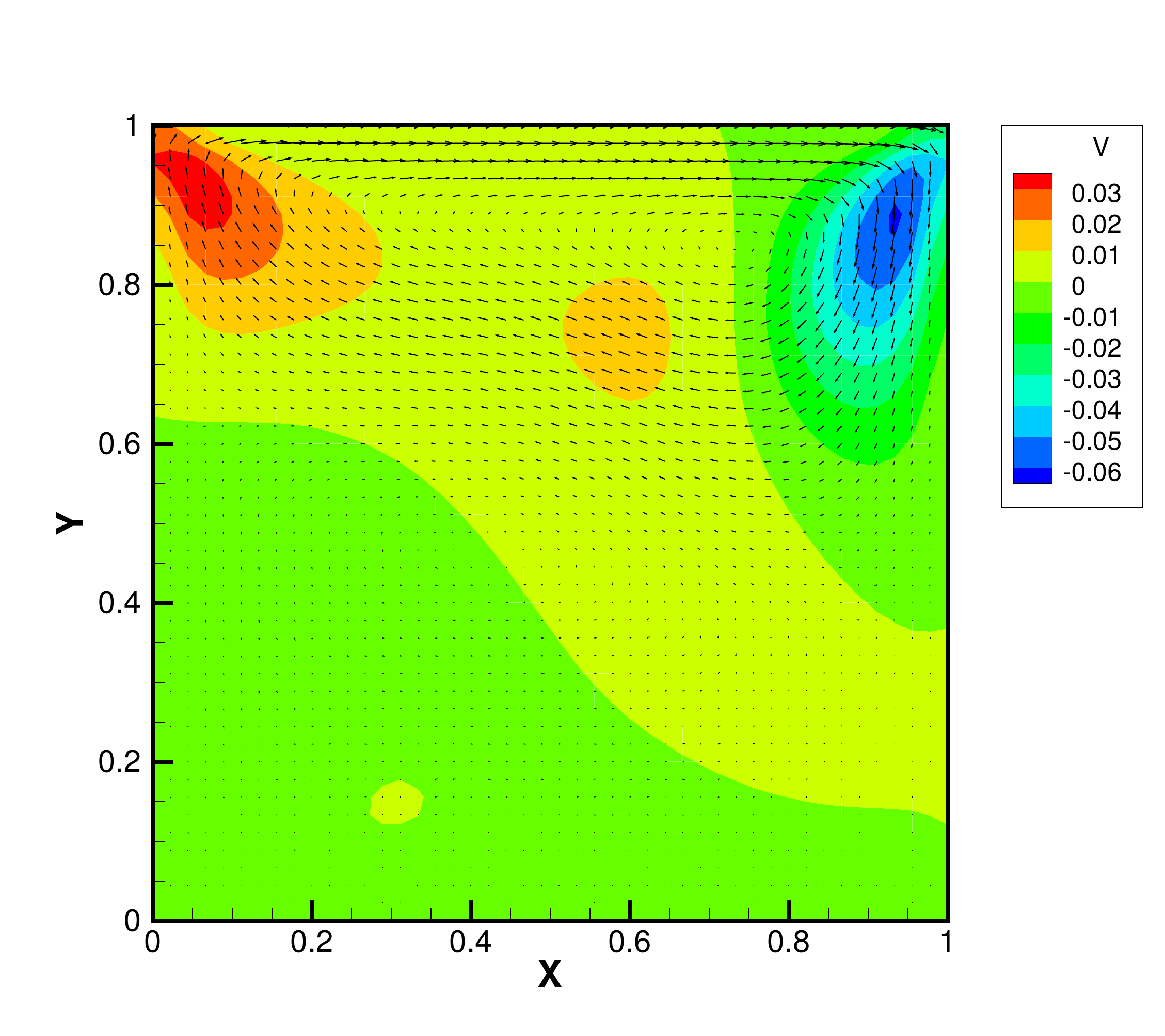}
	}
	\caption{Velocity distribution with $Kn_{ref}=0.001$.}
	\label{cavity vcontour continuum}
\end{figure}

\begin{figure}[htb!]
	\centering
	\subfigure[Density]{
		\includegraphics[width=7cm]{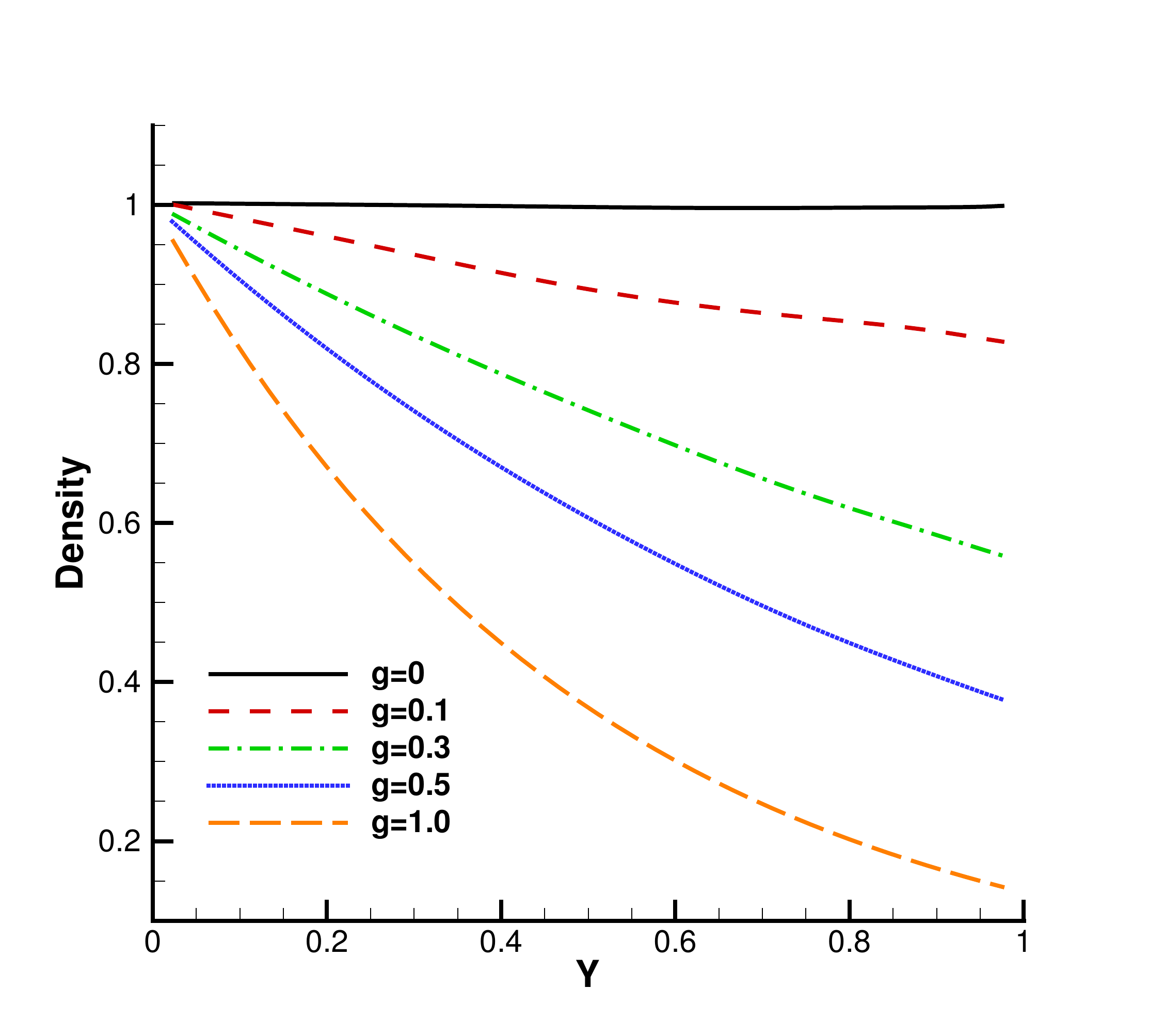}
	}
	\subfigure[Local Knudsen number]{
		\includegraphics[width=7cm]{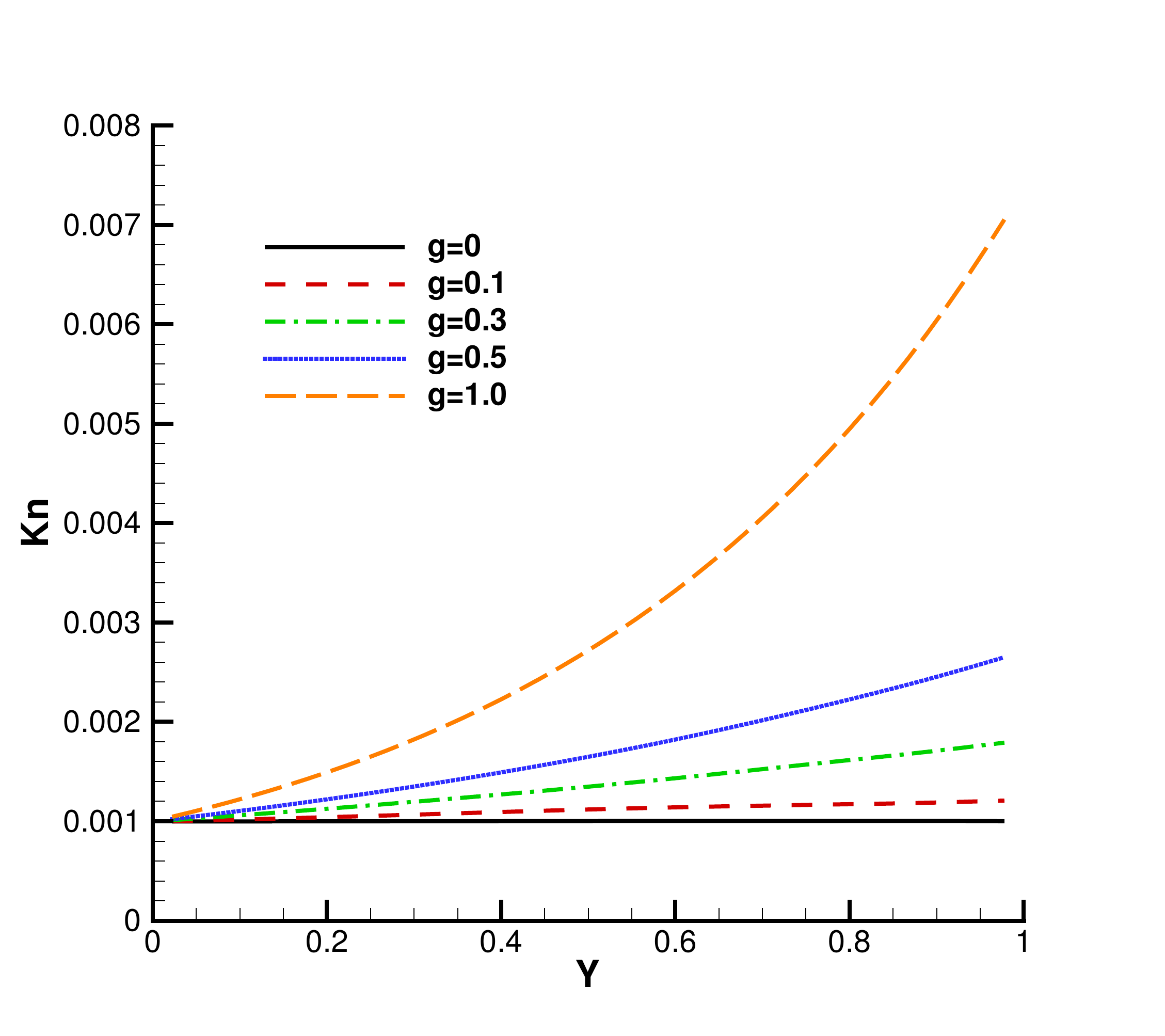}
	}
	\caption{Density and local Knudsen number distribution along the vertical center line with $Kn_{ref}=0.001$.}
	\label{cavity densityline continuum}
\end{figure}

For the gas in the cavity, the movement of upper surface and the external forcing term are two sources for the fluid motion and stability.
In this case, the initial hydrostatic distribution of density and pressure from external force field is perturbed by the upper wall's sudden movement.
Different from the flow dynamics including the viscous dissipation and heat conduction in the absence of external force,
now the external force field participates in the flow and heat transport inside the cavity.
The simple main large eddy topological structure covering the whole cavity domain
may not necessarily appear due to the large density variation and different transport mechanism for different local Knudsen number flow.
In this case, the external forcing acceleration is set up with $\phi_y=0.0, -0.1, -0.3, -0.5, -1.0$ along the negative $y-$direction.
Fig. \ref{cavity vline continuum} shows the velocity distribution along the center lines at $Kn_{ref}=0.001$, and
Fig. \ref{cavity vcontour continuum} chooses typical pictures of velocity contours, vectors and streamlines at $\phi_y=0, -0.1, -0.3$ inside the cavity, in which the flow pattern presents significant differences..
As presented, at a small magnitude of external force, there exists a large main eddy running through the whole cavity domain with two small corner vortices,
and the distribution of $U$-velocity along the vertical center line is a monotonic curve.
However, with the increment of the magnitude of external force, the flow pattern changes significantly.
In spite of the driving effect at the upper surface, under such an external force field the eddy is restricted to the upper half domain of the cavity with the increment of external force,
and the high density region around the bottom forms a weak, inverse running vortex gradually from left corner of the cavity.
An inflexion point appears in the $U$-velocity curve, leaving the lower part flow almost stationary.
In fact, as presented in Fig. \ref{cavity densityline continuum}a, with the relatively large external force,
there is an obvious density variation along the vertical center line, and the flow in the upper region of the cavity stays in the transition regime,
where profound non-equilibrium flow phenomena with a variation of local Knudsen number in Fig. \ref{cavity densityline continuum}b appear in such a gas dynamic system.

\begin{figure}[htb!]
	\centering
	\subfigure[$\phi_y=0$]{
		\includegraphics[width=5cm]{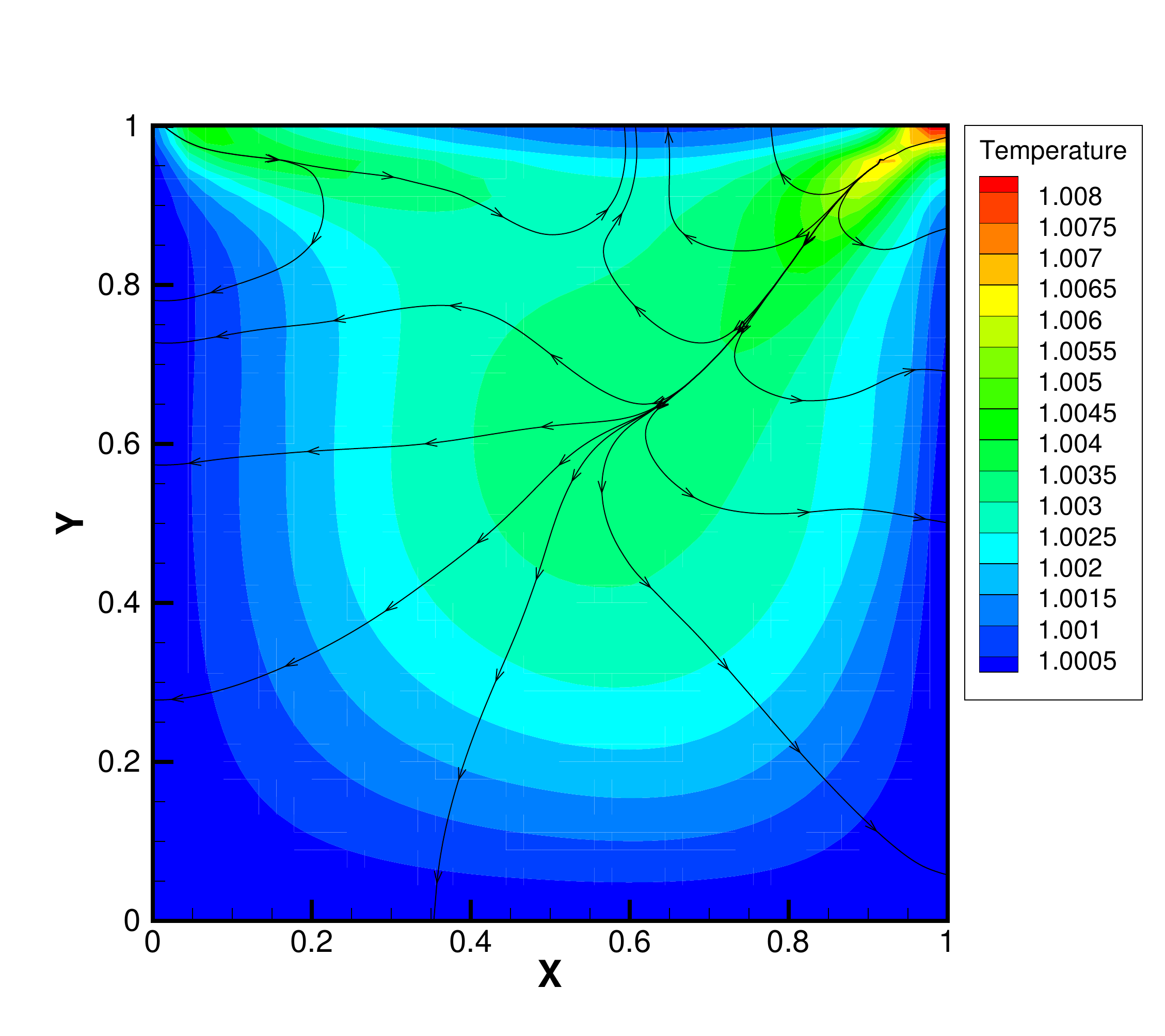}
	}
	\subfigure[$\phi_y=-0.1$, $Fr=0.47$]{
		\includegraphics[width=5cm]{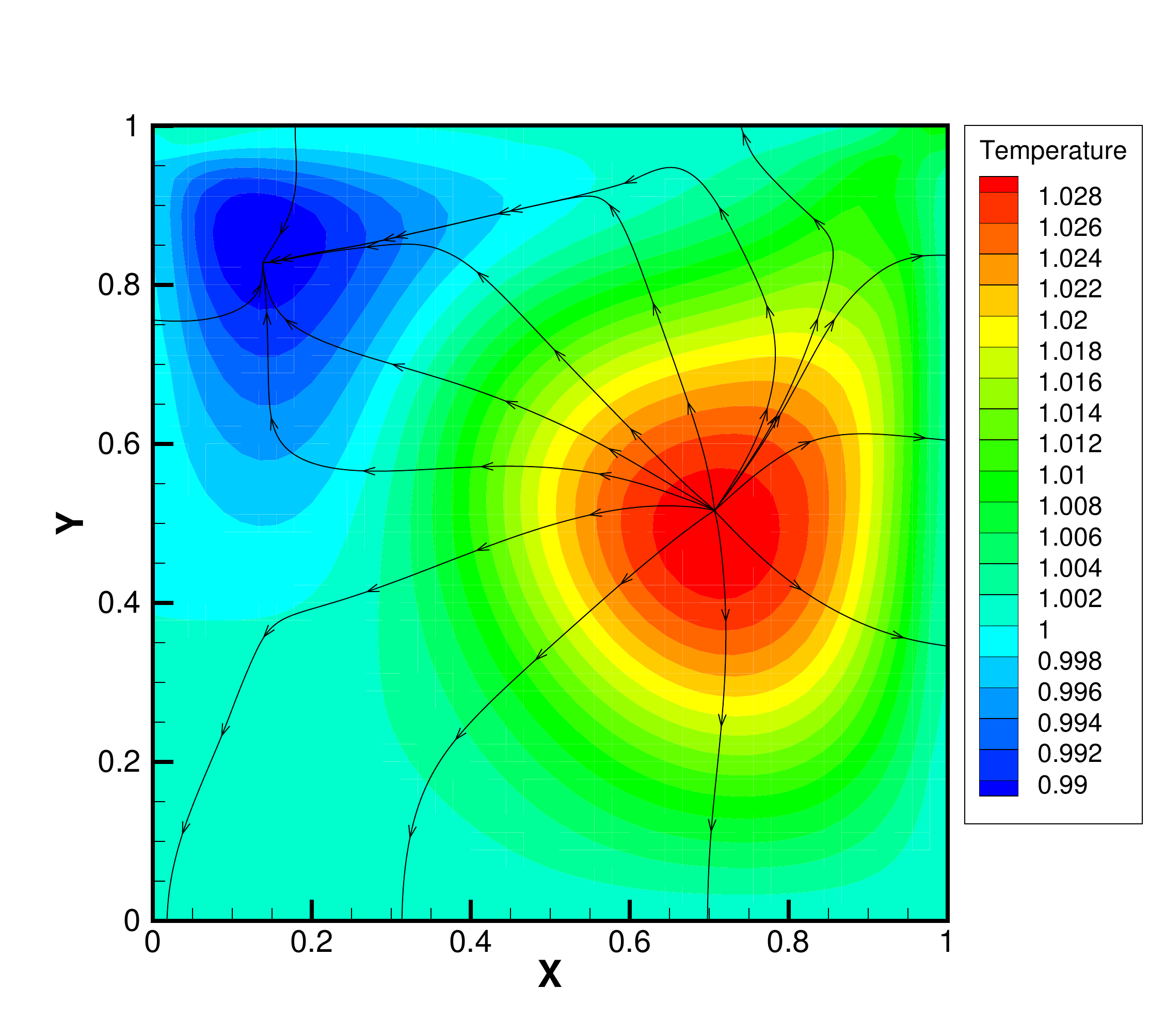}
	}
	\subfigure[$\phi_y=-0.3$,$Fr=0.27$]{
		\includegraphics[width=5cm]{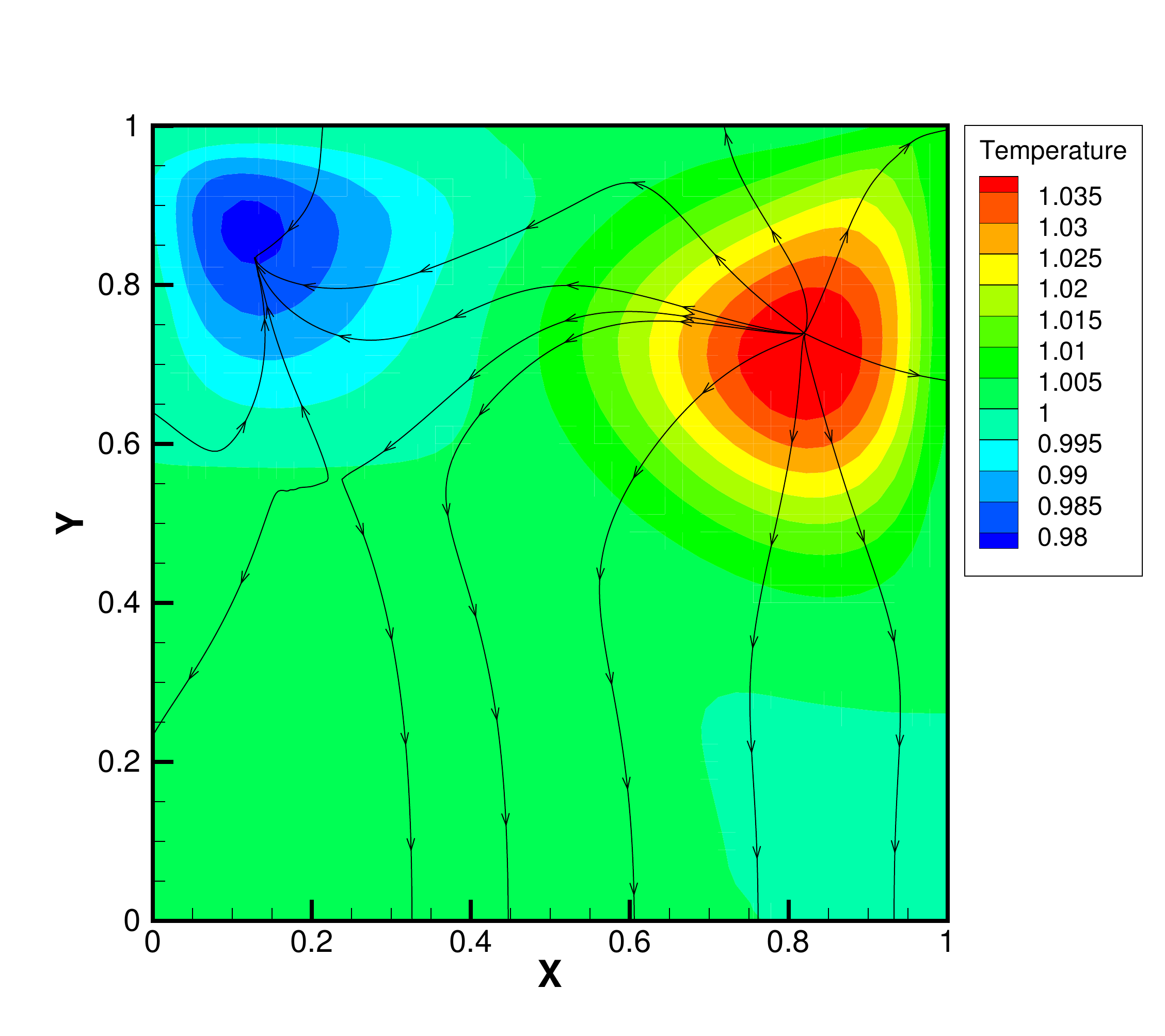}
	}
	\subfigure[$\phi_y=-0.5$,$Fr=0.21$]{
		\includegraphics[width=5cm]{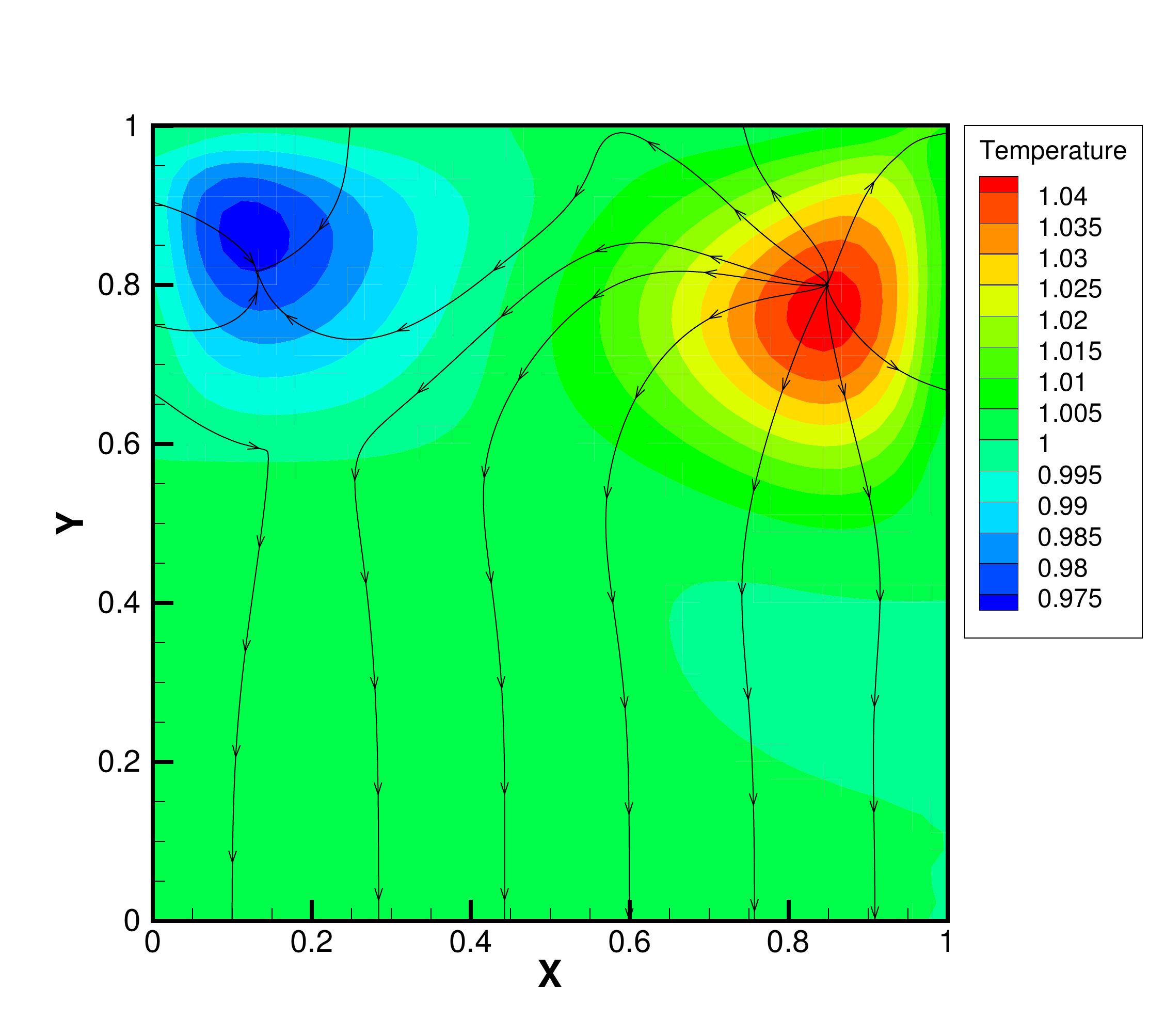}
	}
	\subfigure[$\phi_y=-1.0$,$Fr=0.15$]{
		\includegraphics[width=5cm]{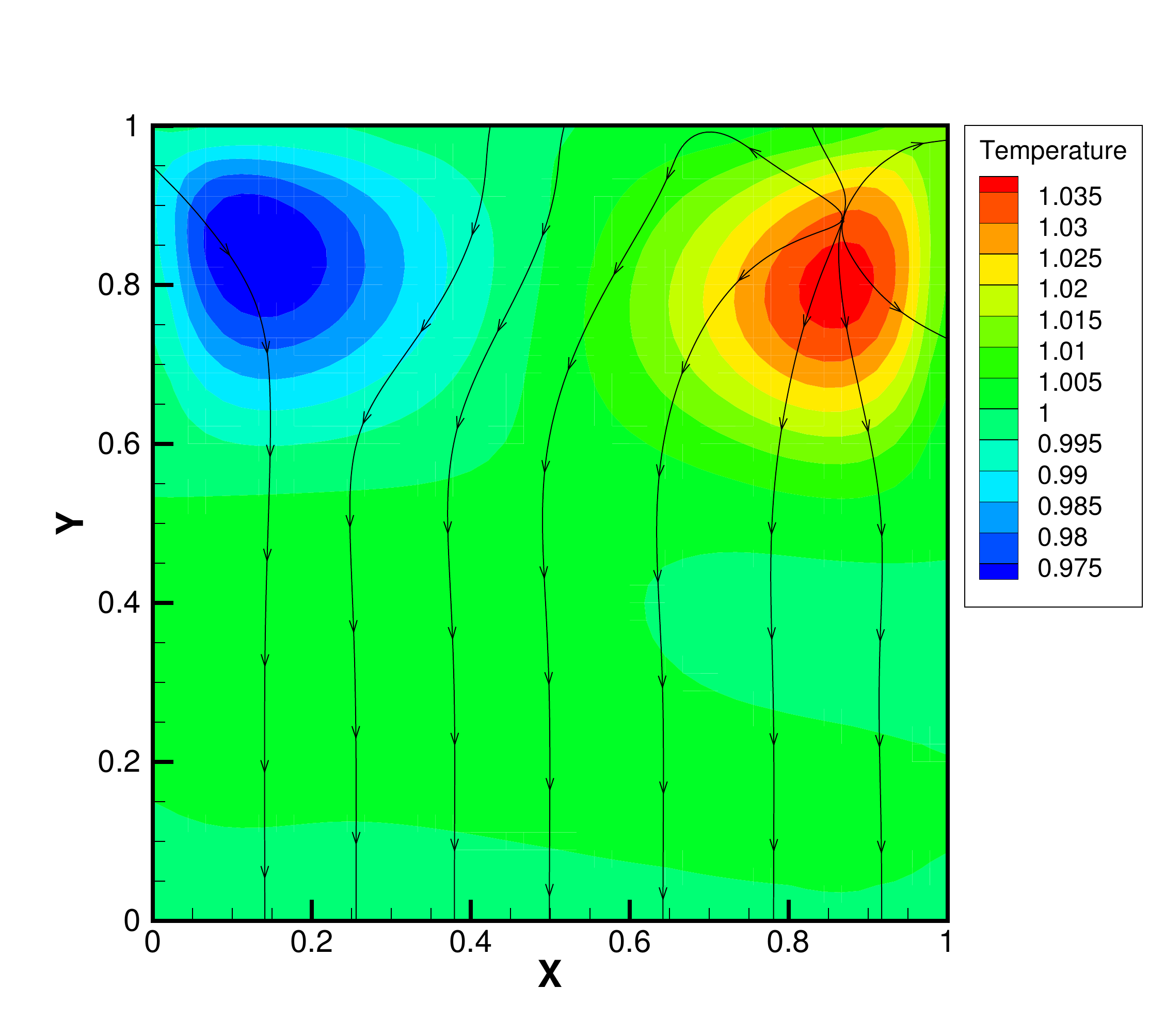}
	}
	\caption{Temperature contour and heat flux with $Kn_{ref}=0.001$.}
	\label{cavity heat continuum}
\end{figure}

The heat transfer inside the cavity is closely coupled with flow transport.
Fig. \ref{cavity heat continuum} presents the temperature contour along with the heat flux
under different external forcing terms at $Kn_{ref}=0.001$
In the absence of external force, particle collisions at the top right corner result in a viscous heating at the macroscopic level,
as shown in Fig. \ref{cavity heat continuum}a.
Due to intensive particle collisions, the expansion cooling at the top left corner is not obvious in this case,
and the temperature around other three boundaries is almost uniform. This is consistent with the NS solutions in the continuum regime \cite{liu2016unified}.
With an increment of external forcing term, the localized hot and cold spots no longer stay at the corner regions,
and propagate into the cavity.
The penetration of the spots is related to the scale of the main eddy.
From the results in Fig. \ref{cavity heat continuum} and \ref{cavity vline continuum},
the center of the hot spot is located around the place where the negative $U$-velocity approaches to its maximum value,
and the center of the cold spot locates a little bit higher than the hot one.
At the current $Kn_{ref}=0.001$, the particle distribution function near the bottom wall will not deviate far from the Maxwellian equilibrium distribution.
As analyzed in Section 2, in the near-equilibrium state, the correlation between the modification of heat flux and the external force is proportional
to the magnitude of the forcing term in Eq. (\ref{analysis q increment}).
In this case, due to the existence of distinct inhomogeneous temperature distribution, the heat flux is mainly aligned with the temperature gradient in the upper domain.
However, in the lower near-static region where there is no significant temperature difference, the heat flux shows the tendency to line up with the direction of the force field.
The adjustment of particle distribution function due to the external forcing term provides a significant mechanism for the non-equilibrium heat transport besides the effect of thermal gradient.

\subsubsection{Transition regime}

\begin{figure}[htb!]
	\centering
	\subfigure[U-velocity along the vertical center line]{
		\includegraphics[width=7cm]{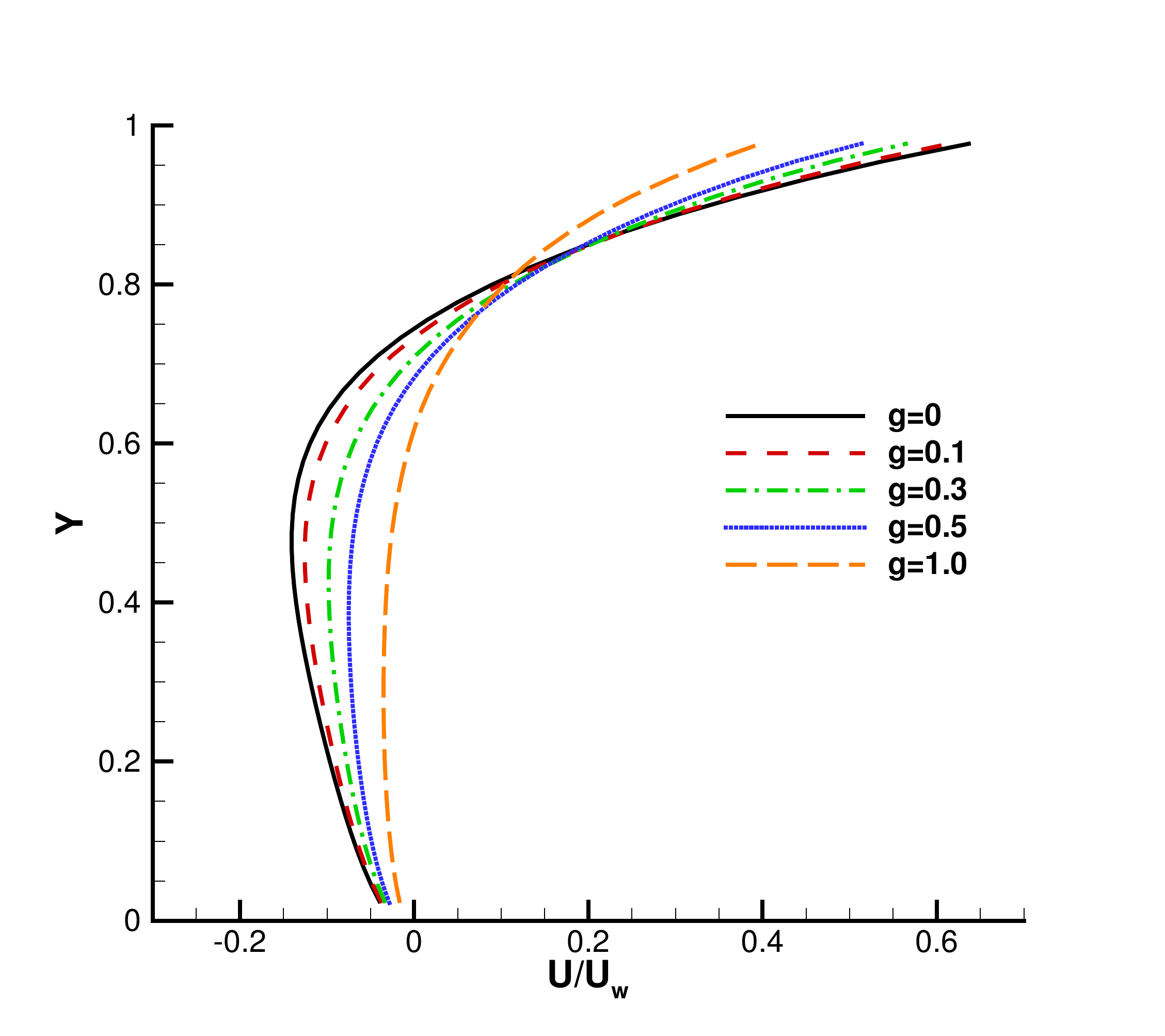}
	}
	\subfigure[V-velocity along the horizontal center line]{
		\includegraphics[width=7cm]{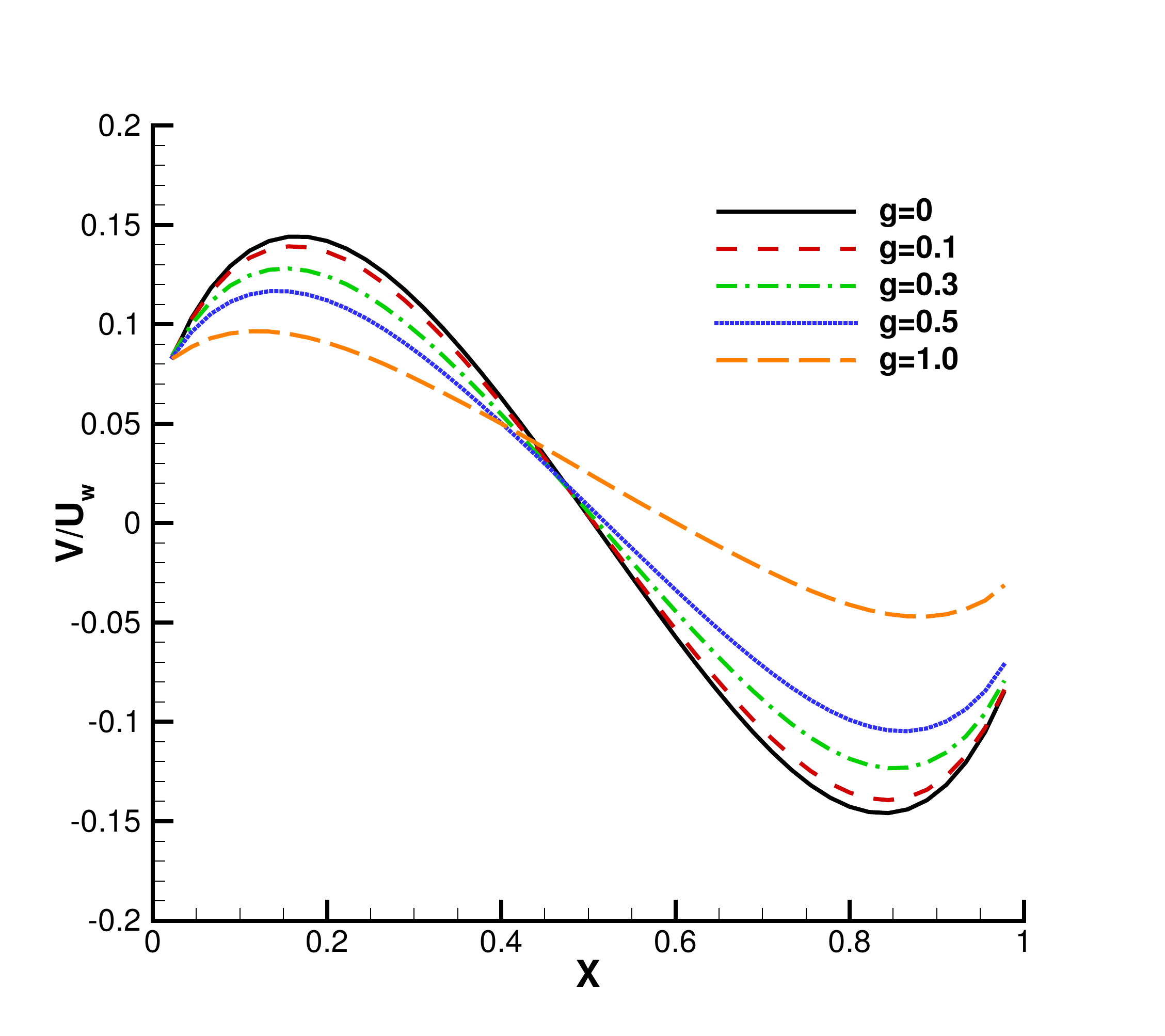}
	}
	\caption{Velocity distribution along the center line with $Kn_{ref}=0.075$.}
	\label{cavity vline transition}
\end{figure}

Now let us turn our attention to the case of full transition regime at $Kn_{ref}=0.075$.
In this case, the external forcing acceleration is set up with $\phi_y=0.0, -0.001, -0.002, -0.003$, $-0.005, -0.01, -0.02, -0.05, -0.1, -0.2, -0.3, -0.5, -1.0$ along the negative $y-$direction.
As shown in Fig. \ref{cavity vline transition}, the particle penetration and efficient mixing generates one large eddy in all cases.
The stabilizing effect due to external force field is to reduce the rotating speed of the vortex.
With the increment of external force, the velocity profile in Fig. \ref{cavity vline transition} is flattened, indicating a weaker vortex motion.

\begin{figure}[htb!]
	\centering
	\subfigure[$\phi_y=0$]{
		\includegraphics[width=5cm]{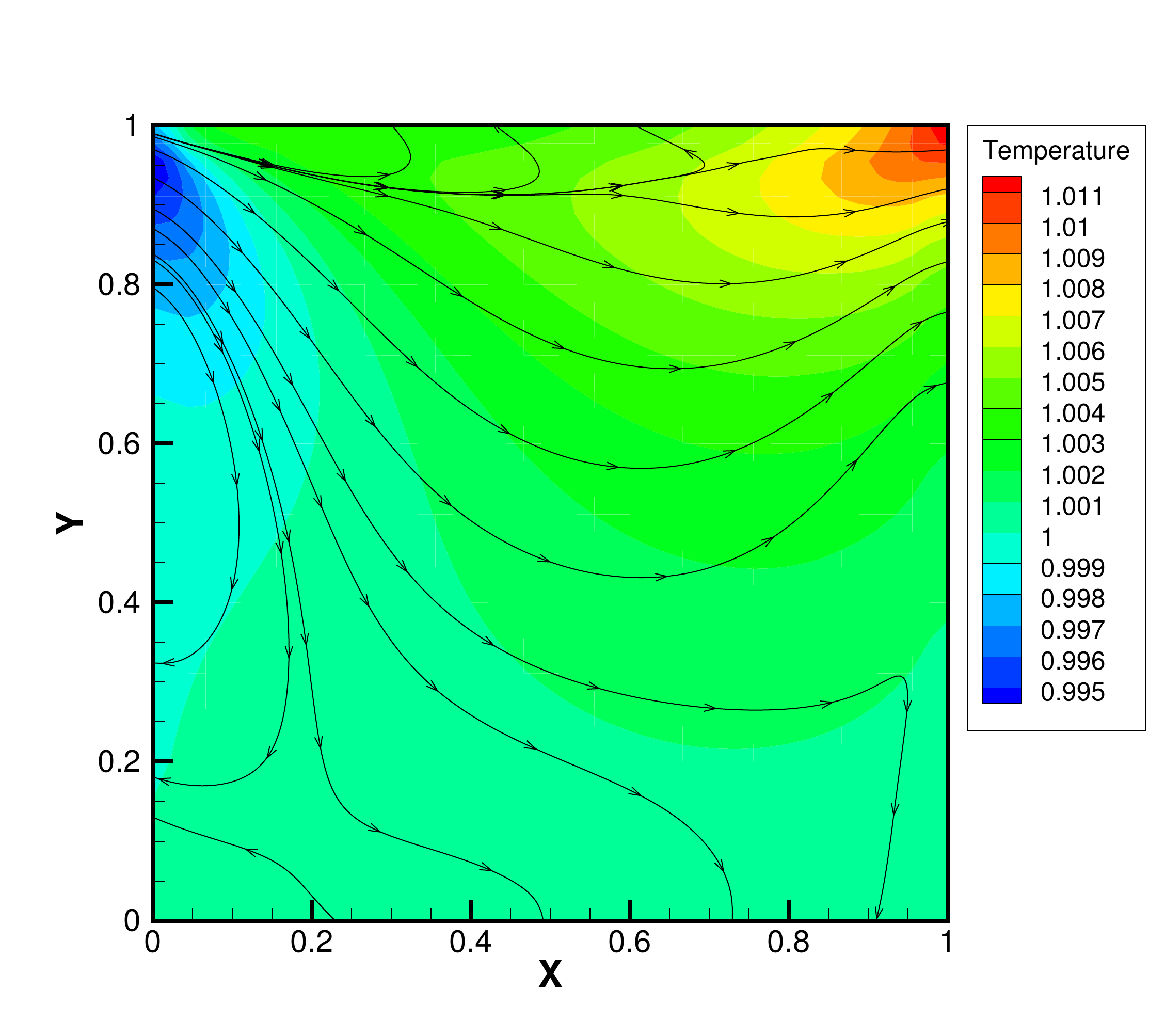}
	}
	\subfigure[$\phi_y=-0.1$,$Fr=0.47$]{
		\includegraphics[width=5cm]{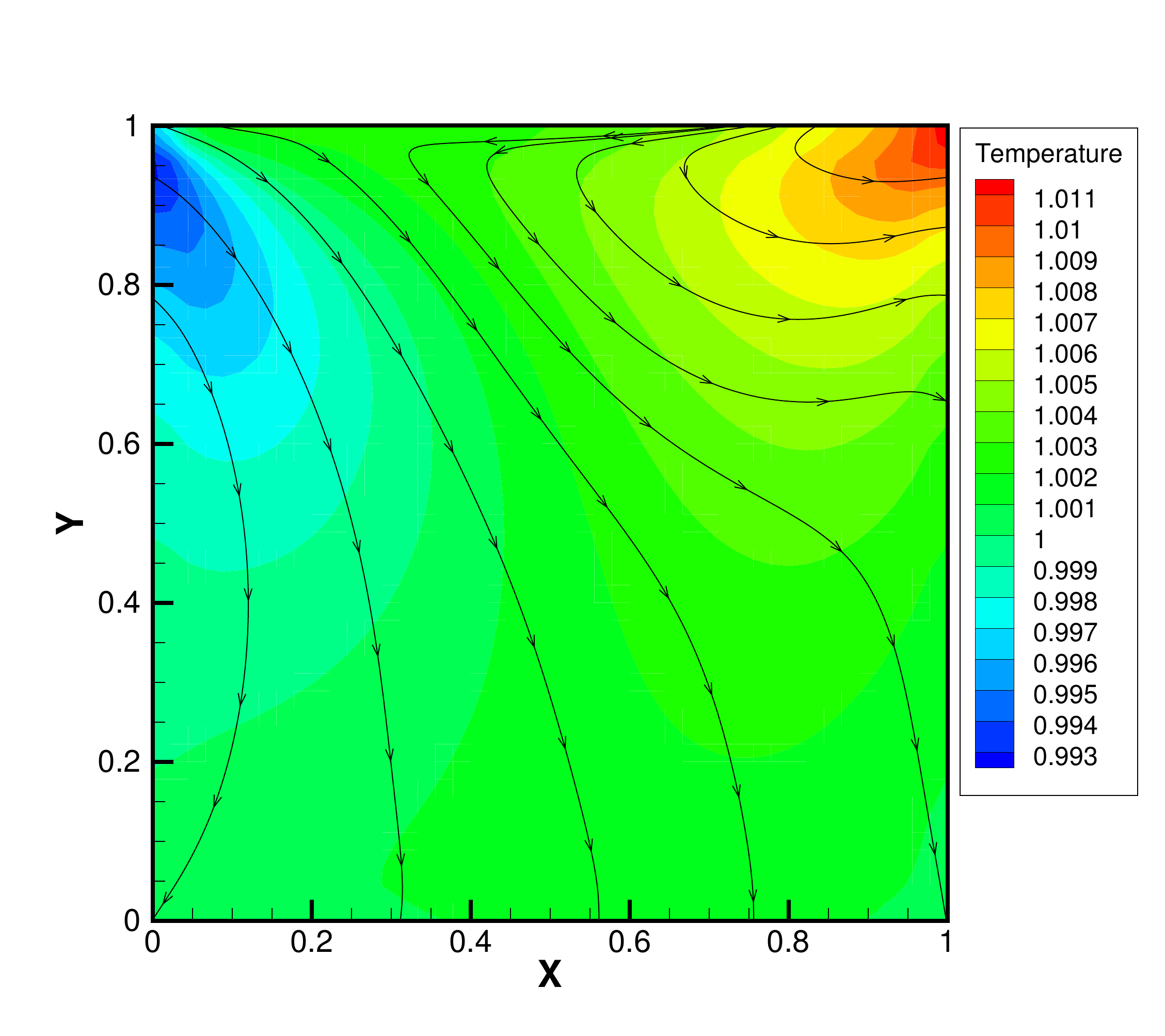}
	}
	\subfigure[$\phi_y=-0.3$,$Fr=0.27$]{
		\includegraphics[width=5cm]{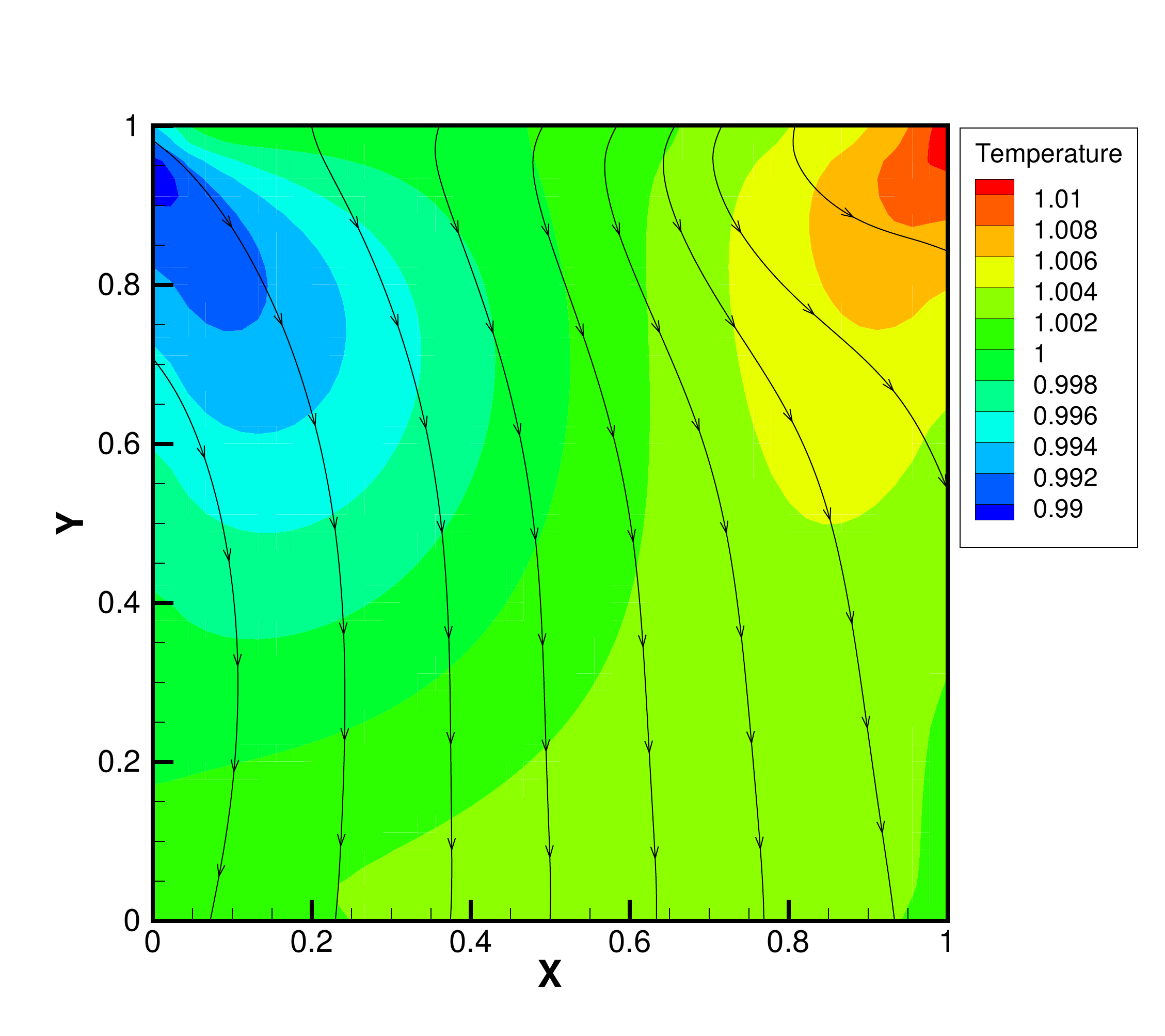}
	}
	\subfigure[$\phi_y=-0.5$,$Fr=0.21$]{
		\includegraphics[width=5cm]{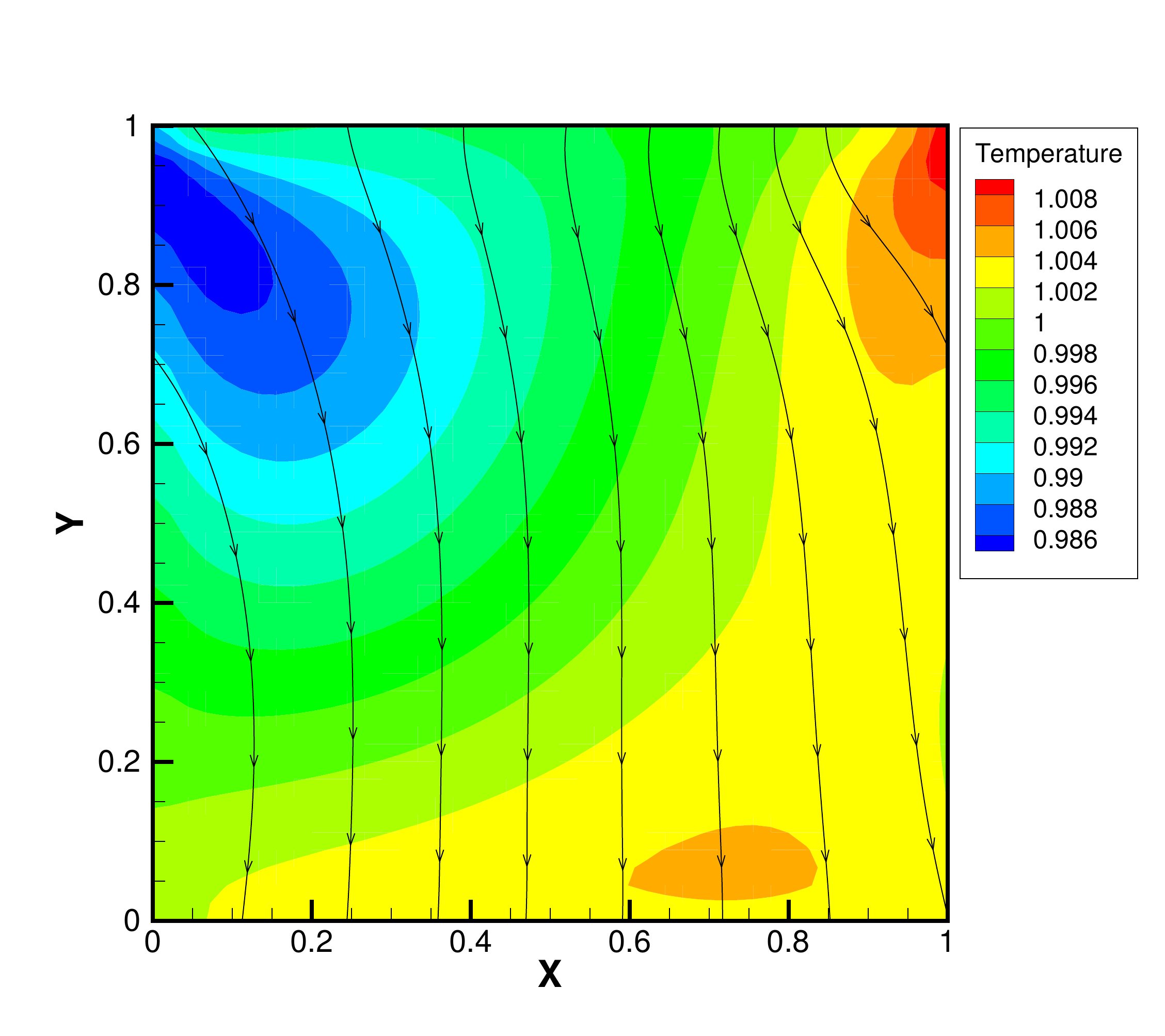}
	}
	\subfigure[$\phi_y=-1.0$,$Fr=0.15$]{
		\includegraphics[width=5cm]{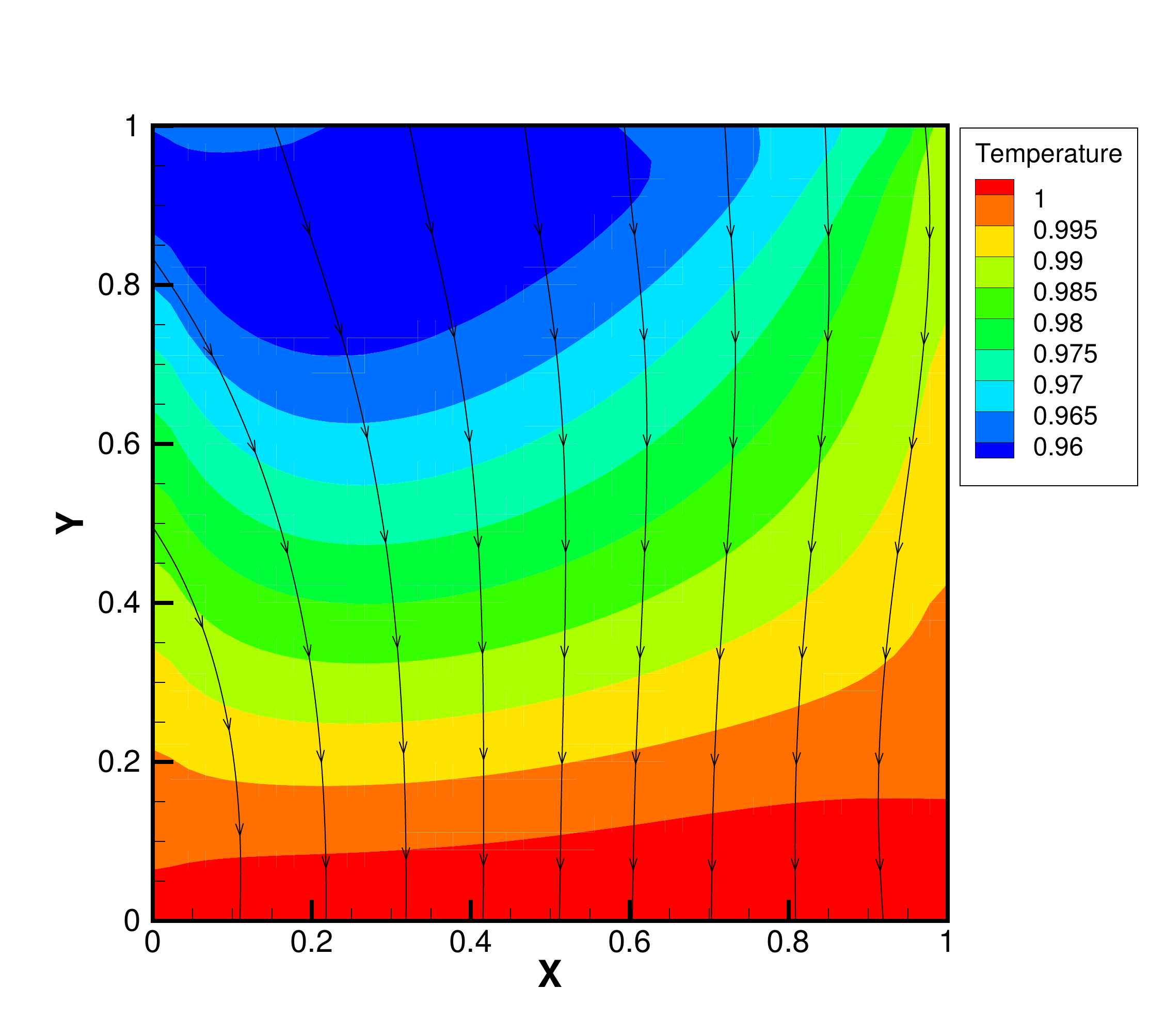}
	}
	\caption{Temperature contour and heat flux with $Kn_{ref}=0.075$.}
	\label{cavity heat transition}
\end{figure}

Even with the similar main vortex structure, in the transition regime the external force field
exerts a greater impact on the heat transfer process.
As presented in Fig. \ref{cavity heat transition}, in the absence of external force field,
the expansion cooling and viscous heating both have distinguishable contribution to the heat flux, which
presents a phenomena for the heat flow from the cold to hot region. This observation is consistent with the DSMC
simulation and unified scheme solution \cite{john2010investigation,liu2016unified}.
With the increment of the external force, the heat transfer gradually turns into the vertical direction along with the
forcing field.
The hot spot becomes wider along the vertical direction, while the cold region expands along the horizontal direction.
As demonstrated, even with the viscous heating from the isothermal upper wall,
the temperature decreases there due to the energy exchange among kinetic, internal and potential one, resulting in the cooling of the upper zone.

\begin{figure}[htb!]
	\centering
	\includegraphics[width=7cm]{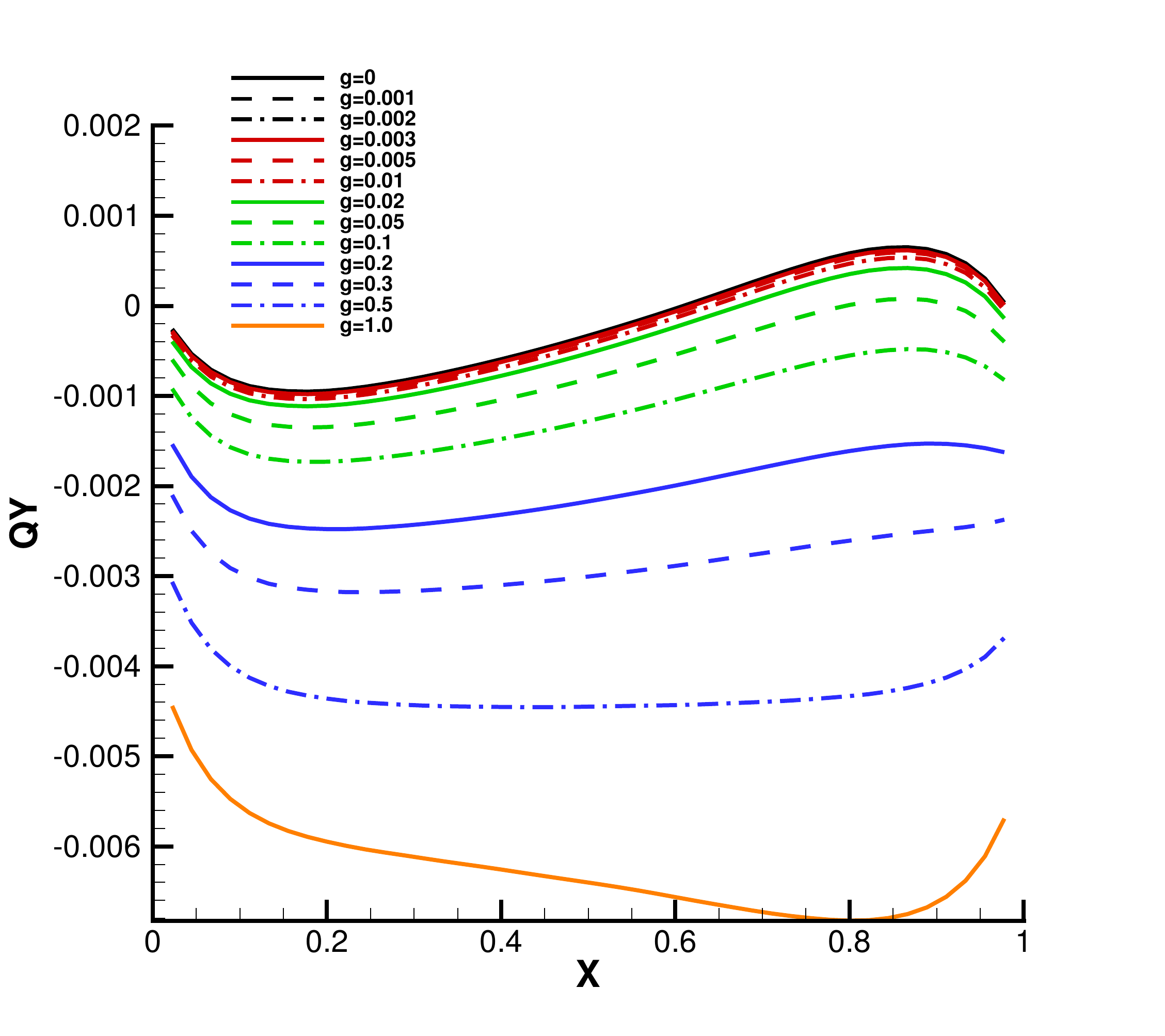}
	\caption{Heat flux distribution along the horizontal center line with $Kn_{ref}=0.075$.}
	\label{cavity qline transition}
\end{figure}

\begin{figure}[htb!]
	\centering
	\subfigure[Horizontal distribution of $q_y$ near the cavity center]{
		\includegraphics[width=7cm]{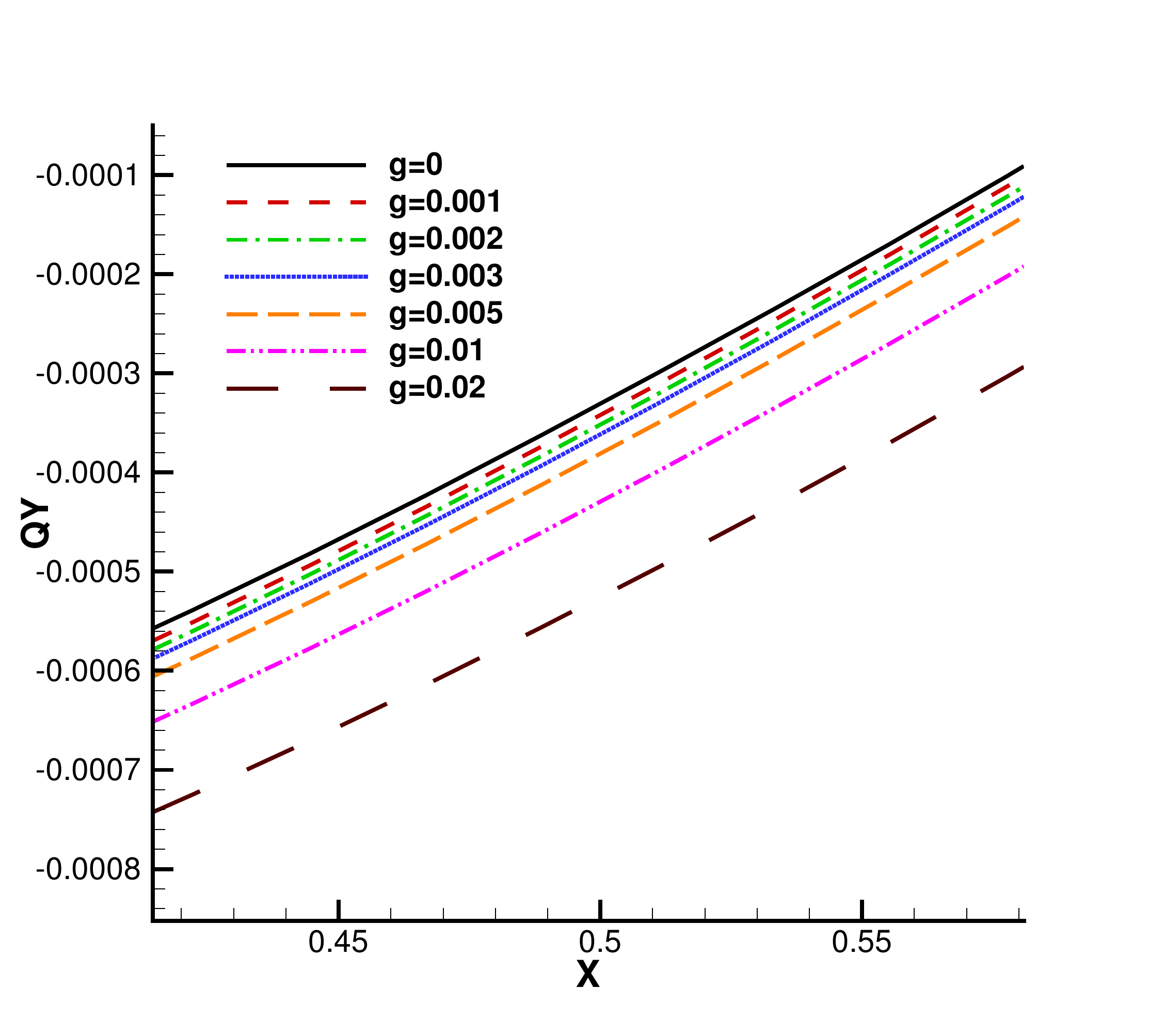}
	}
	\subfigure[Heat flux difference $q_y-q_0$ versus the external force $\phi_x$ at cavity center point]{
		\includegraphics[width=7cm]{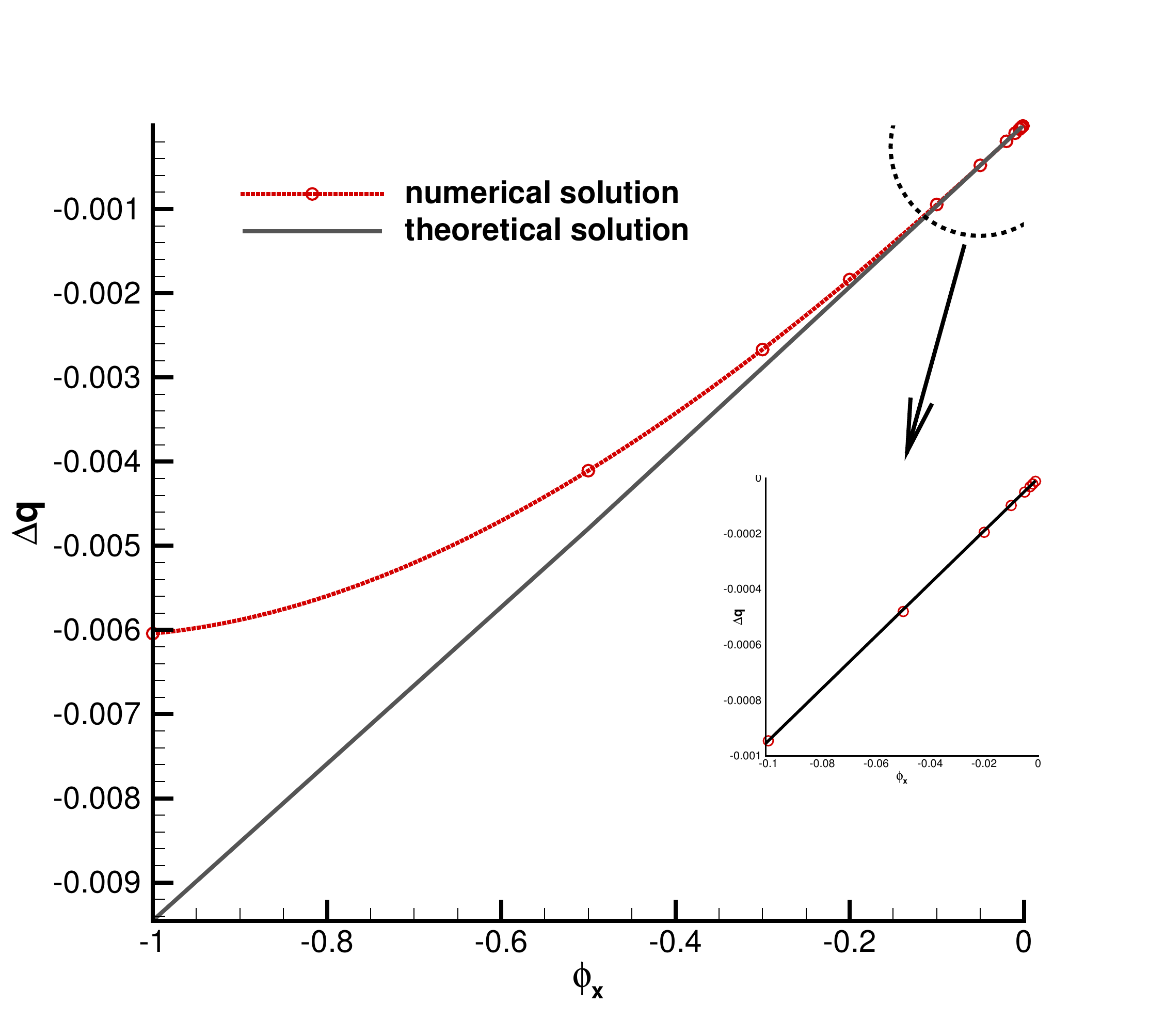}
	}
	\caption{Heat flux distribution near the cavity center with $Kn_{ref}=0.075$.}
	\label{cavity qline transition center}
\end{figure}

Here in Fig. \ref{cavity qline transition} we present the distribution of heat flux in $y$-direction along the horizontal center line.
Since there is no distinct temperature in the vertical direction near the center of the cavity, the modification of can be attributed mainly to the effect of external force field. 
It can be seen that with the increment of external force, the magnitude of heat flux $q_y$ increase as well, along the same direction of the force.
Fig. \ref{cavity qline transition center}a intercepts the horizontal distribution of $q_y$ near the cavity center.
It is obvious that when the magnitude of force is small, all the curves are nearly parallel with each other£¬ and the interval between two lines are determined by the magnitude difference of the force.
In Fig. \ref{cavity qline transition center}b, we plot the heat flux  $q_y-q_0$ at the cavity center $x=y=0.5$, where $q_0$ is the $y$-direction heat flux without external force field.
The numerical results are denoted by red circles, and the red line is the polynomial fitting curve of discretized results.
Based on the theoretical analysis in Section 2, we present in the black line as well, the theoretical modification of heat flux based on local collision time, density and temperature $\Delta q=C_q \tau \phi_y \rho T$,
where the physical parameter takes the same value with the static heat conduction case, such as $C_q=0.0113$.
It can be seen that in the linear region where the force is relatively small, these two solutions are consistent with each other.
With the increment of external force,
due to the vertical thermal stratification and strong non-equilibrium dynamics under the current Knudsen number, the modification of heat flux is no longer coming from the external force.
The flow pattern is totally changed from the no-force case.
In fact, at the current Knudsen number, the gas dynamics presents a
peculiar non-equilibrium manner.
For example, in the case with $\phi_y=-1.0$, the heat flux is almost parallel to the external force direction,
and the heat transport from the upper cold region to the bottom hot region in a non-Fourier way. 
This indicates the thermal instability caused by external force field, where the inhomogeneous distribution of temperature arises.
In the full transition regime, the enlarged degree of freedom for particle motion and strong non-equilibrium effect are expected to appear,
and the correlation of temperature gradient, stress tensor and external force should go beyond the linear theory given in Section 2.
In the transition regime, the external force may play a dominant role in the determination of non-equilibrium heat transport.

\section{Conclusion}

The gas dynamics under external force field is intrinsically a multiple scale flow problem due to large density variation
and a changeable local Knudsen number.
In this paper, based on the unified gas-kinetic scheme we investigate the non-equilibrium flow dynamics
under external forcing in different flow regimes.
For the near equilibrium flow, the contribution of the external force to the heat flux
is analyzed based on the kinetic model equation, and studied numerically as well.
At the same time, a detailed investigation for lid-driven cavity case has been conducted
and the non-equilibrium flow evolution has been quantitatively evaluated.
The dynamic effect of the external force on the flow pattern and heat transfer is presented.
Base on the direct modeling on the mesh size and time step, it is now possible to explore the physics in the non-equilibrium transition flow regime with the UGKS method.
Through the numerical experiments of UGKS, the proportionality between heat flux and external force is quantitatively confirmed. 
The enhanced heat transport from the forcing term may overtake the contribution from the temperature diffusion process, which
determines the heat flow from the upper cold high gravitational potential region to the lower hot low potential region and
triggers the gravity-thermal instability.
The understanding of the multiscale non-equilibrium flow phenomena under external force field will have great help to our understanding to large-scale atmosphere environment.

\section*{Acknowledgement}  
The current research  is supported by Hong Kong research grant council (16207715, 16211014, 16206617), and  National Science Foundation of China (11772281,91530319).

\section*{References}
\bibliography{tbxiao.bib}

\end{document}